\documentclass[%
 reprint,
nofootinbib,
 amsmath,amssymb,
 aps,
 prd,
]{revtex4-2}

\usepackage{graphicx}
\usepackage{dcolumn}
\usepackage{bm}

\usepackage{xcolor}

\begin{document}

\preprint{APS/123-QED}

\title{An analysis and cosmological study of the propagation of \\ gravitational radiation under a graviton of nonzero mass}

\author{Margaret Johnston}
 \email{margaret.johnston@unlv.edu}
\author{Marios Kalomenopoulos}
 \email{marios.kalomenopoulos@unlv.edu}
\author{Carl-Johan Haster}
 \email{carl.haster@unlv.edu}
\affiliation{University of Nevada Las Vegas, Department of Physics and Astronomy}
\affiliation{Nevada Center of Astrophysics}

\date{\today}

\begin{abstract}
Under the assumptions of General Relativity (GR), gravitational waves propagate at the speed of light and their mediation can be represented as a particle through a massless graviton. 
We investigate the impact and observability of the presence of a massive graviton, how such a modification to GR would also modify the propagation of observed gravitational waves from astrophysical sources, and how this effect can be used as an independent measurement of cosmological parameters, focusing on the Hubble parameter $H_0$ and matter energy $\Omega_m$. 
We simulate the impact of a massive graviton on compact binary coalescence observations in a near-future LIGO-Virgo-KAGRA interferometer network through a modification to the gravitational wave phase in the post-Newtonian framework. 
Our analysis finds that if we assume the presence of a graviton with a Compton wavelength of $\lambda_G \approx 5 \times 10^{16}$m, corresponding to a mass $m_G \leq 2.3 \times 10^{-23}$eV/c$^2$, we can utilize a simulated population of 60 binary black hole observations to constrain $H_0$ to a similar precision as current gravitational wave constraints without electromagnetic counterparts (at $90\%$ credible intervals): $H_0 = 58^{+34}_{-19}\,\mathrm{km\; s^{-1}\; Mpc^{-1}}$ and $\Omega_m=0.29^{+0.10}_{-0.08}$. 
More sensitive observatories will be necessary to probe lower values in the graviton mass range and fully exploit this method. 

\begin{description}
\item[Keywords]
cosmology; gravitational waves; post-Newtonian modification; graviton; modified gravity

\end{description}
\end{abstract}

\maketitle

\section{\label{sec:intro} Introduction}

During the past decade of observations, the LIGO
\cite{aLIGO:2020wna}, 
Virgo 
\cite{Virgo:2019juy}, 
and KAGRA 
\cite{KAGRA:2020tym}
(LVK) gravitational wave observatories have enabled new measurements of cosmological parameters which are independent from previous observational methods. 
LVK observes the coalescence of compact-object binaries, comprised of neutron stars or black holes, via their gravitational wave emission during the late stage of their inspiral and eventual merger
~\cite{LIGOScientific:2018mvr, LIGOScientific:2021usb, KAGRA:2021vkt}.
These observations have provided new opportunities to experimental cosmology through independent measurements of cosmological parameters and unprecedented insight into the population of gravitational wave sources in the Universe
\cite{LIGOScientific:2017adf, LIGOScientific:2021aug, KAGRA:2021duu}.
A gravitational wave observation - in the ideal case - can be leveraged to infer cosmological parameters if both a distance and redshift can be associated with its source
\cite{Schutz:1986gp, Holz:2005df}. 

The distance to the gravitational wave source can in principle be measured simply by comparing the amplitude of the strain data and a model waveform for the gravitational wave, calculated directly from the predictions of GR. 
However, the amplitude of the gravitational wave signal depends on both the distance to the source as well as the inclination of the orbital plane of the binary, and disentangling these degenerate contributions can be very difficult
\cite{Mastrogiovanni2021, Salvarese:2024jpq}. 
Under particularly favorable observing conditions this degeneracy may be broken, either by observing spin-orbit precession or by recovering higher-order multipole modes beyond the leading quadrupole contribution from the waveform
\cite{Veitch:2012df, London:2017bcn, Usman:2018imj, Chassande-Mottin:2019nnz}. 
Regardless of observing conditions, minimizing the impact of this inherent uncertainty is important for any observation. 
Because the distance to gravitational wave sources can be reliably measured, they are called ``standard sirens''
\cite{Schutz:1986gp}, 
analogous to ``standard candles'', the variable stars and supernovae which are used to calibrate distances to other galaxies by leveraging their consistent intrinsic brightness
\cite{Riess:2021jrx}. 

The astrophysical parameters describing a compact-object binary -- such as the masses and spins of the binary components, their distance and orientation -- are inferred by comparing the GW data to a set of waveforms, and how well those waveforms match the signal \cite{Sathyaprakash:1991mt, Allen:2005fk, LIGOScientific:2016vbw, Veitch:2014wba, Romero-Shaw:2020owr}.
Additionally, the gravitational wave signal can be decomposed into separate polarization states, where GR only supports the two tensorial polarizations, plus and cross. 
Observing both polarization modes of the signal from a source allows the inclination to be measured more precisely as the two modes have different inclination-dependent amplitude reduction. 
Any amplitude reduction not attributable to the inclination of the binary orbit can therefore be assigned to the distance
\cite{Mastrogiovanni2021}. 

Under GR, the redshift of an individual astrophysical binary source cannot in general be directly measured from a gravitational wave observation, unless additional information about the compact objects' mass or internal structure are taken into account
\cite{Taylor:2011fs, Messenger:2011gi, Messenger:2013fya, DelPozzo:2015bna}.
A more common approach is to use complementary electromagnetic observations, either with direct observation of an electromagnetic counterpart or by utilizing a galaxy catalog. 
If the binary has an electromagnetic counterpart, it is called a ``bright siren''
\cite{Holz:2005df, Schutz:1986gp}
and the EM signal can be used to identify the host galaxy and thus the redshift of the associated GW source. 
Such is the case for the binary neutron star (BNS) GW170817 
\cite{LIGOScientific:2017vwq}. 
At the time of detection, GW170817 was localized to a region of just 31 deg$^2$, allowing for extensive, multiwavelength electromagnetic follow-up and identification of the kilonova AT2017gfo
~\cite{LIGOScientific:2017ync}. 
The binary was inferred to be located relatively close to Earth, with a luminosity distance  $D_L = 43.8^{+2.9}_{-6.9}$ Mpc (maximum posterior value and minimal width $68\%$ credible interval, unless stated otherwise), and was the loudest gravitational wave signal detected to that point. 
The constraint on the Hubble parameter $H_0 = 70^{+12}_{-8}$ km s$^{-1}$ Mpc$^{-1}$ from GW170817 is the strongest cosmological constraint attained from a single GW source so far~\cite{LIGOScientific:2017adf}. 

The inclination may also be constrained from an electromagnetic counterpart if the structure of the source is known. 
Certain cataclysmic systems, including kilonovae, can produce narrow jets which may be identified in electromagnetic observation of the afterglow emission
\cite{Metzger:2011bv, Heinzel:2020qlt, Holmbeck:2023gck}. 
Magnetars -- neutron stars with particularly strong magnetic fields -- have been observed to produce narrowly beamed and short-lived $\gamma$-ray bursts
\cite{Berger:2013jza, Minaev:2024hvi}, 
which can be observed from distant galaxies as well as within the Milky Way. 
GW170817, also observed in $\gamma$-rays as GRB 170817A, confirmed the connection between BNS mergers as the progenitors of short, or type 1, GRBs
\cite{LIGOScientific:2017zic, Lv:2010bz}. 
If these distinctive features are observed, the inclination of the binary orbit may be better constrained, although it is necessary to properly model the generation of the relevant emission components as the predicted opening angles of jets, distribution of expelled material, etc. differ between models and are subject to large uncertainties
\cite{Guidorzi:2017ogy, Chen:2020dyt, Escorial:2022nvp, Palmese:2023beh, Gianfagna:2023cgk, Chen:2023dgw, Muller_Mukherjee_Geoffrey_2024}. 
Moreover, peculiar velocities could affect the redshift determination, although their impact is expected to be smaller \cite{Nicolaou_et_al_2020, Mukherjee_et_al_2021}.

Assigning redshifts to gravitational wave sources via their electromagnetic counterparts is straightforward, and has been successful in producing the most precise single source constraint on $H_0$ \cite{LIGOScientific:2017adf}. 
However, only one such event has been observed so far and this method by construction cannot be applied to the vast majority of the observed LVK sources which as binary black hole systems (BBH) are not expected to produce any observable electromagnetic counterpart\footnote{See~\cite{Graham:2020gwr, Chen:2020gek, Mukherjee_GW190521_2020} for a BBH with a potential counterpart, and its use as a cosmological probe.}. 
While this method is limited by the rarity of bright sirens observed thus far, its precision will continue to improve along with the growth of the population of sources observed with both gravitational and electromagnetic radiation
\cite{Chen:2024gdn}. 

In the more common case where the binary has no electromagnetic counterpart, a ``dark siren'', the sky localization and measured distance for the binary can be used in conjunction with galaxy catalogs to determine a redshift for the system 
\cite{Schutz:1986gp}. 
A localization volume is determined from the distance and sky localization inferred for each GW source, then galaxy catalogs, such as the GLADE catalog
\cite{Dalya:2018cnd}, 
are consulted to determine the galaxies within the localization volume and their associated probability of being the host galaxy. 
Statistically marginalizing over the redshifts for the galaxies within the localization volume yields information on the cosmological parameters
\cite{Schutz:1986gp, DelPozzo:2018dpu, DES:2019_HO_DarkSirens, Bera:2020_DarkSirens_CrossCorr, Mukherjee:2020_DarkSirens_Clustering, LIGOScientific:2021aug,  Palmese:2021_H0_DarkSirens_O3}. 
This method improves with more precise sky localization from gravitational wave detection 
\cite{Mo:2024frl, Mo:2024uim} 
and more complete catalogs. 
Similarly, cosmological constraints could be derived by cross-correlating GW observations with the large-scale structure of galaxies 
\cite{Mukherjee_cross-corr_2024}.
By construction, these methods of determining redshifts incorporate any selection biases from the observational methods which contribute to galaxy catalogs
\cite{Gair:2022zsa}. 
In the case of the GLADE galaxy catalog, created to support gravitational wave observation and to ease identification of host galaxies for gravitational wave sources, the Milky Way's galactic plane obscures background sources resulting in significantly sparser coverage in those regions of the sky
\cite{DelPozzo:2018dpu}. 

This work investigates a new method for measuring redshifts for dark sirens by exploiting a deviation from GR in the form of a modification to the post-Newtonian expansion by which GR can be represented. 
The propagation of gravitational waves can be described with a particle mediator of gravity, the graviton, which in GR is expected to be massless as gravitational waves are predicted to travel at the speed of light
~\cite{Will:1997bb, LIGOScientific:2017zic, Mastrogiovanni2021}.
If we instead live in a universe where the graviton has a small nonzero mass, it will introduce an additional term to the gravitational wave phase which depends on the path length of the gravitational wave emission from the observed binary to the detector, described in detail in Sec. \ref{sec:massive_graviton}. 
Measurement of this effect would allow a redshift to be determined for the event using only gravitational waves observations, without requiring an electromagnetic counterpart or a complementary galaxy catalog. 
This method of redshift measurement would provide new constraints on cosmological parameters, completely independent from existing methods. 

\section{\label{sec:massive_graviton} The Massive Graviton}
To investigate the effects of a graviton of nonzero mass, we introduce a modification to the first order term of the post-Newtonian expansion (1PN)
\cite{Will:1997bb}. 
The post-Newtonian expansion is an analytic approximation of Einstein's field equations in powers of $v/c$, where $v$ is the orbital velocity of the binary source and $c$ the speed of light. 
This expansion yields accurate predictions of the behaviour and evolution of the inspiral of compact-object binaries under gravitational wave radiation and is one of the most frequently used methods for calculating gravitational waveforms due to the computational efficiency and accuracy of the predictions
\cite{Blanchet:2013haa}. 

The modification to the 1PN term, which describes the massive graviton, results in a phase shift $\Psi_{G}$, which modifies the gravitational wave signal. 
The leading order term of this phase shift is given by 

\begin{equation} \label{eq:psi_G}
    \Psi_{G} = -\frac{\pi^{2}D_{\Lambda}\mathcal{M}}{\lambda_{G}^{2}\left(1+z_G\right)}\frac{G}{c^2}u^{-1}
\end{equation}
where $z_G$ is the redshift of the system, $\mathcal{M} = (m_1 m_2)^{3/5} (m_1 +m_2)^{-1/5}$ is the detector-frame chirp mass, with $m_{1,2}$ being the masses of the two binary components
\cite{Will:1997bb}. 
$D_{\Lambda}$ is a redshift-dependent distance measure which we describe below. 
The redshift derived from the effects of the graviton, $z_G$, is here assumed to be an independent parameter separate from the default redshift determined during parameter estimation (which usually is found assuming a cosmological model) and is used to constrain cosmological parameters for this work. 
The frequency dependence of the term is carried by the characteristic velocity $u=\pi G \mathcal{M}f/c^3$, and the graviton Compton wavelength given by $\lambda_G=\frac{h}{m_{G}c}$
with $m_G$ being the graviton's mass.

The cosmological model with the greatest support from a wide set of observations across astrophysics is $\Lambda$CDM
\cite{Grohs:2023voo, Planck:2018vyg, Riess:2021jrx}
which describes the recent evolution of the Universe according to the dominant sources of energy density; the cosmological constant $\Lambda$ and dark matter. 
Over cosmic time, the composition of the energy density of the Universe has changed, originally being radiation dominated, but at present the Universe is $\Lambda$-dominated with a subdominant matter contribution
\cite{Planck:2018vyg}. 
We will be using this model as default, and specifically $(H_0, \Omega_m = (67.9 \mathrm{km\; s^{-1}\; Mpc^{-1}}, 0.3065)$ for our simulated population, and for comparing our cosmological inference. 
For more details on the latter, we refer to Sec. \ref{subsec:bilby}.

Accounting for a $\Lambda$CDM universe, with the presence of both $\Lambda$ and matter, a redshift dependent distance measure can be constructed (following \cite{Will:1997bb}) as 
\begin{equation} \label{eq:d_measure}
    D_{\Lambda} = c(1+z_G) \int_0^{z_f} \frac{dz}{(1+z_G)^2 H(z_G)},
\end{equation}
with the Hubble rate given by 
\begin{equation}
    H(z_G) = H_0 \sqrt{\Omega_m (1+z_G)^3 + \Omega_{\Lambda}},
\end{equation}
where $H_0$ is the Hubble rate at $z=0$, $\Omega_m$ is the energy density of matter in the Universe and $\Omega_{\Lambda}=(1-\Omega_m)$ is the energy density corresponding to the cosmological constant. 
The two terms sum to unity as we assume a flat geometry. 
This distance measure is distinct from the luminosity distance recovered during parameter estimation and from other distance measures conventionally used in cosmology. 

Our treatment of $D_{\Lambda}$ is modified from the original study by Will 
\cite{Will:1997bb} 
which primarily assumes a matter dominated universe, and hence results in a different distance measure used in their equivalent to Eqn.~\eqref{eq:psi_G}. 
Other cosmological measurements 
\cite{Grohs:2023voo, Planck:2018vyg, Blanchard:2022xkk}
have ruled out a presently matter dominated universe. 
The choice of the dominant energy density has implications for the range of redshifts over which this method could produce astrophysically significant constraints. 

In the local Universe, $z \ll 0.1$, the dominant energy density of the cosmological constant, has little effect on the calculation of $H_0$.
However, even at modest redshifts, $z \approx$ 0.3, a matter-dominated universe will underestimate $H_0$ around 12\% compared to a $\Lambda$CDM universe, $\Delta H_0/H_0^{\Lambda {\rm CDM}} \sim 0.12$. 
These continue to diverge more severely as redshift increases, making the choice of cosmology a significant driver of bias. 
By assuming a $\Lambda$CDM universe, we can appropriately describe the resulting phase shift for the universe we observe.

We model any deviations from GR using the parametrized post-Einsteinian (ppE) framework 
\cite{Yunes_Pretorius_ppE_2009}, 
where the first PN ppE parameter is equal to $\Psi_G$ in Eqn. (\ref{eq:psi_G}), and all other ppE parameters are equal to zero. 
Following 
\cite{Agathos_et_al_TIGER_2014, Zimmerman:2019wzo}, 
we define the phase shift as a ratio of the ppE deviation over the equivalent GR PN term. Hence, the final expression for the phase shift introduced by the graviton mass is given by 
\begin{equation} \label{eq:dchi2}
    d\chi_{2}=\frac{\Psi_{G}}{\frac{5}{96}\left(\frac{743}{336}+\frac{11}{4}\eta\right)\eta^{-2/5}u^{-1}}
\end{equation}
where $\eta = m_1 m_2 (m_1 + m_2)^{-2}$ 
is the symmetric mass ratio of the binary
~\cite{Will:1997bb, Zimmerman:2019wzo}. 
This term is equivalent to $\delta \hat{\phi}_2$, as defined in 
\cite{LIGOScientific:2021sio}, 
where the general deviation $\Delta \phi$ in the numerator is substituted with $\Psi_G$.
By measuring $d\chi_2$, we can in turn place constraints on $\lambda_G$, $z$, and $\Omega_m$ as part of a parameter estimation analysis to infer the binary source parameters.

It is this modification that we incorporate into our analysis of both simulated and real gravitational wave events, as described in Sec. \ref{sec:methods}.

\section{\label{sec:mG_constraints} Constraints on \\the mass of the graviton}
While current experiments and observations are unable to provide conclusive evidence in support of a graviton with finite mass, or even whether such a mediator particle is required, they can be used to place meaningful limits on its properties.

There exists a lower limit on the value of $\lambda_{G}$ determined by comparing observations of the solar system with predictions from Kepler's third law. 
The precision with which massive bodies in the solar system can be located allows a constraint of $\lambda_G > 1.2 \times 10^{17}$m to be placed on the graviton Compton wavelength
\cite{Will:2018gku}. 

There exists a strong constraint on the agreement between the speed of light and the speed of gravity from multimessenger coordination and precise timing. 
The near-simultaneous measurement of a binary neutron star merger in gravitational waves and prompt emission in gamma rays, cataloged as GW170817 and GRB170817A respectively, provides a uniquely strong constraint on the speed of gravity
\cite{LIGOScientific:2017zic}. 
These two messengers were detected only 1.7 seconds apart, after having traversed a distance of $\sim 40 \; \mathrm{Mpc}$~\cite{Cantiello:2018ffy}. 
Ignoring any physical mechanism that might introduce a delay between the first instant of contact of the neutron stars' surfaces and the eventual emission of the gamma rays, this pair of observations limit the difference between the speed of light and the speed of gravity to be at most of order one part in $10^{15}$, suggesting that the graviton must be on the same mass scale as the photon. 

Observations from the third catalog of gravitational wave transient sources, GWTC-3
\cite{KAGRA:2021vkt}, 
also contributes a constraint on the graviton of $m_G \leq 1.27 \times 10^{-23}$ eV/c$^2$, corresponding to $\lambda_G \geq 9.22 \times 10^{16}$ m
\cite{LIGOScientific:2021sio}.

Informed by these constraints, we probed a wide range of values of $\lambda_{G}$ to determine what effects might be detectable with a near-future LVK observatory network. 
We found that at smaller values, $\lambda_G \leq 10^{14}$ m, the phase shift produced is sufficiently large to violate our initial assumptions of a small perturbation to the underlying GR signal. 
At large values, $\lambda_G \geq 10^{17}$m, the phase shift produced is too small to be detectable for an individual system given current detector sensitivities, and thus it is very difficult to constrain any parameters of the system related to the introduced phase shift. 
Therefore, to simplify our analysis, we adopt a fiducial value of $\lambda_{G} \approx 5 \times 10^{16}$m, corresponding to $m_G \approx 2 \times 10^{-23}$eV/c$^2$, for this work. 
Although this value is just ruled out by recent observational constraints, as expanded on in Sec. \ref{sec:discussion}, our study highlights the success of our proposed method, and how it can be implemented for future work operating on a larger catalog of observed gravitational wave signals. 

Additionally, there exists a recent constraint specifically on the 1PN term from the population described in GWTC-3. 
Of interest to this work, GW200115~\cite{LIGOScientific:2021qlt}, consistent with a neutron star-black hole merger, is the single event which contributes the individually most stringent constraint on this term
\cite{LIGOScientific:2021sio}. 
We address this system separately, reanalyzing the event using a configuration similar to the one used for the analysis from GWTC-3~\cite{KAGRA:2021vkt, ligo_scientific_collaboration_and_virgo_2021_5546663}, with the addition of the massive graviton effects described here. 
This is done to probe the capabilities of the current set of gravitational wave observations to contribute a constraint on cosmological parameters via this method. 

\section{\label{sec:methods} Methods}

\subsection{\label{subsec:bilby} Astrophysical inference using \textsc{Bilby}}

We utilize the Bayesian inference package, \textsc{Bilby} 
\cite{Ashton:2018jfp, Romero-Shaw:2020owr}, 
to simulate a population of binary black hole systems, the emitted gravitational wave signals, and their expected detectability with near-future configurations of the current ground-based observatory network. 

As seen in the expression for the leading order term of the phase shift, Eqn. \eqref{eq:psi_G}, the result of a graviton of non-zero mass is an accumulated propagation effect. 
Where most previous analyses
\cite{LIGOScientific:2021sio} 
measure graviton effects through inference on the more general $d\chi_2$ parameter, from which model-specific constraints can be derived, our analysis instead models these effects directly using three additional parameters to be estimated; $\lambda_G$, $z_G$, and $\Omega_m$. 
With these parameters, we can implement the phase shift described in Eqn. \eqref{eq:dchi2} and measure the redshift therein, resulting from the effect of the massive graviton, separately from the ``default redshift'' that is commonly derived from the inferred luminosity distance under an assumed cosmology.

We simulate a population of binary black hole (BBH) signals, under the assumption that their propagation is affected by a graviton with mass $m_G \approx 2 \times 10^{-23}$eV/c$^2$.

First, a simulated population consistent with the standard priors used in parameter estimation analysis
\cite{Romero-Shaw:2020owr} 
shown in App. \ref{sec:app_priors} is generated. 
The luminosity distance of a system is chosen consistent with the prior, where the corresponding value of $z_G$ can be calculated assuming 
a cosmology with $H_0=67.9\,\mathrm{km\; s^{-1}\; Mpc^{-1}}$ and
$\Omega_m=0.3065$~\cite{Planck:2015fie}. 
While this choice of cosmology does impose some specific assumptions of the Universe, we do not expect this cosmology to have any distinct advantage over any other cosmology that could have been assumed.

Then, for each generated system, a signal to noise ratio is calculated as a proxy for their observational significance. 
Those systems in the simulated population which have SNR$>$10 are included in our analysis, as they would reasonably be detectable by the network under observing conditions. 
From the initial 1500 systems generated for the simulated population, 152 met the SNR threshold, and 60 of those were analyzed as described in this work. 
We analyze the population of detectable signals using the \textsc{Bilby} parameter inference as observed using a network consisting of LIGO-Hanford, LIGO-Livingston, Virgo, KAGRA, and LIGO-Aundha (HLVKA) and utilizing the improved A+ sensitivity curves
\cite{Miller:2014kma}. 

We utilize the \texttt{IMRPhenomXPHM} waveform for its incorporation of higher-order polarization modes beyond the dominant quadrupole, spin-induced precession of the orbital plane, and its computational efficiency 
\cite{Pratten:2020ceb, Pratten:2020fqn, Garcia-Quiros:2020qpx}. 
Using the HLVKA detector network defined above, we infer the posterior probability distributions describing the source parameters for each event. 
By using this detector network, we can exploit the detection of differing mixes of the two tensor polarization modes, longer network baselines, and higher SNR for each event as compared to a smaller detector network. 
These advantages from the HLVKA network yield more precise parameter estimation, resulting in tighter sky localizations and improved ability to disentangle the distance-inclination degeneracy. 

The formulation of $D_{\Lambda}$ given in Eqn.~\eqref{eq:d_measure} is computationally prohibitive to use with the large number of likelihood evaluations generally necessary to robustly estimate the parameters of the analyzed systems. 
We instead implement an analytical approximation that produces a substantial computational speedup (more than $20$ times faster than the numerical computation of the integral) while not contributing significantly to the error  
of our measurement of $H_0$\footnote{The errors on the distance calculation versus the numerical results are below $1 \%$ for our prior values of $\Omega_m$. In general, the errors reduce as $\Omega_m$ increases, with a perfect fit as $\Omega_m \rightarrow 1$. These errors are well below the statistical uncertainties in the parameter estimation analysis.}. 
By employing a Pad{\'e} approximant 
\cite{wickramasinghe:2010, adachi:2012}, 
we rewrite $D_{\Lambda}$ as
\begin{align}\label{eq:d_measure_pade}
    &D_{\Lambda}=\left(\frac{2c \left(1+z_G\right)}{H_0\Omega_{m}^{1/2}}\right) \times \\
    &\left(\Phi\left[\frac{1-\Omega_{m}}{\Omega_{m}}\right]-\Phi\left[\frac{1-\Omega_{m}}{\Omega_{m}\left(1+z_G\right)^{3}}\right]\left(\frac{1}{1+z_G}\right)^{5/2}\right), \nonumber
\end{align}
where 
\begin{equation}
    \Phi(y)=\frac{1}{2}\left(\frac{0.4+b_{1}y+b_{2}y^{2}+b_{3}y^{3}}{1+c_{1}y+c_{2}y^{2}+c_{3}y^{3}}\right)
\end{equation}
and the constant coefficients are given by 
\begin{equation}
    \left[\begin{array}{c}
b_{1}\\
b_{2}\\
b_{3}\\
c_{1}\\
c_{2}\\
c_{3}
\end{array}\right]=\left[\begin{array}{c}
0.50241404\\
0.15216594\\
0.00656022\\
1.48330782\\
0.60723615\\
0.05874342
\end{array}\right].
\end{equation}

A similar formula can be derived for the luminosity distance $D_L$:
\begin{align}\label{eq:dL_measure_pade}
    &D_{L}=\left(\frac{2c \left(1+z_G\right)}{H_0\Omega_{m}^{1/2}}\right) \times \\
    &\left(\Phi_L\left[\frac{1-\Omega_{m}}{\Omega_{m}}\right]-\Phi_L\left[\frac{1-\Omega_{m}}{\Omega_{m}\left(1+z_G\right)^{3}}\right]\left(\frac{1}{1+z_G}\right)^{1/2}\right) \nonumber
\end{align}
where 
\begin{equation}
    \Phi_L(y)=\frac{1}{2}\left(\frac{2+b^L_{1}y+b^L_{2}y^{2}+b^L_{3}y^{3}}{1+c^L_{1}y+c^L_{2}y^{2}+c^L_{3}y^{3}}\right)
\end{equation}
and the constant coefficients are given by 
\begin{equation}
    \left[\begin{array}{c}
b^L_{1}\\
b^L_{2}\\
b^L_{3}\\
c^L_{1}\\
c^L_{2}\\
c^L_{3}
\end{array}\right]=\left[\begin{array}{c}
2.64086441\\
0.8830444\\
0.05312495\\
1.39186078\\
0.51209467\\
0.03943821
\end{array}\right].
\end{equation}

A ratio of these distance measures will hence be independent of $H_0$ and is given by
\begin{equation}
    \frac{D_{\Lambda}}{D_L} = \frac{\Phi(x_{0})-\Phi(x)/(1+z)^{5/2}}{\Phi_L(x_{0})-\Phi_L(x)/(1+z)^{1/2}},
\end{equation}
with $x = x (z, \Omega_m) = (1-\Omega_m)/(\Omega_m (1+z)^3)$ and $x_0 = x(0, \Omega_m)$. This formula allows fast computation of the distance measure from the $D_L, z$ and $\Omega_m$ samples, during parameter estimation, in order to be used in the $d\chi_2$ calculation in Eqn. (\ref{eq:dchi2}).

In the limit of $\Omega_m \rightarrow 1$, this leads to the standard formula \cite{Will:1997bb, Zimmerman:2019wzo}:
\begin{equation}
    \frac{D_{\Lambda}}{D_L} = \frac{1+(2+z)(1+z+\sqrt{1+z})}{5 (1+z)^2}.
\end{equation}

\subsection{\label{subsec:priors} Priors}
In this work, we use the standard \textsc{Bilby} priors for the majority of the parameters estimated for the systems we analyze 
\cite{Ashton:2018jfp, Romero-Shaw:2020owr}. 
The exception to this is the luminosity distance $D_L$ for which a quadratic power law is assumed for the prior as it is both simple, and therefore easy to remove from the analysis later, and captures the general expectation that the number of sources will be more abundant at larger redshifts. 
For the parameters we introduced to describe the effect of the graviton mass, $\lambda_G$, $z_G$, and $\Omega_m$, we employ a log-uniform prior for the first and uniform prior distributions for the other two, so as to remain agnostic to any particular cosmological model. 
A full description can be found in App. \ref{sec:app_priors}. 

\subsection{\label{subsec:cosmology} Cosmological Measurement}

Traditional cosmological measurements derived from distance and redshift take advantage of the independence of the two measurements. For gravitational wave events which have an electromagnetic counterpart, such as GW170817, the measurement of distance and redshift are independent owing to the differing observational mechanisms and the distinct data and analysis pipelines used in the respective measurements. For events which have no counterpart, the GWs strain data in conjunction with galaxy catalogs is used to probabilistically assign a redshift, which again have separate contributing observatories and biases. 

With our analysis, we are deriving both the distance and redshift from a single gravitational wave signal. 
Therefore, we must pay particular attention to understanding the dependencies between the two measurements during the parameter estimation process as they are subject to the same potential observatory and pipeline systematics. 
However, because we utilize different portions of the signal to determine the two quantities, with $D_L$ inferred solely from the amplitude and $z_G$ solely from the phase, and they are handled consistently during parameter estimation, we can treat these parameters as independent and therefore correlated only through the cosmology we are trying to measure. 
We confirm this with the analysis of the posterior samples. A representative set of inferred parameters is shown in App. \ref{sec:app_population_PE}. 

We perform the cosmological analysis at post-processing, using the PE samples of $z_G,\ \Omega_m$ and $D_L$. Each sample draw $S$ has the triplet ($z^S_G, \Omega^S_m$, $D^S_L$). When considering the full population of observations, the impact of the single-event prior probabilities in PE will need to be corrected for before combining them into a single constraint \cite{Zimmerman:2019wzo}. In order to determine the likelihood distribution from the luminosity distance samples, the posterior is resampled to be consistent with a uniform prior for the luminosity distance to account for the quadratic power-law prior used in parameter estimation. For the other variables of cosmological interest, the priors were uniform in their range (see App. \ref{sec:app_priors}), so we use them directly. 

For a flat, $\Lambda$CDM universe, the luminosity distance is given by
\begin{equation}
    D_L(z) = c (1+z) \int_0^z \frac{dz'}{H_0 \sqrt{\Omega_m(1+z')^3+1-\Omega_m}}.
\end{equation}
Therefore each sample triplet leads to a precise $H^S_0$ determination, i.e. a specific sample value of $H_0$. Therefore, for each simulated system, we keep $N_s\ H_0$ samples. 
For $H_0$ we assume a prior range in [$20, 120$] $\mathrm{km\; s^{-1}\; Mpc^{-1}}$. 
For each event, we keep $N_s=2000$ samples as the default \footnote{We investigated the effect of sample size, and for $N_s>2000$ our results were converging to the same distributions. Moreover, we examined the effect of random seed when drawing the samples with the distribution being robust irrespective of choice.}. 
We fit the $H^S_0$ samples with a gaussian Kernel Density Estimator (KDE) \cite{ScottKDE, SilvermanKDE, Scipy} and obtain the $H_0$ likelihood for each event. KDEs are known to be sensitive to the choice of bandwidth (BW) \cite{BWSelection_Turlach, BWSelection_Bashtannyk_Hyndman}. To avoid an arbitrary selection and achieve a better bias-variance trade-off, we optimize the BW using K-fold cross-validation \cite{KFoldCV, StatisticalLearning, scikit-learn}, with $K=5$ and with a test dataset of size $20\%$ of $N_s$. This procedure produces an array of optimised BWs, one for each event, for their respective $H_0$ samples. A similar procedure is followed for $\Omega_m$, with the only difference that we use the \emph{bounded} KDE options of \verb|PESUMMARY| \cite{PESUMMARY}, which allow an accurate density estimation on samples bounded in a given range.

Lastly, once an $H_0$ likelihood is calculated for each simulated event, these individual constraints are multiplied together to derive a joint, population-level constraint \cite{Mastrogiovanni2021}, 
\begin{equation}
    p(H_0|D) \sim \pi(H_0) \prod_i^{N_s} \mathcal{L}(D|H_0),
\end{equation}
with $\pi(H_0)$ our prior, and $\mathcal{L}$ the KDE likelihood estimated for each event.

Finally, using the joint samples of ($H_0, \Omega_m$), we also perform a 2D KDE fit for each event for simultaneous constraints on both parameters. Here, we modify slightly the default procedure of \verb|Scipy|, which employs a 2D gaussian kernel with a BW proportional to the covariance of the two datasets
\begin{equation}
    \hat{f}(\mathbf{x}) \sim \exp \left(-\frac{1}{2}(\mathbf{x}-\mathbf{x}_i)^T \mathbf{H}^{-1}(\mathbf{x}-\mathbf{x}_i) \right),
\end{equation}
where $\mathbf{x}$ is the grid point to estimate the density, $\mathbf{x}_i$ is an observed datapoint/sample, and $\mathbf{H}$ is a bandwidth matrix proportional to the covariance matrix of the dataset $\mathbf{H} \sim \mathbf{\Sigma}$. As a result, this method can be quite sensitive to outliers. Indeed, for some events we have extreme $H_0$ samples, of the order of $\mathcal{O}(1000)$ $\mathrm{km\; s^{-1}\; Mpc^{-1}}$. To overcome this limitation, we develop a method where we treat the two datasets as uncorrelated, i.e. setting the inverse covariance matrix between them as a diagonal one\footnote{We confirm that this procedure is robust by a number of tests: 1) we calculate the distribution of the \emph{Pearson correlation coefficient} with all the samples and with samples within our prior ranges for ($H_0, \Omega_m$) and observe that the omission of the outliers samples returns a distribution of correlations consistent with zero, 2) we compare the 1D KDE fits, with their respective equivalents after marginalizing the 2D KDE fit, and we verify that these coincide only in the case where the 2D correlations are removed.}. Our final 2D KDE uses the cross-validated BWs for each event and a bounded fit for the $\Omega_m^s$ samples.

By taking the product of all the event-level constraints, we recover the probable range of $H_0$ and ($H_0, \Omega_m$) which is simultaneously consistent with all the events in the population.

\subsubsection{Selection Effects} 

Selection effects can be crucial when analyzing GWs observations \cite{Messenger:2012_SelEf, Mandel:2018_SelEf, Vitale:2020_selection_effects_review}. There are many physical mechanisms that can impact the SNR of a GW signal. Higher mass systems have larger amplitudes, but are more short-lived than lower-mass systems for binaries at the same distances. Binaries with orbits that are viewed edge-on will have smaller signal amplitudes, but their inclinations may be well constrained if spin-orbit precession is observed. In general, selection effects penalize parameters related to the probability of detection of the signal we are considering. 

For a very basic example, in the dark sirens scenario, one considers which galaxies enter the sky localization volume. This is obtained by starting from galaxy redshifts and calculating their distances based on cosmological parameters. For a given distance threshold, i.e. a maximum distance that the detectors would be able to observe a signal, some galaxies would not be observable depending on the value of the cosmological parameters \cite{Gair:2022zsa}. For example, a large value of $H_0$ would make the distance of a galaxy smaller, and could in principle make it detectable. In other words, for larger $H_0$ more galaxies would be detectable, and contribute to the $H_0$ inference. To correct for this extra contribution, selection effects are considered.

In our case, the cosmological analysis does not require any external information, and the input sources are all individually detected. Moreover, the prior ranges of distances and redshifts have been chosen to be compatible with the $H_0$ range we consider, and samples leading to $H_0$ values outside the prior range are eventually discarded. For this reason, we are not including any additional selection effects in this work.






\section{\label{sec:results} Results}
\subsubsection{\label{subsec:mock_population} The Simulated Population}

For each of the 60 systems analyzed in this work, we infer the distributions of astrophysical parameters consistent with the observed gravitational wave signal. 
We are able to constrain some parameters well: chirp mass and luminosity distance. 
Other parameters, including  the redshift inferred from the massive graviton phase term, are more poorly constrained.

\begin{figure*}
    \includegraphics[trim={0 0 0 5cm},clip, width=0.85\textwidth]{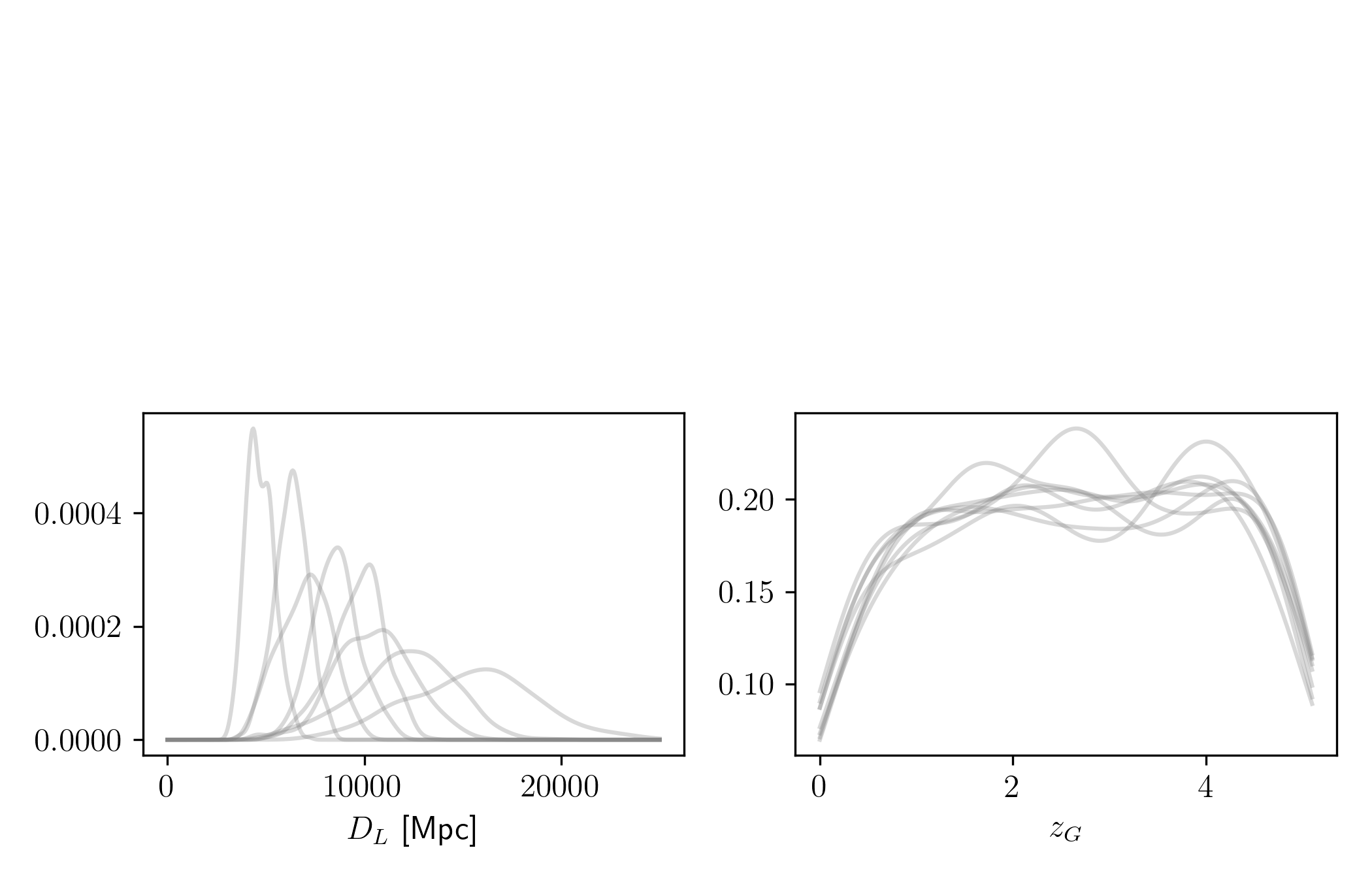}
    \caption{The distance and redshift gaussian KDEs from parameter estimation for a subset of the simulated population. From $D_L$ and $z_G$ likelihoods, the corresponding posterior PDFs for $H_0$ are calculated and shown in Fig \ref{fig:H0_posterior_population}.}
    \label{fig:H0_posterior_event}
\end{figure*}

For each system, the likelihoods for the luminosity distance $D_L$ and the redshift $z_G$ are used to infer a posterior probability density function (PDF) for $H_0$. 
Although the distance likelihood is generally well constrained, the wider likelihood on $z_G$ results in a similarly wide $H_0$ posterior PDF for individual events. 
An illustrative subset of the population is shown in Fig. \ref{fig:H0_posterior_event}. 

\begin{figure}[h]
    \includegraphics[width=0.49\textwidth]{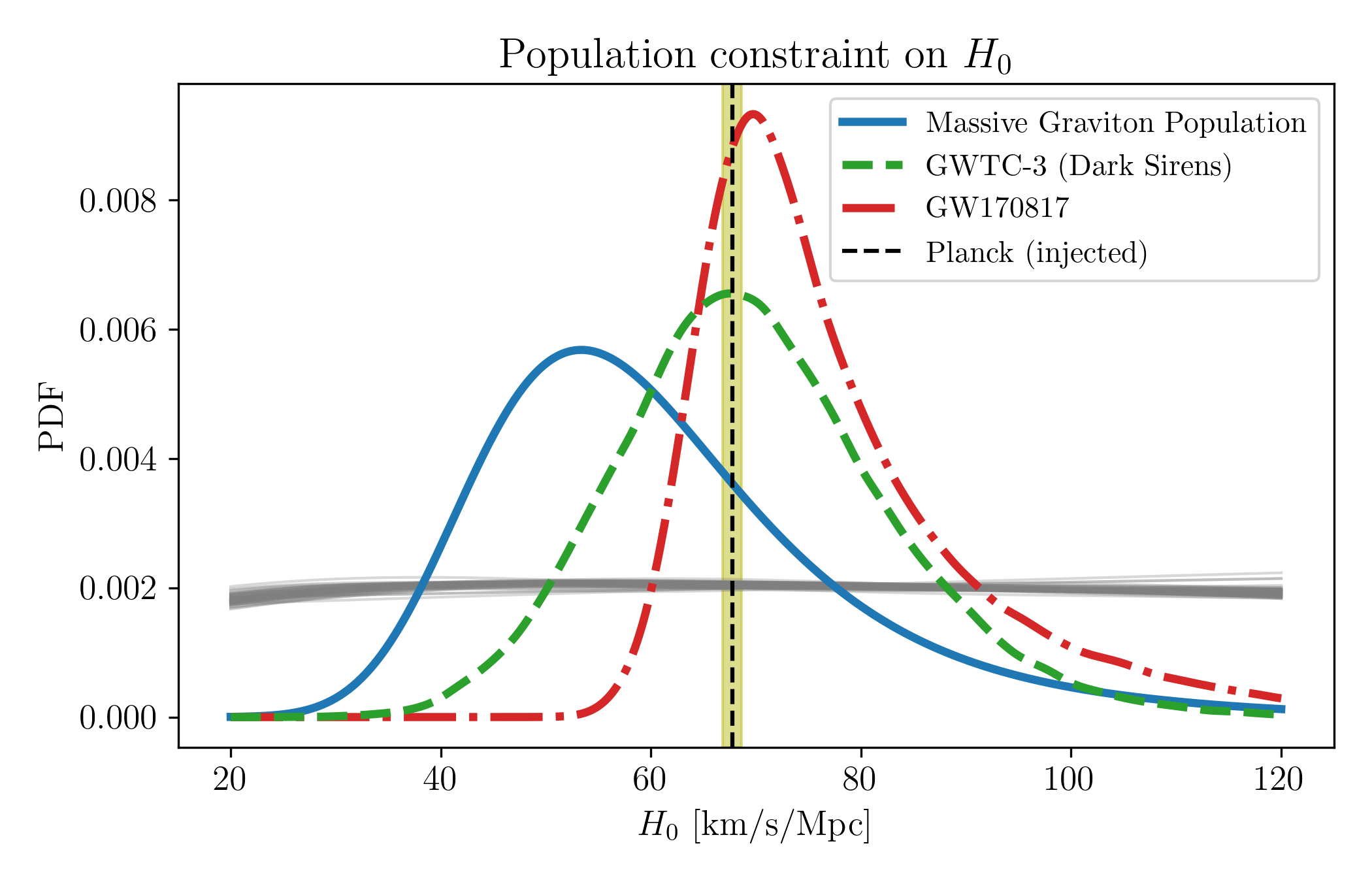}
    \caption{Constraint on $H_{0}$ determined from a simulated population of 60 observed binary black hole signals (blue, solid). 
    The $H_0$ pdf for each individual event is shown in gray (solid). The black, dashed line and the yellow band show the Planck 2015 result and its uncertainty \cite{Planck:2015fie}. For comparison, we also include the $H_0$ posterior from the only GWs event with an EM counterpart, GW10817 \cite{LIGOScientific:2017adf} (red, dot-dashed), and from the GWTC-3 ``dark sirens'' result (green, dashed) \cite{LIGOScientific:2021aug, GWTC-3_LVK_Cosmic_Expansion}.}
    \label{fig:H0_posterior_population}
\end{figure}

The individual $H_0$ posterior PDFs are then combined to obtain a population-level constraint, shown in Fig. \ref{fig:H0_posterior_population}. 

For a population of 60 binaries, this method can recover the constraint (at $90\%$ credible intervals) of $H_0 = 58^{+34}_{-19}\,\mathrm{km\; s^{-1}\; Mpc^{-1}}$ and $\Omega_m=0.29^{+0.10}_{-0.08}$, if the graviton Compton wavelength is $\lambda_G \approx 5 \times 10^{16}$m, corresponding to a mass of $m_G \approx 2 \times 10^{-23}$eV/c$^2$. 
This work is consistent with both the  $68\%$ credible intervals of the Planck value $H_0=67.8\pm 0.9\,\mathrm{km\; s^{-1}\; Mpc^{-1}}$
\cite{Planck:2015fie} 
as well as the SH0ES value $H_0 = 73.04^{+1.04}_{-1.04}\,\mathrm{km\; s^{-1}\; Mpc^{-1}}$
\cite{Riess:2021jrx}. 
The constraint derived from GW170817 $H_0 = 70.0^{+12.0}_{-8.0}\,\mathrm{km\; s^{-1}\; Mpc^{-1}}$
\cite{LIGOScientific:2017adf}
has better precision to this work when leveraging a population of 60 sources. 

\begin{figure*}
    \includegraphics[width=0.99\textwidth]{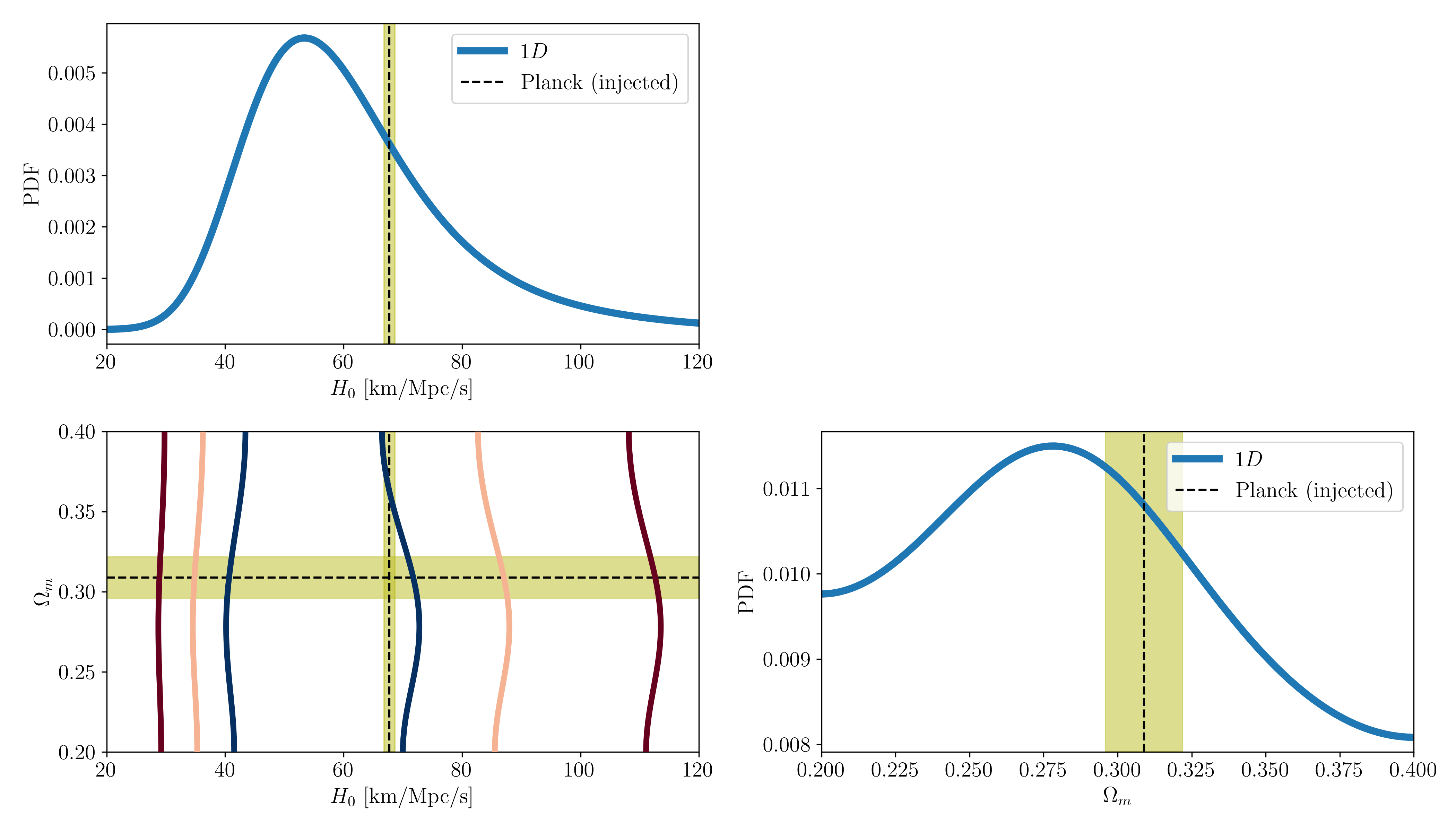}
    \caption{The combined posteriors for $H_{0}$ and $\Omega_m$ for 60 simulated BBHs. The final constraints are consistent with the input cosmology. The 2D contours correspond to ($68\%, 90\%, 99\%$) confidence intervals. The 1D PDF on $H_{0}$ shown here is also seen in Fig. \ref{fig:H0_posterior_population}. The $5\%-95\%$ percentiles correspond to $H_0 = 57.88^{+34.06}_{-19.24}$ km/s/Mpc and $\Omega_m=0.29^{+0.10}_{-0.08}$, i.e. the $\Omega_m$ distribution does not practically exclude any domain of the prior range. The black, dashed lines and the yellow bands show the Planck 2015 results and their uncertainty \cite{Planck:2015fie}.}
    \label{fig:2D_H0_Om_posterior_population}
\end{figure*}

We find no significant correlations between the value of $\lambda_G$ and the other physical parameters estimated in the simulated population, indicating that this measurement may suffer little observational bias but similarly there are no proxy parameters that can assist in determining constraints through the inference, and potential breaking, of such correlations. 

\subsubsection{\label{subsec:GW200115} GW200115}

For the observed binary system GW200115~\cite{LIGOScientific:2021qlt}, the astrophysical parameters follow the trends of the simulated population: 
$\mathcal{M}_C$ and $D_L$ are once again well constrained in this analysis, while $\Omega_m$ broadly recovers only the prior. 
Additionally, we generally recover the priors for both $\lambda_G$ and $z_G$. 
The 
posterior PDFs for these parameters can be seen in App. \ref{sec:app_GW200115_PE}. 

The nondetection of the effects of the massive graviton allow for an upper bound to be placed on the graviton mass and a one-sided constraint to be placed on $H_0$, yielding an upper limit of $H_0 \lesssim 6150$ km/s/Mpc ($90 \%$ credible interval). 
In practice, the one-sided constraint determined by this analysis of GW200115 is uninformative as the limit lies outside our physically informed prior range, which in turn was chosen to be wide compared to the current range of $H_0$ measured with other methods. 
The upper limit above is based on the analysis of the PE samples of GW2001115. An analytical derivation for an upper limit estimate can be found in App. \ref{sec:app_Ho_upper_limit}.

Therefore, an analysis of the most promising single observation out of the currently detected events with our methodology, cannot yet provide any informative constraint regarding $H_0$, or the existence of a massive graviton.


\section{\label{sec:discussion} Discussion}

\subsubsection{Leveraging a population of events }
Deriving a constraint on $H_0$ from the propagation effects of a massive graviton leverages a large population of events rather than relying on the extraordinary precision of any particular event. 
The recent fourth catalog of gravitational wave transients, GWTC-4.0~\cite{LIGOScientific:2025slb}, has increased the number of observed BBH events to over 200, and with the fourth observing run of the LVK network currently still underway, this number is expected to further increase significantly. 
All of these events could in principle contribute to cosmological constraints if analyzed with the method presented in this work, as it does not have any dependence on knowledge about either an electromagnetic counterpart or the likely host galaxy of the binary source. 

Cosmological constraints derived from the effects of a massive graviton cannot compete on an event-by-event basis with the multimessenger approach to assigning a redshift to observed binaries, which may localize a binary to a single host galaxy. 
However, there is at present only a single multimessenger event which provides a constraint on $H_0$
\cite{LIGOScientific:2017vwq, LIGOScientific:2017adf}. 
By contrast, this method utilizes the entire set of GW events from a given catalog, with the majority being undetectable through the multimessenger approach. 

While galaxy catalogs can in principle be used with all the events in GWTC-4.0,, they are not consistently complete in all parts of the sky or at all redshifts
\cite{DelPozzo:2018dpu, Mo:2024uim}, 
and also require well-localised sources to provide informative constraints. 
The method explored in this work instead applies consistently for all sources included in the analyzed population and does not require any external dataset. 
Additionally, while other GW cosmology methods will reduce in applicability and usefulness for sources at high redshifts as galaxy catalogs become too incomplete or where an EM counterpart becomes too faint, the method presented here has no such limitation.
It will therefore in principle be more capable of exploring the impact from cosmological parameters beyond simply $H_0$ and $\Omega_m$, as their impact will become more dominant at increasing redshifts
\cite{Chen:2020gek}.

\subsubsection{The choice of $\lambda_G$}
In order to simplify our analysis and to conform to the current physical limitations of today's ground-based observatories, we adopt an unphysical value for $\lambda_G$ throughout this work. 
Our adopted value is roughly consistent with earlier constraints
\cite{Will:1997bb}, but the most recent upper limit for the graviton mass based on the GTWC-3 catalog is $m_G \leq 1.27 \times 10^{-23}$ eV/c$^2$, corresponding to $\lambda_G \gtrsim 9.2 \times 10^{16} \:\mathrm{m}$ 
\cite{LIGOScientific:2021sio}, slightly larger than the value explored in this work. 

For higher values of $\lambda_G$, the analysis described in this work remains valid, it is simply the case that achieving comparable precision will require larger populations of sources from which to draw the joint constraints. 
As the population of observed gravitational wave sources grows, the cosmological constraints recovered from this method will improve as well. 


As future GW detectors come online, such as Cosmic Explorer
\cite{Evans:2021gyd} 
or the Einstein Telescope
\cite{Maggiore:2019uih}, 
cosmological measurements using the method presented here will improve even further, as a result both of the increased population of observed systems and with the more precise estimation of parameters, including $\lambda_G$.
We leave a study of the specific scientific potential of such observations, and the computational cost they would incur, for future work.
Additionally, waveform systematics are projected to be a significant source of bias and new waveform models will need to be developed in order to take advantage of the improved observing capabilities
\cite{Purrer:2019jcp}.

\subsubsection{Correlations between measurement of $\lambda_G$ and other parameters of the binaries }
During our analysis, we searched our population for correlations between $\lambda_G$ and any other parameters of the binary including chirp mass, spin magnitudes of the primaries, luminosity distance and the inclination of the orbital plane. 
We find no significant correlations. 
While there are no proxy parameters that can be utilized to improve the constraints determined in this work, we can be reasonably confident that the constraint itself is unbiased.

\subsubsection{A proof of concept for other beyond-GR effects}
Theories of gravity which include a massive graviton may also include a massless mode
\cite{Hassan:2011zd}.
Although this work does not address such theories directly, they need only implement a mixing fraction between the modified waveforms produced here and the unmodified waveforms usually generated for gravitational wave analysis. 
In fact, any parameterized model which describes propagation effects can utilize the framework described throughout this work. 

Any deviation from GR which imparts a propagation effect can in principle be simulated and studied in a similar way for use as a cosmological probe. 
While the impact from the non-zero graviton mass is the focus of this work, the applications of the analysis methods employed are not limited to this problem only. 

Finally, we note that the assumption of a massive graviton implies a modification of GR, which in principle can also impact the dynamics and GWs signal produced in the merger. 
Will \cite{Will:1997bb} argues that for velocities of the order $v/c>10^{-2}$ and for $\lambda_G > 10^{15}$ m, the GR approximation of the merger dynamics is satisfied. 
In our work, we are within this limit for the graviton's wavelengths considered, so we safely neglect any ``production'' effects on the waveforms considered.

\subsubsection{Constraints from GW200115}

As we effectively recover the priors from the parameters connected with a massive graviton in the analysis of the observation of GW200115, we have little constraining power to inform cosmological measurements, like $H_0$. 
Since this most promising event reveals little, a detailed reproduction of the full population of events from the current catalog can be assumed to be even less informative and require detector sensitivities of next generation instruments. 

The deviation from GR that we derive with this method, $d\chi_2$, is consistent with the constraints from GWTC-3 (see Fig. 6 of \cite{LIGOScientific:2021sio}), and therefore is consistent with standard GR. 
While such a nondetection can place one-sided constraints on cosmological parameters, observatories with greater sensitivities will be required to extract cosmological constraints from this method. 

\begin{figure}[h]
    \includegraphics[width=0.49\textwidth]{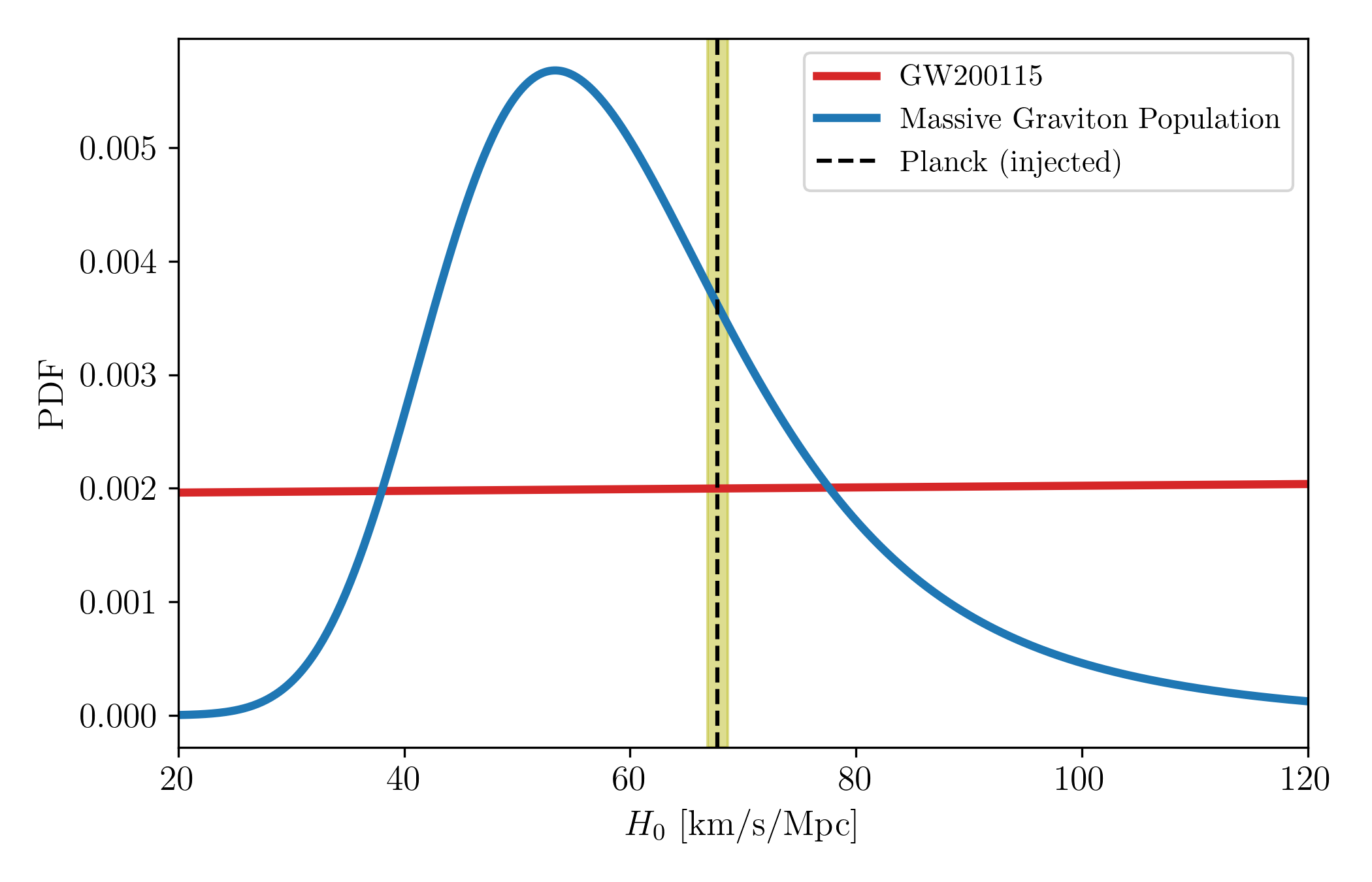}
    \caption{$H_0$ posterior for GW200015 (red line) vs the population constraint of our $60$ simulated BBHs events (blue, solid). As expected for a single event, the cosmological constraint is uninformative. The black, dashed line and the yellow band show the Planck 2015 result and its uncertainty \cite{Planck:2015fie}.}
    \label{fig:H0_posterior_GW20015}
\end{figure}

\section{\label{sec:conclusion} Conclusion}
If we live in a universe where the mass of the graviton is nonzero, its effect on the phase evolution of a gravitational wave signal traversing the Universe could be measured and used to inform cosmological inference. 
We show how distance and redshift measurements, as inferred from the observed impact of the GW phase caused by the finite mass, can be extracted from GW observation. 
By analyzing a population of compact-object binaries generated from reasonable source distributions, these distance and redshift measurements can thus be combined to make precise constraints on cosmological parameters. 

Assuming a graviton with mass $m_G \approx 2.3 \times 10^{-23}\mathrm{eV/c}^2$ and a population consisting of 60 events, we constrain (at $90\%$ credible intervals) $H_0 = 58^{+34}_{-19}\,\mathrm{km\; s^{-1}\; Mpc^{-1}}$ and $\Omega_m=0.29^{+0.10}_{-0.08}$ as observed using a near-future HLVKA gravitational wave observatory network. 
We successfully recover the $H_0$ value from the fiducial cosmology we have assumed for our simulations, but the inferred precision is not able, on its own, to adjudicate the tension in $H_0$ as measured from standard candles or the cosmic microwave background from the SH0ES
\cite{Riess:2021jrx}
and Planck
\cite{Planck:2015fie}
collaborations respectively. 
Such precision would require larger contributing populations of compact-object binary observations than are currently public, but will certainly be achieved within the operating lifetimes of current and future observatories. 

Strictly, a smaller graviton mass than that explored in this work is consistent with the most current constraints. 
In order to recover similarly informative constraints from the subtler signal, next generation detectors, significantly larger contributing populations, or ideally both would be necessary. 
Hence, the amount and fidelity of the data used in the analysis would need to increase, but the method of inference itself would remain robust.
In case the graviton mass would be exactly zero, or more likely if it is too small to robustly measure through either GW observations or other means, the cosmological inference method presented here will still be able to place limits on what cosmological parameters can be supported. 
It is however important to reiterate the lack of a dependence for either an EM counterpart or a galaxy catalog for the feasibility or success of this method. 
Hence, it will always be able to be applied to the entire set of observed GW events, and especially so for high-redshift events where other methods are expected to lose constraining power.

In addition to the specific physical application explored here, a universe with a massive graviton, these methods are applicable to any parameterized model which imparts propagation effects onto the gravitational waveform. 

\begin{acknowledgments}
The authors thank Hsin-Yu Chen and Tom Callister for their helpful conversation and collaboration while this work was ongoing. 
We also thank Antonio Enea Romano for a careful review of a draft version of the paper and useful comments.
CJH acknowledges the support from the Nevada Center for Astrophysics, from NASA Grant No. 80NSSC23M0104, and the NSF through the Award No.~PHY-2409727.
The authors are grateful for computational resources provided by the LIGO Laboratory and supported by National Science Foundation Grants PHY-0757058 and PHY-0823459.
This material is based upon work supported by NSF's LIGO Laboratory which is a major facility fully funded by the National Science Foundation.
This work carries LIGO Document Number LIGO-P2500494.

All authors contributed equally to the LIGO Scientific collaboration.
MJ was responsible for writing software, running simulations, and drafting and submitting the manuscript. 
MK provided cosmological analysis and edited the manuscript.
CJH obtained funding, presented the initial research concept, and edited the manuscript. 
\end{acknowledgments}

\appendix

\section{Priors used for parameter estimation \label{sec:app_priors}}

\begin{tabular}{c|c|c}
    \hline
    Quantity & Prior Distribution & Bounds \\
    \hline
    \hline
    $m_1$ [$M_{\odot}$] & Uniform & [5, 200] \\
    $q$ & \begin{tabular}{c}UniformInComponents \\ MassRatio\end{tabular} & [0.1, 1.0] \\
    $a_1$ & Uniform & [0, 0.99] \\
    $a_2$ & Uniform & [0, 0.99] \\
    $D_L$ [Mpc] & PowerLaw & [100, 25000] \\
    dec & Cosine  & [0, $\pi$] \\
    ra & Uniform  & [0, 2$\pi$] \\
    $\theta_{jn}$ & Sine & [0, $\pi$] \\
    $\psi$ & Uniform & [0, $\pi$] \\
    phase & Uniform & [0, 2$\pi$] \\
    tilt$_1$ & Sine & [0, $\pi$] \\
    tilt$_2$ & Sine & [0, $\pi$] \\
    $\phi_{12}$ & Uniform & [0, 2$\pi$] \\
    $\phi_{jl}$ & Uniform & [0, 2$\pi$] \\
    $\lambda_G$ [m] & LogUniform & [5$\times 10^{14}$, 5$\times10^{18}$] \\
    $z_G$ & Uniform & [0, 5.1] \\
    $\Omega_m$ & Uniform & [0.2, 0.4] \\
    \hline
\end{tabular}\\

The redshift priors selected in this work are chosen to be consistent with the luminosity distance priors, and to be able to give support to the priors on the cosmological parameters ($H_0, \Omega_m$). 
To be more specific, given the prior range of $H_0$ between [$20, 120$] km/s/Mpc and of $\Omega_m$ between [$0.2, 0.4$], the upper and lower limits of the distance priors lead to upper and lower limits for $z$ between [$0.007, 5.05$]. 
Our prior range for redshift is chosen to be consistent with these limits.

\begin{figure}[h]
    \includegraphics[width=0.49\textwidth]{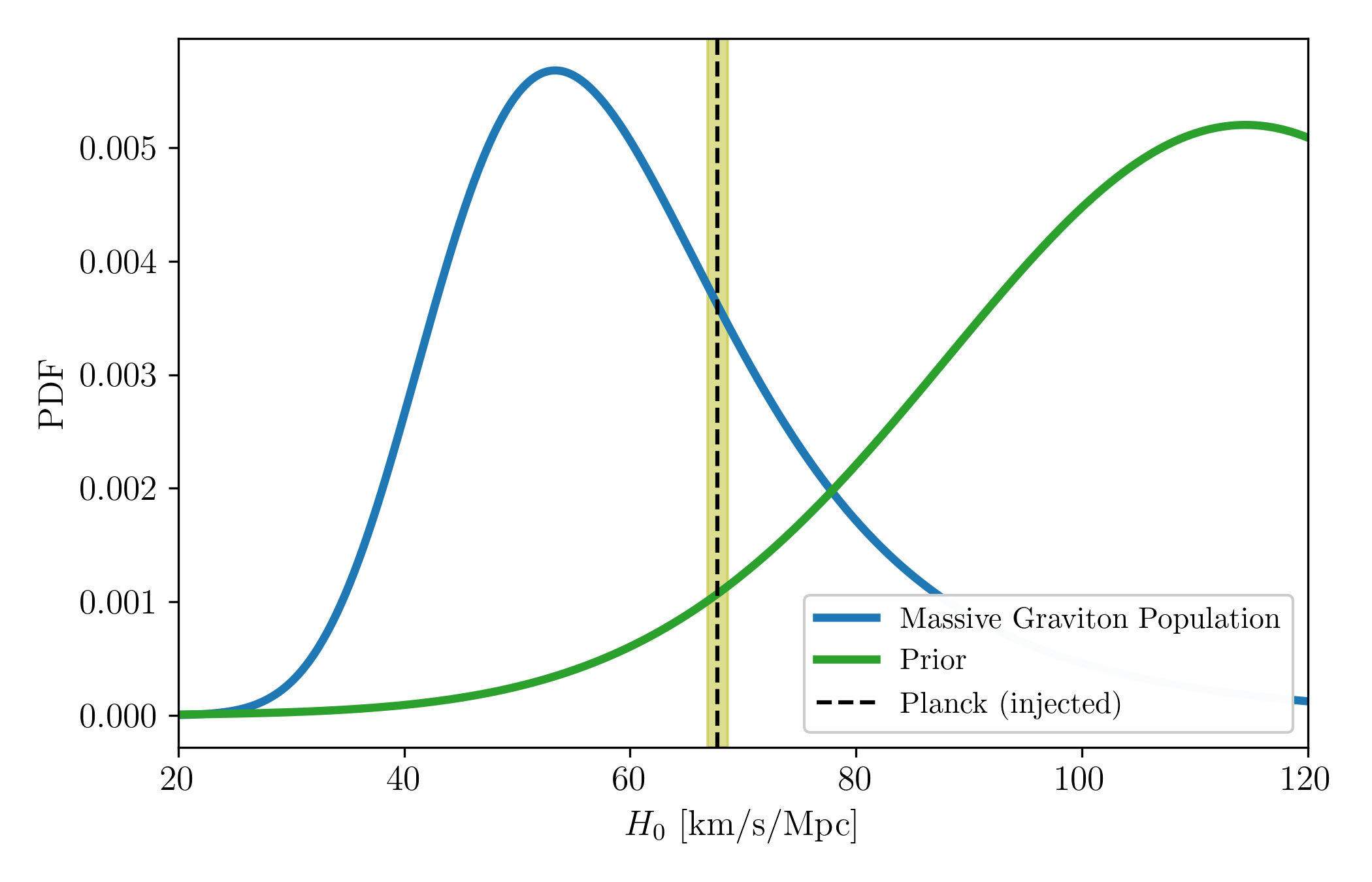}
    \caption{$H_0$ posterior for $60$ events with ($z^S_G, \Omega^S_m$, $D^S_L$) samples drawn from their uniform prior ranges (green, solid line) vs the population constraint of our $60$ simulated BBHs events (blue, solid line). This demonstrates that our result is not driven by the prior choices. Moreover, it emphasizes the effect of better distance determination of our realistic population on the final $H_0$ inference. The black, dashed line and the yellow band show the Planck 2015 result and its uncertainty \cite{Planck:2015fie}.}
    \label{fig:prior_effects}
\end{figure}

To better understand the effect of the priors on our result, we run an analysis where are the samples ($z^S_G, \Omega^S_m$, $D^S_L$) are drawn from their prior ranges. Then, using the analysis described in Sec. \ref{subsec:cosmology}, we infer $H_0$ for $60$ `prior events`. The result is shown in Fig \ref{fig:prior_effects}, where we also compare with the result of our realistic population. It is clear that our method yields informative results, i.e. distinct from the $H_0$ posterior based on prior draws. We find that the main difference comes from the power to determine the luminosity distance $d_L$ in a realistic analysis.

The prior range for $\lambda_G$ was chosen to cover the input value of $\lambda_G^{\rm sim} \simeq 5 \times 10^{16}$ m employed in our analysis, without being tightly constraining. This choice is justified by the example posterior results shown in Figs. \ref{fig:PE_representative} and \ref{fig:PE_GW200115}, where $\lambda_G$ is only setting a lower-limit around the input value.

\section{Parameter estimation for an \\ event in the simulated population \label{sec:app_population_PE}}

\begin{figure*}
    \centering
    \includegraphics[width=\textwidth]{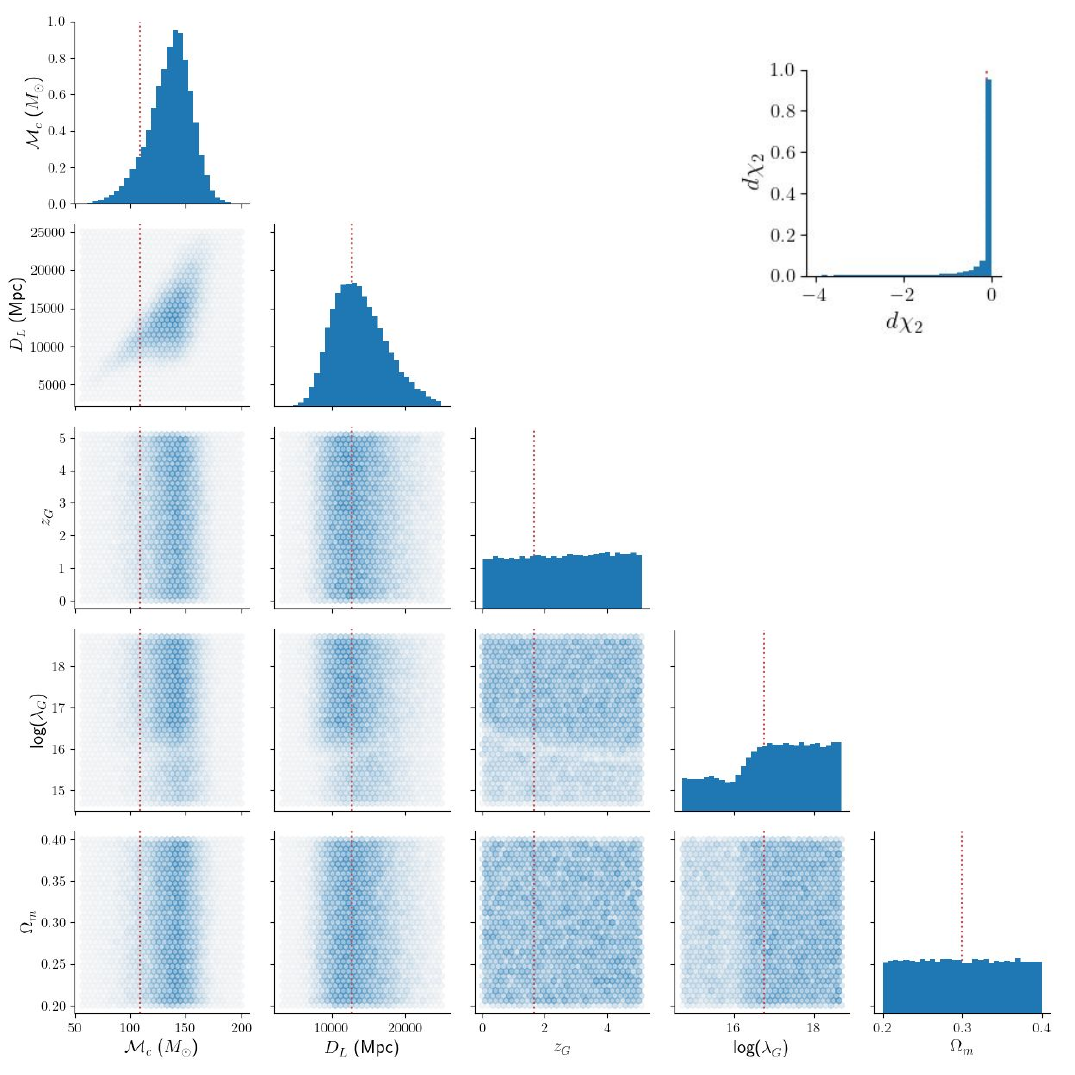}
    \caption{Showing the posterior probability distributions for a subset of the inferred parametrers for a representative system in the simulated population. 1D posteriors for the included parameters are on the diagonal and the 2D combined posteriors populate the rest of the corner plot. The 1D posterior for $d\chi_2$ is included as an inset. The injected parameter values are shown in red and reveal which parameters are more accurately recovered during the parameter estimation process.}
    \label{fig:PE_representative}
\end{figure*}

We show a subset of the parameters estimated for one system in the generated simulated population in Fig. \ref{fig:PE_representative}. 

\section{Parameter estimation for \\ GW200115 \label{sec:app_GW200115_PE}}

\begin{figure*}
    \centering
    \includegraphics[width=\textwidth]{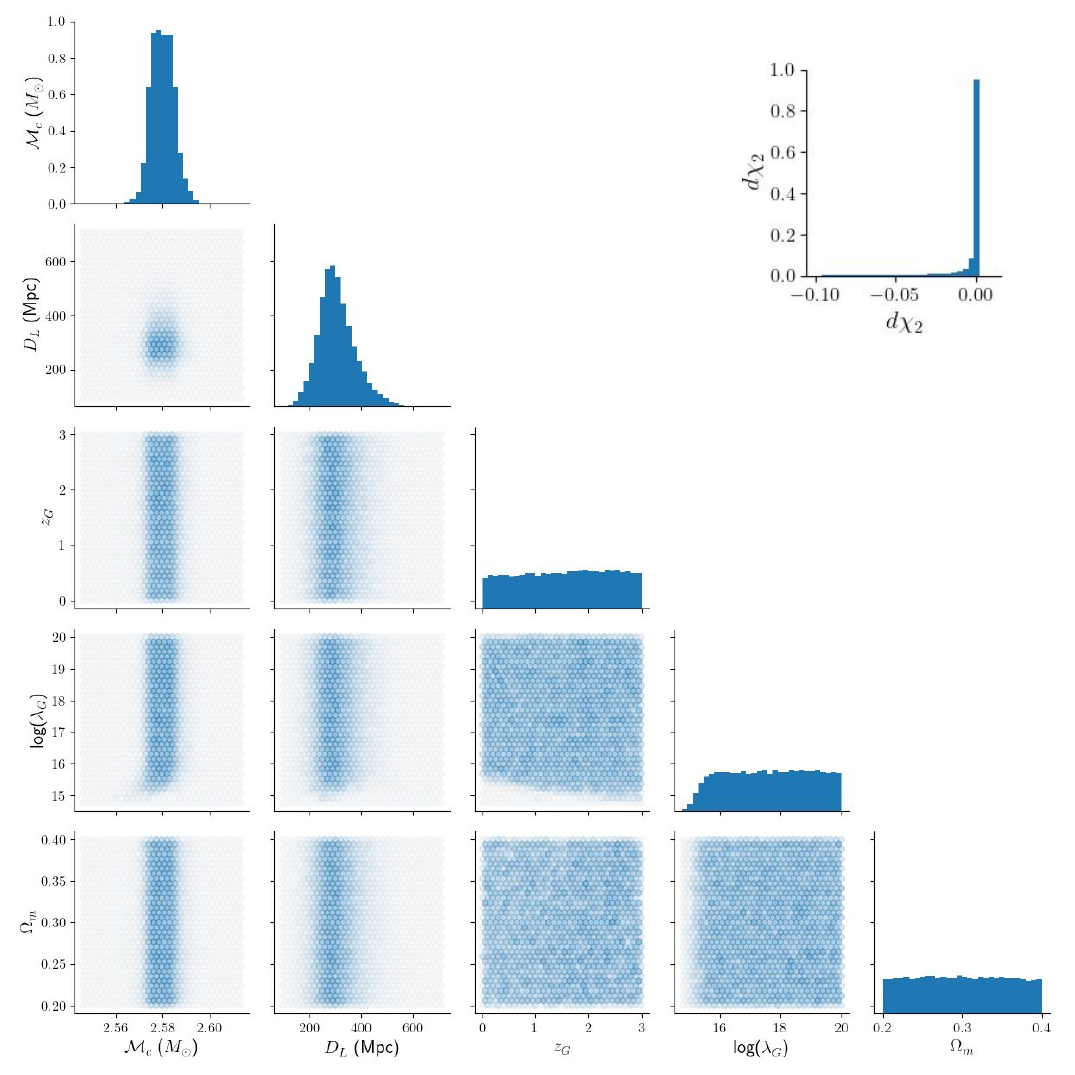}
    \caption{Showing the posterior probability distributions for a subset of the inferred parametrers  for GW200115. As in Fig. \ref{fig:PE_representative}, the 1D posteriors for each parameter and the 2D combined posteriors populate the corner plot. The 1D posterior for $d\chi_2$ is included as an inset. As there is no known ``true'' value for these parameters, no injection values are indicated.}
    \label{fig:PE_GW200115}
\end{figure*}

We show a subset of the parameters estimated for the simulated detection of GW200115, as seen in Fig. \ref{fig:PE_GW200115}.

\section{An upper limit formula for $H_0$ \label{sec:app_Ho_upper_limit}}

\subsection{General result}

The phase shift formula due to the presence of a massive graviton - Eqn. (\ref{eq:dchi2}) - can be inverted to give an upper limit on the Hubble parameter.

We start from
\begin{equation}
    d\chi_{2}=\frac{\Psi_{G}}{\frac{5}{96}\left(\frac{743}{336}+\frac{11}{4}\eta\right)\eta^{-2/5}u^{-1}},
\end{equation}
and introduce Eqn. (\ref{eq:psi_G}) and simplify, to get
\begin{equation}
    d\chi_{2}=\frac{- \pi^2 \mathcal{M} D_\Lambda}{\frac{5}{96}\left(\frac{743}{336}+\frac{11}{4}\eta\right)\eta^{-2/5}}\frac{G}{c^2 \lambda_G^2(1+z_G)}.
\end{equation}
We now introduce the function of the distance measure - Eqn.(\ref{eq:d_measure}) and solve for $H_0$.
\begin{align}\label{eq:app_Ho_from_dchi2}
    H_0=&\frac{1}{d\chi_{2}}\frac{- \pi^2 \mathcal{M}}{\frac{5}{96}\left(\frac{743}{336}+\frac{11}{4}\eta\right)\eta^{-2/5}}\frac{G}{c \lambda_G^2} \times \\ \nonumber
    &\int_0^{z_f}\frac{dz}{(1+z_G)^2\sqrt{\Omega_m (1+z_G)^3+\Omega_\Lambda}}.
\end{align}
Under fairly general assumptions, Eqn. (\ref{eq:app_Ho_from_dchi2}) leads to an upper limit for $H_0$, i.e. $H_0 \lesssim H_0^{\rm max}$. Under a flat, $\Lambda$CDM universe it can be verified that the redshift integral gives $\sim 0.5$ for redshifts $z>2$. In parallel, taking into account that the symmetric mass ratio satisfies $\eta \leq 1/4$, the $\eta$ part is bounded from above:
\begin{equation}
    \frac{\eta^{2/5}}{\frac{5}{96}\left(\frac{743}{336}+\frac{11}{4}\eta\right)} \simeq \frac{\eta^{2/5}}{0.1(1+\eta)} \lesssim 5.
\end{equation}
Substituting in Eqn. (\ref{eq:app_Ho_from_dchi2}) yields
\begin{equation}
    H_0 \lesssim \frac{1}{d\chi_{2}}\frac{- \pi^2 \mathcal{M}G}{c \lambda_G^2} \frac{5}{2}.
\end{equation}
We simplify this equation by substituting the numerical constants. This gives
\begin{equation}
    H_0 \lesssim 5.5\cdot 10^{-18} \frac{\mathcal{M}}{\lambda_G^2|d\chi_{2}|}.
\end{equation}
Normalizing for the upper limit of chirp masses considered in this work, for the current constraints on $\lambda_G$, $\lambda_G \geq 10^{17}$ m, and transforming to the standard units of $H_0$, we can prove that:
\begin{equation}
    H_0 \lesssim 6.6 \left(\frac{\mathcal{M}}{200 M_\odot}\right)\left(\frac{10^{34} {\rm m}}{\lambda_G^2}\right)\frac{1}{|d\chi_{2}|} \ {\rm km/s/Mpc}.
\end{equation}
Hence, any precise GW measurement of $d\chi_2$ would allow an upper limit on $H_0$ (assuming an independent value of $\lambda_g$ as input).




\clearpage

\bibliography{mj_aps} 

\providecommand{\noopsort}[1]{}\providecommand{\singleletter}[1]{#1}%
\begin{thebibliography}{98}%
\makeatletter
\providecommand \@ifxundefined [1]{%
 \@ifx{#1\undefined}
}%
\providecommand \@ifnum [1]{%
 \ifnum #1\expandafter \@firstoftwo
 \else \expandafter \@secondoftwo
 \fi
}%
\providecommand \@ifx [1]{%
 \ifx #1\expandafter \@firstoftwo
 \else \expandafter \@secondoftwo
 \fi
}%
\providecommand \natexlab [1]{#1}%
\providecommand \enquote  [1]{``#1''}%
\providecommand \bibnamefont  [1]{#1}%
\providecommand \bibfnamefont [1]{#1}%
\providecommand \citenamefont [1]{#1}%
\providecommand \href@noop [0]{\@secondoftwo}%
\providecommand \href [0]{\begingroup \@sanitize@url \@href}%
\providecommand \@href[1]{\@@startlink{#1}\@@href}%
\providecommand \@@href[1]{\endgroup#1\@@endlink}%
\providecommand \@sanitize@url [0]{\catcode `\\12\catcode `\$12\catcode
  `\&12\catcode `\#12\catcode `\^12\catcode `\_12\catcode `\%12\relax}%
\providecommand \@@startlink[1]{}%
\providecommand \@@endlink[0]{}%
\providecommand \url  [0]{\begingroup\@sanitize@url \@url }%
\providecommand \@url [1]{\endgroup\@href {#1}{\urlprefix }}%
\providecommand \urlprefix  [0]{URL }%
\providecommand \Eprint [0]{\href }%
\providecommand \doibase [0]{https://doi.org/}%
\providecommand \selectlanguage [0]{\@gobble}%
\providecommand \bibinfo  [0]{\@secondoftwo}%
\providecommand \bibfield  [0]{\@secondoftwo}%
\providecommand \translation [1]{[#1]}%
\providecommand \BibitemOpen [0]{}%
\providecommand \bibitemStop [0]{}%
\providecommand \bibitemNoStop [0]{.\EOS\space}%
\providecommand \EOS [0]{\spacefactor3000\relax}%
\providecommand \BibitemShut  [1]{\csname bibitem#1\endcsname}%
\let\auto@bib@innerbib\@empty
\bibitem [{\citenamefont {Buikema}\ \emph {et~al.}(2020)\citenamefont {Buikema}
  \emph {et~al.}}]{aLIGO:2020wna}%
  \BibitemOpen
  \bibfield  {author} {\bibinfo {author} {\bibfnamefont {A.}~\bibnamefont
  {Buikema}} \emph {et~al.} (\bibinfo {collaboration} {aLIGO}),\ }\bibfield
  {title} {\bibinfo {title} {{Sensitivity and performance of the Advanced LIGO
  detectors in the third observing run}},\ }\href
  {https://doi.org/10.1103/PhysRevD.102.062003} {\bibfield  {journal} {\bibinfo
   {journal} {Phys. Rev. D}\ }\textbf {\bibinfo {volume} {102}},\ \bibinfo
  {pages} {062003} (\bibinfo {year} {2020})},\ \Eprint
  {https://arxiv.org/abs/2008.01301} {arXiv:2008.01301 [astro-ph.IM]}
  \BibitemShut {NoStop}%
\bibitem [{\citenamefont {Acernese}\ \emph {et~al.}(2019)\citenamefont
  {Acernese} \emph {et~al.}}]{Virgo:2019juy}%
  \BibitemOpen
  \bibfield  {author} {\bibinfo {author} {\bibfnamefont {F.}~\bibnamefont
  {Acernese}} \emph {et~al.} (\bibinfo {collaboration} {Virgo}),\ }\bibfield
  {title} {\bibinfo {title} {{Increasing the Astrophysical Reach of the
  Advanced Virgo Detector via the Application of Squeezed Vacuum States of
  Light}},\ }\href {https://doi.org/10.1103/PhysRevLett.123.231108} {\bibfield
  {journal} {\bibinfo  {journal} {Phys. Rev. Lett.}\ }\textbf {\bibinfo
  {volume} {123}},\ \bibinfo {pages} {231108} (\bibinfo {year}
  {2019})}\BibitemShut {NoStop}%
\bibitem [{\citenamefont {Akutsu}\ \emph {et~al.}(2021)\citenamefont {Akutsu}
  \emph {et~al.}}]{KAGRA:2020tym}%
  \BibitemOpen
  \bibfield  {author} {\bibinfo {author} {\bibfnamefont {T.}~\bibnamefont
  {Akutsu}} \emph {et~al.} (\bibinfo {collaboration} {KAGRA}),\ }\bibfield
  {title} {\bibinfo {title} {{Overview of KAGRA: Detector design and
  construction history}},\ }\href {https://doi.org/10.1093/ptep/ptaa125}
  {\bibfield  {journal} {\bibinfo  {journal} {PTEP}\ }\textbf {\bibinfo
  {volume} {2021}},\ \bibinfo {pages} {05A101} (\bibinfo {year} {2021})},\
  \Eprint {https://arxiv.org/abs/2005.05574} {arXiv:2005.05574
  [physics.ins-det]} \BibitemShut {NoStop}%
\bibitem [{\citenamefont {Abbott}\ \emph {et~al.}(2019)\citenamefont {Abbott}
  \emph {et~al.}}]{LIGOScientific:2018mvr}%
  \BibitemOpen
  \bibfield  {author} {\bibinfo {author} {\bibfnamefont {B.~P.}\ \bibnamefont
  {Abbott}} \emph {et~al.} (\bibinfo {collaboration} {LIGO Scientific,
  Virgo}),\ }\bibfield  {title} {\bibinfo {title} {{GWTC-1: A
  Gravitational-Wave Transient Catalog of Compact Binary Mergers Observed by
  LIGO and Virgo during the First and Second Observing Runs}},\ }\href
  {https://doi.org/10.1103/PhysRevX.9.031040} {\bibfield  {journal} {\bibinfo
  {journal} {Phys. Rev. X}\ }\textbf {\bibinfo {volume} {9}},\ \bibinfo {pages}
  {031040} (\bibinfo {year} {2019})},\ \Eprint
  {https://arxiv.org/abs/1811.12907} {arXiv:1811.12907 [astro-ph.HE]}
  \BibitemShut {NoStop}%
\bibitem [{\citenamefont {Abbott}\ \emph {et~al.}(2024)\citenamefont {Abbott}
  \emph {et~al.}}]{LIGOScientific:2021usb}%
  \BibitemOpen
  \bibfield  {author} {\bibinfo {author} {\bibfnamefont {R.}~\bibnamefont
  {Abbott}} \emph {et~al.} (\bibinfo {collaboration} {LIGO Scientific,
  VIRGO}),\ }\bibfield  {title} {\bibinfo {title} {{GWTC-2.1: Deep extended
  catalog of compact binary coalescences observed by LIGO and Virgo during the
  first half of the third observing run}},\ }\href
  {https://doi.org/10.1103/PhysRevD.109.022001} {\bibfield  {journal} {\bibinfo
   {journal} {Phys. Rev. D}\ }\textbf {\bibinfo {volume} {109}},\ \bibinfo
  {pages} {022001} (\bibinfo {year} {2024})},\ \Eprint
  {https://arxiv.org/abs/2108.01045} {arXiv:2108.01045 [gr-qc]} \BibitemShut
  {NoStop}%
\bibitem [{\citenamefont {Abbott}\ \emph
  {et~al.}(2023{\natexlab{a}})\citenamefont {Abbott} \emph
  {et~al.}}]{KAGRA:2021vkt}%
  \BibitemOpen
  \bibfield  {author} {\bibinfo {author} {\bibfnamefont {R.}~\bibnamefont
  {Abbott}} \emph {et~al.} (\bibinfo {collaboration} {KAGRA, VIRGO, LIGO
  Scientific}),\ }\bibfield  {title} {\bibinfo {title} {{GWTC-3: Compact Binary
  Coalescences Observed by LIGO and Virgo during the Second Part of the Third
  Observing Run}},\ }\href {https://doi.org/10.1103/PhysRevX.13.041039}
  {\bibfield  {journal} {\bibinfo  {journal} {Phys. Rev. X}\ }\textbf {\bibinfo
  {volume} {13}},\ \bibinfo {pages} {041039} (\bibinfo {year}
  {2023}{\natexlab{a}})},\ \Eprint {https://arxiv.org/abs/2111.03606}
  {arXiv:2111.03606 [gr-qc]} \BibitemShut {NoStop}%
\bibitem [{\citenamefont {Abbott}\ \emph
  {et~al.}(2017{\natexlab{a}})\citenamefont {Abbott} \emph
  {et~al.}}]{LIGOScientific:2017adf}%
  \BibitemOpen
  \bibfield  {author} {\bibinfo {author} {\bibfnamefont {B.~P.}\ \bibnamefont
  {Abbott}} \emph {et~al.} (\bibinfo {collaboration} {LIGO Scientific, Virgo,
  1M2H, Dark Energy Camera GW-E, DES, DLT40, Las Cumbres Observatory, VINROUGE,
  MASTER}),\ }\bibfield  {title} {\bibinfo {title} {{A gravitational-wave
  standard siren measurement of the Hubble constant}},\ }\href
  {https://doi.org/10.1038/nature24471} {\bibfield  {journal} {\bibinfo
  {journal} {Nature}\ }\textbf {\bibinfo {volume} {551}},\ \bibinfo {pages}
  {85} (\bibinfo {year} {2017}{\natexlab{a}})},\ \Eprint
  {https://arxiv.org/abs/1710.05835} {arXiv:1710.05835 [astro-ph.CO]}
  \BibitemShut {NoStop}%
\bibitem [{\citenamefont {Abbott}\ \emph
  {et~al.}(2023{\natexlab{b}})\citenamefont {Abbott} \emph
  {et~al.}}]{LIGOScientific:2021aug}%
  \BibitemOpen
  \bibfield  {author} {\bibinfo {author} {\bibfnamefont {R.}~\bibnamefont
  {Abbott}} \emph {et~al.} (\bibinfo {collaboration} {LIGO Scientific, Virgo,
  KAGRA}),\ }\bibfield  {title} {\bibinfo {title} {{Constraints on the Cosmic
  Expansion History from GWTC\textendash{}3}},\ }\href
  {https://doi.org/10.3847/1538-4357/ac74bb} {\bibfield  {journal} {\bibinfo
  {journal} {Astrophys. J.}\ }\textbf {\bibinfo {volume} {949}},\ \bibinfo
  {pages} {76} (\bibinfo {year} {2023}{\natexlab{b}})},\ \Eprint
  {https://arxiv.org/abs/2111.03604} {arXiv:2111.03604 [astro-ph.CO]}
  \BibitemShut {NoStop}%
\bibitem [{\citenamefont {Abbott}\ \emph
  {et~al.}(2023{\natexlab{c}})\citenamefont {Abbott} \emph
  {et~al.}}]{KAGRA:2021duu}%
  \BibitemOpen
  \bibfield  {author} {\bibinfo {author} {\bibfnamefont {R.}~\bibnamefont
  {Abbott}} \emph {et~al.} (\bibinfo {collaboration} {KAGRA, VIRGO, LIGO
  Scientific}),\ }\bibfield  {title} {\bibinfo {title} {{Population of Merging
  Compact Binaries Inferred Using Gravitational Waves through GWTC-3}},\ }\href
  {https://doi.org/10.1103/PhysRevX.13.011048} {\bibfield  {journal} {\bibinfo
  {journal} {Phys. Rev. X}\ }\textbf {\bibinfo {volume} {13}},\ \bibinfo
  {pages} {011048} (\bibinfo {year} {2023}{\natexlab{c}})},\ \Eprint
  {https://arxiv.org/abs/2111.03634} {arXiv:2111.03634 [astro-ph.HE]}
  \BibitemShut {NoStop}%
\bibitem [{\citenamefont {Schutz}(1986)}]{Schutz:1986gp}%
  \BibitemOpen
  \bibfield  {author} {\bibinfo {author} {\bibfnamefont {B.~F.}\ \bibnamefont
  {Schutz}},\ }\bibfield  {title} {\bibinfo {title} {{Determining the Hubble
  Constant from Gravitational Wave Observations}},\ }\href
  {https://doi.org/10.1038/323310a0} {\bibfield  {journal} {\bibinfo  {journal}
  {Nature}\ }\textbf {\bibinfo {volume} {323}},\ \bibinfo {pages} {310}
  (\bibinfo {year} {1986})}\BibitemShut {NoStop}%
\bibitem [{\citenamefont {Holz}\ and\ \citenamefont
  {Hughes}(2005)}]{Holz:2005df}%
  \BibitemOpen
  \bibfield  {author} {\bibinfo {author} {\bibfnamefont {D.~E.}\ \bibnamefont
  {Holz}}\ and\ \bibinfo {author} {\bibfnamefont {S.~A.}\ \bibnamefont
  {Hughes}},\ }\bibfield  {title} {\bibinfo {title} {{Using gravitational-wave
  standard sirens}},\ }\href {https://doi.org/10.1086/431341} {\bibfield
  {journal} {\bibinfo  {journal} {Astrophys. J.}\ }\textbf {\bibinfo {volume}
  {629}},\ \bibinfo {pages} {15} (\bibinfo {year} {2005})},\ \Eprint
  {https://arxiv.org/abs/astro-ph/0504616} {arXiv:astro-ph/0504616}
  \BibitemShut {NoStop}%
\bibitem [{\citenamefont {Mastrogiovanni}\ and\ \citenamefont
  {Steer}(2021)}]{Mastrogiovanni2021}%
  \BibitemOpen
  \bibfield  {author} {\bibinfo {author} {\bibfnamefont {S.}~\bibnamefont
  {Mastrogiovanni}}\ and\ \bibinfo {author} {\bibfnamefont {D.~A.}\
  \bibnamefont {Steer}},\ }\bibinfo {title} {Measuring cosmological parameters
  with gravitational waves},\ in\ \href
  {https://doi.org/10.1007/978-981-15-4702-7_48-1} {\emph {\bibinfo {booktitle}
  {Handbook of Gravitational Wave Astronomy}}}\ (\bibinfo  {publisher}
  {Springer Singapore},\ \bibinfo {year} {2021})\ p.\ \bibinfo {pages}
  {1–51}\BibitemShut {NoStop}%
\bibitem [{\citenamefont {Salvarese}\ and\ \citenamefont
  {Chen}(2024)}]{Salvarese:2024jpq}%
  \BibitemOpen
  \bibfield  {author} {\bibinfo {author} {\bibfnamefont {A.}~\bibnamefont
  {Salvarese}}\ and\ \bibinfo {author} {\bibfnamefont {H.-Y.}\ \bibnamefont
  {Chen}},\ }\bibfield  {title} {\bibinfo {title} {{Mitigating the Binary
  Viewing Angle Bias for Standard Sirens}},\ }\href
  {https://doi.org/10.3847/2041-8213/ad7bbc} {\bibfield  {journal} {\bibinfo
  {journal} {Astrophys. J. Lett.}\ }\textbf {\bibinfo {volume} {974}},\
  \bibinfo {pages} {L16} (\bibinfo {year} {2024})},\ \Eprint
  {https://arxiv.org/abs/2406.11126} {arXiv:2406.11126 [astro-ph.CO]}
  \BibitemShut {NoStop}%
\bibitem [{\citenamefont {Veitch}\ \emph {et~al.}(2012)\citenamefont {Veitch},
  \citenamefont {Mandel}, \citenamefont {Aylott}, \citenamefont {Farr},
  \citenamefont {Raymond}, \citenamefont {Rodriguez}, \citenamefont {van~der
  Sluys}, \citenamefont {Kalogera},\ and\ \citenamefont
  {Vecchio}}]{Veitch:2012df}%
  \BibitemOpen
  \bibfield  {author} {\bibinfo {author} {\bibfnamefont {J.}~\bibnamefont
  {Veitch}}, \bibinfo {author} {\bibfnamefont {I.}~\bibnamefont {Mandel}},
  \bibinfo {author} {\bibfnamefont {B.}~\bibnamefont {Aylott}}, \bibinfo
  {author} {\bibfnamefont {B.}~\bibnamefont {Farr}}, \bibinfo {author}
  {\bibfnamefont {V.}~\bibnamefont {Raymond}}, \bibinfo {author} {\bibfnamefont
  {C.}~\bibnamefont {Rodriguez}}, \bibinfo {author} {\bibfnamefont
  {M.}~\bibnamefont {van~der Sluys}}, \bibinfo {author} {\bibfnamefont
  {V.}~\bibnamefont {Kalogera}},\ and\ \bibinfo {author} {\bibfnamefont
  {A.}~\bibnamefont {Vecchio}},\ }\bibfield  {title} {\bibinfo {title}
  {{Estimating parameters of coalescing compact binaries with proposed advanced
  detector networks}},\ }\href {https://doi.org/10.1103/PhysRevD.85.104045}
  {\bibfield  {journal} {\bibinfo  {journal} {Phys. Rev. D}\ }\textbf {\bibinfo
  {volume} {85}},\ \bibinfo {pages} {104045} (\bibinfo {year} {2012})},\
  \Eprint {https://arxiv.org/abs/1201.1195} {arXiv:1201.1195 [astro-ph.HE]}
  \BibitemShut {NoStop}%
\bibitem [{\citenamefont {London}\ \emph {et~al.}(2018)\citenamefont {London},
  \citenamefont {Khan}, \citenamefont {Fauchon-Jones}, \citenamefont
  {Garc{\'\i}a}, \citenamefont {Hannam}, \citenamefont {Husa}, \citenamefont
  {Jim{\'e}nez-Forteza}, \citenamefont {Kalaghatgi}, \citenamefont {Ohme},\
  and\ \citenamefont {Pannarale}}]{London:2017bcn}%
  \BibitemOpen
  \bibfield  {author} {\bibinfo {author} {\bibfnamefont {L.}~\bibnamefont
  {London}}, \bibinfo {author} {\bibfnamefont {S.}~\bibnamefont {Khan}},
  \bibinfo {author} {\bibfnamefont {E.}~\bibnamefont {Fauchon-Jones}}, \bibinfo
  {author} {\bibfnamefont {C.}~\bibnamefont {Garc{\'\i}a}}, \bibinfo {author}
  {\bibfnamefont {M.}~\bibnamefont {Hannam}}, \bibinfo {author} {\bibfnamefont
  {S.}~\bibnamefont {Husa}}, \bibinfo {author} {\bibfnamefont {X.}~\bibnamefont
  {Jim{\'e}nez-Forteza}}, \bibinfo {author} {\bibfnamefont {C.}~\bibnamefont
  {Kalaghatgi}}, \bibinfo {author} {\bibfnamefont {F.}~\bibnamefont {Ohme}},\
  and\ \bibinfo {author} {\bibfnamefont {F.}~\bibnamefont {Pannarale}},\
  }\bibfield  {title} {\bibinfo {title} {{First higher-multipole model of
  gravitational waves from spinning and coalescing black-hole binaries}},\
  }\href {https://doi.org/10.1103/PhysRevLett.120.161102} {\bibfield  {journal}
  {\bibinfo  {journal} {Phys. Rev. Lett.}\ }\textbf {\bibinfo {volume} {120}},\
  \bibinfo {pages} {161102} (\bibinfo {year} {2018})},\ \Eprint
  {https://arxiv.org/abs/1708.00404} {arXiv:1708.00404 [gr-qc]} \BibitemShut
  {NoStop}%
\bibitem [{\citenamefont {Usman}\ \emph {et~al.}(2019)\citenamefont {Usman},
  \citenamefont {Mills},\ and\ \citenamefont {Fairhurst}}]{Usman:2018imj}%
  \BibitemOpen
  \bibfield  {author} {\bibinfo {author} {\bibfnamefont {S.~A.}\ \bibnamefont
  {Usman}}, \bibinfo {author} {\bibfnamefont {J.~C.}\ \bibnamefont {Mills}},\
  and\ \bibinfo {author} {\bibfnamefont {S.}~\bibnamefont {Fairhurst}},\
  }\bibfield  {title} {\bibinfo {title} {{Constraining the Inclinations of
  Binary Mergers from Gravitational-wave Observations}},\ }\href
  {https://doi.org/10.3847/1538-4357/ab0b3e} {\bibfield  {journal} {\bibinfo
  {journal} {Astrophys. J.}\ }\textbf {\bibinfo {volume} {877}},\ \bibinfo
  {pages} {82} (\bibinfo {year} {2019})},\ \Eprint
  {https://arxiv.org/abs/1809.10727} {arXiv:1809.10727 [gr-qc]} \BibitemShut
  {NoStop}%
\bibitem [{\citenamefont {Chassande-Mottin}\ \emph {et~al.}(2019)\citenamefont
  {Chassande-Mottin}, \citenamefont {Leyde}, \citenamefont {Mastrogiovanni},\
  and\ \citenamefont {Steer}}]{Chassande-Mottin:2019nnz}%
  \BibitemOpen
  \bibfield  {author} {\bibinfo {author} {\bibfnamefont {E.}~\bibnamefont
  {Chassande-Mottin}}, \bibinfo {author} {\bibfnamefont {K.}~\bibnamefont
  {Leyde}}, \bibinfo {author} {\bibfnamefont {S.}~\bibnamefont
  {Mastrogiovanni}},\ and\ \bibinfo {author} {\bibfnamefont {D.~A.}\
  \bibnamefont {Steer}},\ }\bibfield  {title} {\bibinfo {title} {{Gravitational
  wave observations, distance measurement uncertainties, and cosmology}},\
  }\href {https://doi.org/10.1103/PhysRevD.100.083514} {\bibfield  {journal}
  {\bibinfo  {journal} {Phys. Rev. D}\ }\textbf {\bibinfo {volume} {100}},\
  \bibinfo {pages} {083514} (\bibinfo {year} {2019})},\ \Eprint
  {https://arxiv.org/abs/1906.02670} {arXiv:1906.02670 [astro-ph.CO]}
  \BibitemShut {NoStop}%
\bibitem [{\citenamefont {Riess}\ \emph {et~al.}(2022)\citenamefont {Riess}
  \emph {et~al.}}]{Riess:2021jrx}%
  \BibitemOpen
  \bibfield  {author} {\bibinfo {author} {\bibfnamefont {A.~G.}\ \bibnamefont
  {Riess}} \emph {et~al.},\ }\bibfield  {title} {\bibinfo {title} {{A
  Comprehensive Measurement of the Local Value of the Hubble Constant with 1 km
  s$^{-1}$ Mpc$^{-1}$ Uncertainty from the Hubble Space Telescope and the SH0ES
  Team}},\ }\href {https://doi.org/10.3847/2041-8213/ac5c5b} {\bibfield
  {journal} {\bibinfo  {journal} {Astrophys. J. Lett.}\ }\textbf {\bibinfo
  {volume} {934}},\ \bibinfo {pages} {L7} (\bibinfo {year} {2022})},\ \Eprint
  {https://arxiv.org/abs/2112.04510} {arXiv:2112.04510 [astro-ph.CO]}
  \BibitemShut {NoStop}%
\bibitem [{\citenamefont {Sathyaprakash}\ and\ \citenamefont
  {Dhurandhar}(1991)}]{Sathyaprakash:1991mt}%
  \BibitemOpen
  \bibfield  {author} {\bibinfo {author} {\bibfnamefont {B.~S.}\ \bibnamefont
  {Sathyaprakash}}\ and\ \bibinfo {author} {\bibfnamefont {S.~V.}\ \bibnamefont
  {Dhurandhar}},\ }\bibfield  {title} {\bibinfo {title} {{Choice of filters for
  the detection of gravitational waves from coalescing binaries}},\ }\href
  {https://doi.org/10.1103/PhysRevD.44.3819} {\bibfield  {journal} {\bibinfo
  {journal} {Phys. Rev. D}\ }\textbf {\bibinfo {volume} {44}},\ \bibinfo
  {pages} {3819} (\bibinfo {year} {1991})}\BibitemShut {NoStop}%
\bibitem [{\citenamefont {Allen}\ \emph {et~al.}(2012)\citenamefont {Allen},
  \citenamefont {Anderson}, \citenamefont {Brady}, \citenamefont {Brown},\ and\
  \citenamefont {Creighton}}]{Allen:2005fk}%
  \BibitemOpen
  \bibfield  {author} {\bibinfo {author} {\bibfnamefont {B.}~\bibnamefont
  {Allen}}, \bibinfo {author} {\bibfnamefont {W.~G.}\ \bibnamefont {Anderson}},
  \bibinfo {author} {\bibfnamefont {P.~R.}\ \bibnamefont {Brady}}, \bibinfo
  {author} {\bibfnamefont {D.~A.}\ \bibnamefont {Brown}},\ and\ \bibinfo
  {author} {\bibfnamefont {J.~D.~E.}\ \bibnamefont {Creighton}},\ }\bibfield
  {title} {\bibinfo {title} {{FINDCHIRP: An Algorithm for detection of
  gravitational waves from inspiraling compact binaries}},\ }\href
  {https://doi.org/10.1103/PhysRevD.85.122006} {\bibfield  {journal} {\bibinfo
  {journal} {Phys. Rev. D}\ }\textbf {\bibinfo {volume} {85}},\ \bibinfo
  {pages} {122006} (\bibinfo {year} {2012})},\ \Eprint
  {https://arxiv.org/abs/gr-qc/0509116} {arXiv:gr-qc/0509116} \BibitemShut
  {NoStop}%
\bibitem [{\citenamefont {Abbott}\ \emph {et~al.}(2016)\citenamefont {Abbott}
  \emph {et~al.}}]{LIGOScientific:2016vbw}%
  \BibitemOpen
  \bibfield  {author} {\bibinfo {author} {\bibfnamefont {B.~P.}\ \bibnamefont
  {Abbott}} \emph {et~al.} (\bibinfo {collaboration} {LIGO Scientific,
  Virgo}),\ }\bibfield  {title} {\bibinfo {title} {{GW150914: First results
  from the search for binary black hole coalescence with Advanced LIGO}},\
  }\href {https://doi.org/10.1103/PhysRevD.93.122003} {\bibfield  {journal}
  {\bibinfo  {journal} {Phys. Rev. D}\ }\textbf {\bibinfo {volume} {93}},\
  \bibinfo {pages} {122003} (\bibinfo {year} {2016})},\ \Eprint
  {https://arxiv.org/abs/1602.03839} {arXiv:1602.03839 [gr-qc]} \BibitemShut
  {NoStop}%
\bibitem [{\citenamefont {Veitch}\ \emph {et~al.}(2015)\citenamefont {Veitch}
  \emph {et~al.}}]{Veitch:2014wba}%
  \BibitemOpen
  \bibfield  {author} {\bibinfo {author} {\bibfnamefont {J.}~\bibnamefont
  {Veitch}} \emph {et~al.},\ }\bibfield  {title} {\bibinfo {title} {{Parameter
  estimation for compact binaries with ground-based gravitational-wave
  observations using the LALInference software library}},\ }\href
  {https://doi.org/10.1103/PhysRevD.91.042003} {\bibfield  {journal} {\bibinfo
  {journal} {Phys. Rev. D}\ }\textbf {\bibinfo {volume} {91}},\ \bibinfo
  {pages} {042003} (\bibinfo {year} {2015})},\ \Eprint
  {https://arxiv.org/abs/1409.7215} {arXiv:1409.7215 [gr-qc]} \BibitemShut
  {NoStop}%
\bibitem [{\citenamefont {Romero-Shaw}\ \emph {et~al.}(2020)\citenamefont
  {Romero-Shaw} \emph {et~al.}}]{Romero-Shaw:2020owr}%
  \BibitemOpen
  \bibfield  {author} {\bibinfo {author} {\bibfnamefont {I.~M.}\ \bibnamefont
  {Romero-Shaw}} \emph {et~al.},\ }\bibfield  {title} {\bibinfo {title}
  {{Bayesian inference for compact binary coalescences with bilby: validation
  and application to the first LIGO\textendash{}Virgo gravitational-wave
  transient catalogue}},\ }\href {https://doi.org/10.1093/mnras/staa2850}
  {\bibfield  {journal} {\bibinfo  {journal} {Mon. Not. Roy. Astron. Soc.}\
  }\textbf {\bibinfo {volume} {499}},\ \bibinfo {pages} {3295} (\bibinfo {year}
  {2020})},\ \Eprint {https://arxiv.org/abs/2006.00714} {arXiv:2006.00714
  [astro-ph.IM]} \BibitemShut {NoStop}%
\bibitem [{\citenamefont {Taylor}\ \emph {et~al.}(2012)\citenamefont {Taylor},
  \citenamefont {Gair},\ and\ \citenamefont {Mandel}}]{Taylor:2011fs}%
  \BibitemOpen
  \bibfield  {author} {\bibinfo {author} {\bibfnamefont {S.~R.}\ \bibnamefont
  {Taylor}}, \bibinfo {author} {\bibfnamefont {J.~R.}\ \bibnamefont {Gair}},\
  and\ \bibinfo {author} {\bibfnamefont {I.}~\bibnamefont {Mandel}},\
  }\bibfield  {title} {\bibinfo {title} {{Hubble without the Hubble: Cosmology
  using advanced gravitational-wave detectors alone}},\ }\href
  {https://doi.org/10.1103/PhysRevD.85.023535} {\bibfield  {journal} {\bibinfo
  {journal} {Phys. Rev. D}\ }\textbf {\bibinfo {volume} {85}},\ \bibinfo
  {pages} {023535} (\bibinfo {year} {2012})},\ \Eprint
  {https://arxiv.org/abs/1108.5161} {arXiv:1108.5161 [gr-qc]} \BibitemShut
  {NoStop}%
\bibitem [{\citenamefont {Messenger}\ and\ \citenamefont
  {Read}(2012)}]{Messenger:2011gi}%
  \BibitemOpen
  \bibfield  {author} {\bibinfo {author} {\bibfnamefont {C.}~\bibnamefont
  {Messenger}}\ and\ \bibinfo {author} {\bibfnamefont {J.}~\bibnamefont
  {Read}},\ }\bibfield  {title} {\bibinfo {title} {{Measuring a cosmological
  distance-redshift relationship using only gravitational wave observations of
  binary neutron star coalescences}},\ }\href
  {https://doi.org/10.1103/PhysRevLett.108.091101} {\bibfield  {journal}
  {\bibinfo  {journal} {Phys. Rev. Lett.}\ }\textbf {\bibinfo {volume} {108}},\
  \bibinfo {pages} {091101} (\bibinfo {year} {2012})},\ \Eprint
  {https://arxiv.org/abs/1107.5725} {arXiv:1107.5725 [gr-qc]} \BibitemShut
  {NoStop}%
\bibitem [{\citenamefont {Messenger}\ \emph {et~al.}(2014)\citenamefont
  {Messenger}, \citenamefont {Takami}, \citenamefont {Gossan}, \citenamefont
  {Rezzolla},\ and\ \citenamefont {Sathyaprakash}}]{Messenger:2013fya}%
  \BibitemOpen
  \bibfield  {author} {\bibinfo {author} {\bibfnamefont {C.}~\bibnamefont
  {Messenger}}, \bibinfo {author} {\bibfnamefont {K.}~\bibnamefont {Takami}},
  \bibinfo {author} {\bibfnamefont {S.}~\bibnamefont {Gossan}}, \bibinfo
  {author} {\bibfnamefont {L.}~\bibnamefont {Rezzolla}},\ and\ \bibinfo
  {author} {\bibfnamefont {B.~S.}\ \bibnamefont {Sathyaprakash}},\ }\bibfield
  {title} {\bibinfo {title} {{Source Redshifts from Gravitational-Wave
  Observations of Binary Neutron Star Mergers}},\ }\href
  {https://doi.org/10.1103/PhysRevX.4.041004} {\bibfield  {journal} {\bibinfo
  {journal} {Phys. Rev. X}\ }\textbf {\bibinfo {volume} {4}},\ \bibinfo {pages}
  {041004} (\bibinfo {year} {2014})},\ \Eprint
  {https://arxiv.org/abs/1312.1862} {arXiv:1312.1862 [gr-qc]} \BibitemShut
  {NoStop}%
\bibitem [{\citenamefont {Del~Pozzo}\ \emph {et~al.}(2017)\citenamefont
  {Del~Pozzo}, \citenamefont {Li},\ and\ \citenamefont
  {Messenger}}]{DelPozzo:2015bna}%
  \BibitemOpen
  \bibfield  {author} {\bibinfo {author} {\bibfnamefont {W.}~\bibnamefont
  {Del~Pozzo}}, \bibinfo {author} {\bibfnamefont {T.~G.~F.}\ \bibnamefont
  {Li}},\ and\ \bibinfo {author} {\bibfnamefont {C.}~\bibnamefont
  {Messenger}},\ }\bibfield  {title} {\bibinfo {title} {{Cosmological inference
  using only gravitational wave observations of binary neutron stars}},\ }\href
  {https://doi.org/10.1103/PhysRevD.95.043502} {\bibfield  {journal} {\bibinfo
  {journal} {Phys. Rev. D}\ }\textbf {\bibinfo {volume} {95}},\ \bibinfo
  {pages} {043502} (\bibinfo {year} {2017})},\ \Eprint
  {https://arxiv.org/abs/1506.06590} {arXiv:1506.06590 [gr-qc]} \BibitemShut
  {NoStop}%
\bibitem [{\citenamefont {Abbott}\ \emph
  {et~al.}(2017{\natexlab{b}})\citenamefont {Abbott} \emph
  {et~al.}}]{LIGOScientific:2017vwq}%
  \BibitemOpen
  \bibfield  {author} {\bibinfo {author} {\bibfnamefont {B.~P.}\ \bibnamefont
  {Abbott}} \emph {et~al.} (\bibinfo {collaboration} {LIGO Scientific,
  Virgo}),\ }\bibfield  {title} {\bibinfo {title} {{GW170817: Observation of
  Gravitational Waves from a Binary Neutron Star Inspiral}},\ }\href
  {https://doi.org/10.1103/PhysRevLett.119.161101} {\bibfield  {journal}
  {\bibinfo  {journal} {Phys. Rev. Lett.}\ }\textbf {\bibinfo {volume} {119}},\
  \bibinfo {pages} {161101} (\bibinfo {year} {2017}{\natexlab{b}})},\ \Eprint
  {https://arxiv.org/abs/1710.05832} {arXiv:1710.05832 [gr-qc]} \BibitemShut
  {NoStop}%
\bibitem [{\citenamefont {Abbott}\ \emph
  {et~al.}(2017{\natexlab{c}})\citenamefont {Abbott} \emph
  {et~al.}}]{LIGOScientific:2017ync}%
  \BibitemOpen
  \bibfield  {author} {\bibinfo {author} {\bibfnamefont {B.~P.}\ \bibnamefont
  {Abbott}} \emph {et~al.} (\bibinfo {collaboration} {LIGO Scientific, Virgo,
  Fermi GBM, INTEGRAL, IceCube, AstroSat Cadmium Zinc Telluride Imager Team,
  IPN, Insight-Hxmt, ANTARES, Swift, AGILE Team, 1M2H Team, Dark Energy Camera
  GW-EM, DES, DLT40, GRAWITA, Fermi-LAT, ATCA, ASKAP, Las Cumbres Observatory
  Group, OzGrav, DWF (Deeper Wider Faster Program), AST3, CAASTRO, VINROUGE,
  MASTER, J-GEM, GROWTH, JAGWAR, CaltechNRAO, TTU-NRAO, NuSTAR, Pan-STARRS,
  MAXI Team, TZAC Consortium, KU, Nordic Optical Telescope, ePESSTO, GROND,
  Texas Tech University, SALT Group, TOROS, BOOTES, MWA, CALET, IKI-GW
  Follow-up, H.E.S.S., LOFAR, LWA, HAWC, Pierre Auger, ALMA, Euro VLBI Team, Pi
  of Sky, Chandra Team at McGill University, DFN, ATLAS Telescopes, High Time
  Resolution Universe Survey, RIMAS, RATIR, SKA South Africa/MeerKAT}),\
  }\bibfield  {title} {\bibinfo {title} {{Multi-messenger Observations of a
  Binary Neutron Star Merger}},\ }\href
  {https://doi.org/10.3847/2041-8213/aa91c9} {\bibfield  {journal} {\bibinfo
  {journal} {Astrophys. J. Lett.}\ }\textbf {\bibinfo {volume} {848}},\
  \bibinfo {pages} {L12} (\bibinfo {year} {2017}{\natexlab{c}})},\ \Eprint
  {https://arxiv.org/abs/1710.05833} {arXiv:1710.05833 [astro-ph.HE]}
  \BibitemShut {NoStop}%
\bibitem [{\citenamefont {Metzger}\ and\ \citenamefont
  {Berger}(2012)}]{Metzger:2011bv}%
  \BibitemOpen
  \bibfield  {author} {\bibinfo {author} {\bibfnamefont {B.~D.}\ \bibnamefont
  {Metzger}}\ and\ \bibinfo {author} {\bibfnamefont {E.}~\bibnamefont
  {Berger}},\ }\bibfield  {title} {\bibinfo {title} {{What is the Most
  Promising Electromagnetic Counterpart of a Neutron Star Binary Merger?}},\
  }\href {https://doi.org/10.1088/0004-637X/746/1/48} {\bibfield  {journal}
  {\bibinfo  {journal} {Astrophys. J.}\ }\textbf {\bibinfo {volume} {746}},\
  \bibinfo {pages} {48} (\bibinfo {year} {2012})},\ \Eprint
  {https://arxiv.org/abs/1108.6056} {arXiv:1108.6056 [astro-ph.HE]}
  \BibitemShut {NoStop}%
\bibitem [{\citenamefont {Heinzel}\ \emph {et~al.}(2021)\citenamefont
  {Heinzel}, \citenamefont {Coughlin}, \citenamefont {Dietrich}, \citenamefont
  {Bulla}, \citenamefont {Antier}, \citenamefont {Christensen}, \citenamefont
  {Coulter}, \citenamefont {Foley}, \citenamefont {Issa},\ and\ \citenamefont
  {Khetan}}]{Heinzel:2020qlt}%
  \BibitemOpen
  \bibfield  {author} {\bibinfo {author} {\bibfnamefont {J.}~\bibnamefont
  {Heinzel}}, \bibinfo {author} {\bibfnamefont {M.~W.}\ \bibnamefont
  {Coughlin}}, \bibinfo {author} {\bibfnamefont {T.}~\bibnamefont {Dietrich}},
  \bibinfo {author} {\bibfnamefont {M.}~\bibnamefont {Bulla}}, \bibinfo
  {author} {\bibfnamefont {S.}~\bibnamefont {Antier}}, \bibinfo {author}
  {\bibfnamefont {N.}~\bibnamefont {Christensen}}, \bibinfo {author}
  {\bibfnamefont {D.~A.}\ \bibnamefont {Coulter}}, \bibinfo {author}
  {\bibfnamefont {R.~J.}\ \bibnamefont {Foley}}, \bibinfo {author}
  {\bibfnamefont {L.}~\bibnamefont {Issa}},\ and\ \bibinfo {author}
  {\bibfnamefont {N.}~\bibnamefont {Khetan}},\ }\bibfield  {title} {\bibinfo
  {title} {{Comparing inclination dependent analyses of kilonova transients}},\
  }\href {https://doi.org/10.1093/mnras/stab221} {\bibfield  {journal}
  {\bibinfo  {journal} {Mon. Not. Roy. Astron. Soc.}\ }\textbf {\bibinfo
  {volume} {502}},\ \bibinfo {pages} {3057} (\bibinfo {year} {2021})},\ \Eprint
  {https://arxiv.org/abs/2010.10746} {arXiv:2010.10746 [astro-ph.HE]}
  \BibitemShut {NoStop}%
\bibitem [{\citenamefont {Holmbeck}\ \emph {et~al.}(2023)\citenamefont
  {Holmbeck}, \citenamefont {Barnes}, \citenamefont {Lund}, \citenamefont
  {Sprouse}, \citenamefont {McLaughlin},\ and\ \citenamefont
  {Mumpower}}]{Holmbeck:2023gck}%
  \BibitemOpen
  \bibfield  {author} {\bibinfo {author} {\bibfnamefont {E.~M.}\ \bibnamefont
  {Holmbeck}}, \bibinfo {author} {\bibfnamefont {J.}~\bibnamefont {Barnes}},
  \bibinfo {author} {\bibfnamefont {K.~A.}\ \bibnamefont {Lund}}, \bibinfo
  {author} {\bibfnamefont {T.~M.}\ \bibnamefont {Sprouse}}, \bibinfo {author}
  {\bibfnamefont {G.~C.}\ \bibnamefont {McLaughlin}},\ and\ \bibinfo {author}
  {\bibfnamefont {M.~R.}\ \bibnamefont {Mumpower}},\ }\bibfield  {title}
  {\bibinfo {title} {{Superheavy Elements in Kilonovae}},\ }\href
  {https://doi.org/10.3847/2041-8213/acd9cb} {\bibfield  {journal} {\bibinfo
  {journal} {Astrophys. J. Lett.}\ }\textbf {\bibinfo {volume} {951}},\
  \bibinfo {pages} {L13} (\bibinfo {year} {2023})},\ \Eprint
  {https://arxiv.org/abs/2304.02125} {arXiv:2304.02125 [astro-ph.HE]}
  \BibitemShut {NoStop}%
\bibitem [{\citenamefont {Berger}(2014)}]{Berger:2013jza}%
  \BibitemOpen
  \bibfield  {author} {\bibinfo {author} {\bibfnamefont {E.}~\bibnamefont
  {Berger}},\ }\bibfield  {title} {\bibinfo {title} {{Short-Duration Gamma-Ray
  Bursts}},\ }\href {https://doi.org/10.1146/annurev-astro-081913-035926}
  {\bibfield  {journal} {\bibinfo  {journal} {Ann. Rev. Astron. Astrophys.}\
  }\textbf {\bibinfo {volume} {52}},\ \bibinfo {pages} {43} (\bibinfo {year}
  {2014})},\ \Eprint {https://arxiv.org/abs/1311.2603} {arXiv:1311.2603
  [astro-ph.HE]} \BibitemShut {NoStop}%
\bibitem [{\citenamefont {Minaev}\ \emph {et~al.}(2024)\citenamefont {Minaev},
  \citenamefont {Pozanenko}, \citenamefont {Grebenev}, \citenamefont
  {Chelovekov}, \citenamefont {Pankov}, \citenamefont {Khabibullin},
  \citenamefont {Inasaridze},\ and\ \citenamefont
  {Novichonok}}]{Minaev:2024hvi}%
  \BibitemOpen
  \bibfield  {author} {\bibinfo {author} {\bibfnamefont {P.~Y.}\ \bibnamefont
  {Minaev}}, \bibinfo {author} {\bibfnamefont {A.~S.}\ \bibnamefont
  {Pozanenko}}, \bibinfo {author} {\bibfnamefont {S.~A.}\ \bibnamefont
  {Grebenev}}, \bibinfo {author} {\bibfnamefont {I.~V.}\ \bibnamefont
  {Chelovekov}}, \bibinfo {author} {\bibfnamefont {N.~S.}\ \bibnamefont
  {Pankov}}, \bibinfo {author} {\bibfnamefont {A.~A.}\ \bibnamefont
  {Khabibullin}}, \bibinfo {author} {\bibfnamefont {R.~Y.}\ \bibnamefont
  {Inasaridze}},\ and\ \bibinfo {author} {\bibfnamefont {A.~O.}\ \bibnamefont
  {Novichonok}},\ }\bibfield  {title} {\bibinfo {title} {{GRB
  231115A\textemdash{}a Magnetar Giant Flare in the M82 Galaxy}},\ }\href
  {https://doi.org/10.1134/S1063773724600152} {\bibfield  {journal} {\bibinfo
  {journal} {Astron. Lett.}\ }\textbf {\bibinfo {volume} {50}},\ \bibinfo
  {pages} {1} (\bibinfo {year} {2024})},\ \Eprint
  {https://arxiv.org/abs/2402.08623} {arXiv:2402.08623 [astro-ph.HE]}
  \BibitemShut {NoStop}%
\bibitem [{\citenamefont {Abbott}\ \emph
  {et~al.}(2017{\natexlab{d}})\citenamefont {Abbott} \emph
  {et~al.}}]{LIGOScientific:2017zic}%
  \BibitemOpen
  \bibfield  {author} {\bibinfo {author} {\bibfnamefont {B.~P.}\ \bibnamefont
  {Abbott}} \emph {et~al.} (\bibinfo {collaboration} {LIGO Scientific, Virgo,
  Fermi-GBM, INTEGRAL}),\ }\bibfield  {title} {\bibinfo {title} {{Gravitational
  Waves and Gamma-rays from a Binary Neutron Star Merger: GW170817 and GRB
  170817A}},\ }\href {https://doi.org/10.3847/2041-8213/aa920c} {\bibfield
  {journal} {\bibinfo  {journal} {Astrophys. J. Lett.}\ }\textbf {\bibinfo
  {volume} {848}},\ \bibinfo {pages} {L13} (\bibinfo {year}
  {2017}{\natexlab{d}})},\ \Eprint {https://arxiv.org/abs/1710.05834}
  {arXiv:1710.05834 [astro-ph.HE]} \BibitemShut {NoStop}%
\bibitem [{\citenamefont {Lv}\ \emph {et~al.}(2010)\citenamefont {Lv},
  \citenamefont {Liang}, \citenamefont {Zhang},\ and\ \citenamefont
  {Zhang}}]{Lv:2010bz}%
  \BibitemOpen
  \bibfield  {author} {\bibinfo {author} {\bibfnamefont {H.}~\bibnamefont
  {Lv}}, \bibinfo {author} {\bibfnamefont {E.}~\bibnamefont {Liang}}, \bibinfo
  {author} {\bibfnamefont {B.}~\bibnamefont {Zhang}},\ and\ \bibinfo {author}
  {\bibfnamefont {B.}~\bibnamefont {Zhang}},\ }\bibfield  {title} {\bibinfo
  {title} {{A New Classification Method for Gamma-Ray Bursts}},\ }\href
  {https://doi.org/10.1088/0004-637X/725/2/1965} {\bibfield  {journal}
  {\bibinfo  {journal} {Astrophys. J.}\ }\textbf {\bibinfo {volume} {725}},\
  \bibinfo {pages} {1965} (\bibinfo {year} {2010})},\ \Eprint
  {https://arxiv.org/abs/1001.0598} {arXiv:1001.0598 [astro-ph.HE]}
  \BibitemShut {NoStop}%
\bibitem [{\citenamefont {Guidorzi}\ \emph {et~al.}(2017)\citenamefont
  {Guidorzi} \emph {et~al.}}]{Guidorzi:2017ogy}%
  \BibitemOpen
  \bibfield  {author} {\bibinfo {author} {\bibfnamefont {C.}~\bibnamefont
  {Guidorzi}} \emph {et~al.},\ }\bibfield  {title} {\bibinfo {title} {{Improved
  Constraints on $H_0$ from a Combined Analysis of Gravitational-wave and
  Electromagnetic Emission from GW170817}},\ }\href
  {https://doi.org/10.3847/2041-8213/aaa009} {\bibfield  {journal} {\bibinfo
  {journal} {Astrophys. J. Lett.}\ }\textbf {\bibinfo {volume} {851}},\
  \bibinfo {pages} {L36} (\bibinfo {year} {2017})},\ \Eprint
  {https://arxiv.org/abs/1710.06426} {arXiv:1710.06426 [astro-ph.CO]}
  \BibitemShut {NoStop}%
\bibitem [{\citenamefont {Chen}(2020)}]{Chen:2020dyt}%
  \BibitemOpen
  \bibfield  {author} {\bibinfo {author} {\bibfnamefont {H.-Y.}\ \bibnamefont
  {Chen}},\ }\bibfield  {title} {\bibinfo {title} {{Systematic Uncertainty of
  Standard Sirens from the Viewing Angle of Binary Neutron Star Inspirals}},\
  }\href {https://doi.org/10.1103/PhysRevLett.125.201301} {\bibfield  {journal}
  {\bibinfo  {journal} {Phys. Rev. Lett.}\ }\textbf {\bibinfo {volume} {125}},\
  \bibinfo {pages} {201301} (\bibinfo {year} {2020})},\ \Eprint
  {https://arxiv.org/abs/2006.02779} {arXiv:2006.02779 [astro-ph.HE]}
  \BibitemShut {NoStop}%
\bibitem [{\citenamefont {Escorial}\ \emph {et~al.}(2023)\citenamefont
  {Escorial} \emph {et~al.}}]{Escorial:2022nvp}%
  \BibitemOpen
  \bibfield  {author} {\bibinfo {author} {\bibfnamefont {A.~R.}\ \bibnamefont
  {Escorial}} \emph {et~al.},\ }\bibfield  {title} {\bibinfo {title} {{The Jet
  Opening Angle and Event Rate Distributions of Short Gamma-Ray Bursts from
  Late-time X-Ray Afterglows}},\ }\href
  {https://doi.org/10.3847/1538-4357/acf830} {\bibfield  {journal} {\bibinfo
  {journal} {Astrophys. J.}\ }\textbf {\bibinfo {volume} {959}},\ \bibinfo
  {pages} {13} (\bibinfo {year} {2023})},\ \Eprint
  {https://arxiv.org/abs/2210.05695} {arXiv:2210.05695 [astro-ph.HE]}
  \BibitemShut {NoStop}%
\bibitem [{\citenamefont {Palmese}\ \emph {et~al.}(2024)\citenamefont
  {Palmese}, \citenamefont {Kaur}, \citenamefont {Hajela}, \citenamefont
  {Margutti}, \citenamefont {McDowell},\ and\ \citenamefont
  {MacFadyen}}]{Palmese:2023beh}%
  \BibitemOpen
  \bibfield  {author} {\bibinfo {author} {\bibfnamefont {A.}~\bibnamefont
  {Palmese}}, \bibinfo {author} {\bibfnamefont {R.}~\bibnamefont {Kaur}},
  \bibinfo {author} {\bibfnamefont {A.}~\bibnamefont {Hajela}}, \bibinfo
  {author} {\bibfnamefont {R.}~\bibnamefont {Margutti}}, \bibinfo {author}
  {\bibfnamefont {A.}~\bibnamefont {McDowell}},\ and\ \bibinfo {author}
  {\bibfnamefont {A.}~\bibnamefont {MacFadyen}},\ }\bibfield  {title} {\bibinfo
  {title} {{Standard siren measurement of the Hubble constant using GW170817
  and the latest observations of the electromagnetic counterpart afterglow}},\
  }\href {https://doi.org/10.1103/PhysRevD.109.063508} {\bibfield  {journal}
  {\bibinfo  {journal} {Phys. Rev. D}\ }\textbf {\bibinfo {volume} {109}},\
  \bibinfo {pages} {063508} (\bibinfo {year} {2024})},\ \Eprint
  {https://arxiv.org/abs/2305.19914} {arXiv:2305.19914 [astro-ph.CO]}
  \BibitemShut {NoStop}%
\bibitem [{\citenamefont {Gianfagna}\ \emph {et~al.}(2024)\citenamefont
  {Gianfagna}, \citenamefont {Piro}, \citenamefont {Pannarale}, \citenamefont
  {Van~Eerten}, \citenamefont {Ricci},\ and\ \citenamefont
  {Ryan}}]{Gianfagna:2023cgk}%
  \BibitemOpen
  \bibfield  {author} {\bibinfo {author} {\bibfnamefont {G.}~\bibnamefont
  {Gianfagna}}, \bibinfo {author} {\bibfnamefont {L.}~\bibnamefont {Piro}},
  \bibinfo {author} {\bibfnamefont {F.}~\bibnamefont {Pannarale}}, \bibinfo
  {author} {\bibfnamefont {H.}~\bibnamefont {Van~Eerten}}, \bibinfo {author}
  {\bibfnamefont {F.}~\bibnamefont {Ricci}},\ and\ \bibinfo {author}
  {\bibfnamefont {G.}~\bibnamefont {Ryan}},\ }\bibfield  {title} {\bibinfo
  {title} {{Potential biases and prospects for the Hubble constant estimation
  via electromagnetic and gravitational-wave joint analyses}},\ }\href
  {https://doi.org/10.1093/mnras/stae198} {\bibfield  {journal} {\bibinfo
  {journal} {Mon. Not. Roy. Astron. Soc.}\ }\textbf {\bibinfo {volume} {528}},\
  \bibinfo {pages} {2600} (\bibinfo {year} {2024})},\ \Eprint
  {https://arxiv.org/abs/2309.17073} {arXiv:2309.17073 [astro-ph.HE]}
  \BibitemShut {NoStop}%
\bibitem [{\citenamefont {Chen}\ \emph
  {et~al.}(2024{\natexlab{a}})\citenamefont {Chen}, \citenamefont {Talbot},\
  and\ \citenamefont {Chase}}]{Chen:2023dgw}%
  \BibitemOpen
  \bibfield  {author} {\bibinfo {author} {\bibfnamefont {H.-Y.}\ \bibnamefont
  {Chen}}, \bibinfo {author} {\bibfnamefont {C.}~\bibnamefont {Talbot}},\ and\
  \bibinfo {author} {\bibfnamefont {E.~A.}\ \bibnamefont {Chase}},\ }\bibfield
  {title} {\bibinfo {title} {{Mitigating the Counterpart Selection Effect for
  Standard Sirens}},\ }\href {https://doi.org/10.1103/PhysRevLett.132.191003}
  {\bibfield  {journal} {\bibinfo  {journal} {Phys. Rev. Lett.}\ }\textbf
  {\bibinfo {volume} {132}},\ \bibinfo {pages} {191003} (\bibinfo {year}
  {2024}{\natexlab{a}})},\ \Eprint {https://arxiv.org/abs/2307.10402}
  {arXiv:2307.10402 [astro-ph.CO]} \BibitemShut {NoStop}%
\bibitem [{\citenamefont {{M{\"u}ller}}\ \emph {et~al.}(2024)\citenamefont
  {{M{\"u}ller}}, \citenamefont {{Mukherjee}},\ and\ \citenamefont
  {{Ryan}}}]{Muller_Mukherjee_Geoffrey_2024}%
  \BibitemOpen
  \bibfield  {author} {\bibinfo {author} {\bibfnamefont {M.}~\bibnamefont
  {{M{\"u}ller}}}, \bibinfo {author} {\bibfnamefont {S.}~\bibnamefont
  {{Mukherjee}}},\ and\ \bibinfo {author} {\bibfnamefont {G.}~\bibnamefont
  {{Ryan}}},\ }\bibfield  {title} {\bibinfo {title} {{Be careful in
  multi-messenger inference of the Hubble constant: A path forward for robust
  inference}},\ }\href {https://doi.org/10.48550/arXiv.2406.11965} {\bibfield
  {journal} {\bibinfo  {journal} {arXiv e-prints}\ ,\ \bibinfo {eid}
  {arXiv:2406.11965}} (\bibinfo {year} {2024})},\ \Eprint
  {https://arxiv.org/abs/2406.11965} {arXiv:2406.11965 [astro-ph.CO]}
  \BibitemShut {NoStop}%
\bibitem [{\citenamefont {{Nicolaou}}\ \emph {et~al.}(2020)\citenamefont
  {{Nicolaou}}, \citenamefont {{Lahav}}, \citenamefont {{Lemos}}, \citenamefont
  {{Hartley}},\ and\ \citenamefont {{Braden}}}]{Nicolaou_et_al_2020}%
  \BibitemOpen
  \bibfield  {author} {\bibinfo {author} {\bibfnamefont {C.}~\bibnamefont
  {{Nicolaou}}}, \bibinfo {author} {\bibfnamefont {O.}~\bibnamefont {{Lahav}}},
  \bibinfo {author} {\bibfnamefont {P.}~\bibnamefont {{Lemos}}}, \bibinfo
  {author} {\bibfnamefont {W.}~\bibnamefont {{Hartley}}},\ and\ \bibinfo
  {author} {\bibfnamefont {J.}~\bibnamefont {{Braden}}},\ }\bibfield  {title}
  {\bibinfo {title} {{The impact of peculiar velocities on the estimation of
  the Hubble constant from gravitational wave standard sirens}},\ }\href@noop
  {} {\bibfield  {journal} {\bibinfo  {journal} {Mon. Not. Roy. Astron. Soc.}\
  }\textbf {\bibinfo {volume} {495}},\ \bibinfo {pages} {90} (\bibinfo {year}
  {2020})}\BibitemShut {NoStop}%
\bibitem [{\citenamefont {Mukherjee}\ \emph
  {et~al.}(2021{\natexlab{a}})\citenamefont {Mukherjee}, \citenamefont
  {Lavaux}, \citenamefont {Bouchet}, \citenamefont {Jasche}, \citenamefont
  {Wandelt}, \citenamefont {Nissanke}, \citenamefont {Leclercq},\ and\
  \citenamefont {Hotokezaka}}]{Mukherjee_et_al_2021}%
  \BibitemOpen
  \bibfield  {author} {\bibinfo {author} {\bibfnamefont {S.}~\bibnamefont
  {Mukherjee}}, \bibinfo {author} {\bibfnamefont {G.}~\bibnamefont {Lavaux}},
  \bibinfo {author} {\bibfnamefont {F.~R.}\ \bibnamefont {Bouchet}}, \bibinfo
  {author} {\bibfnamefont {J.}~\bibnamefont {Jasche}}, \bibinfo {author}
  {\bibfnamefont {B.~D.}\ \bibnamefont {Wandelt}}, \bibinfo {author}
  {\bibfnamefont {S.~M.}\ \bibnamefont {Nissanke}}, \bibinfo {author}
  {\bibfnamefont {F.}~\bibnamefont {Leclercq}},\ and\ \bibinfo {author}
  {\bibfnamefont {K.}~\bibnamefont {Hotokezaka}},\ }\bibfield  {title}
  {\bibinfo {title} {{Velocity correction for Hubble constant measurements from
  standard sirens}},\ }\href {https://doi.org/10.1051/0004-6361/201936724}
  {\bibfield  {journal} {\bibinfo  {journal} {Astron. Astrophys.}\ }\textbf
  {\bibinfo {volume} {646}},\ \bibinfo {pages} {A65} (\bibinfo {year}
  {2021}{\natexlab{a}})},\ \Eprint {https://arxiv.org/abs/1909.08627}
  {arXiv:1909.08627 [astro-ph.CO]} \BibitemShut {NoStop}%
\bibitem [{\citenamefont {Graham}\ \emph {et~al.}(2020)\citenamefont {Graham}
  \emph {et~al.}}]{Graham:2020gwr}%
  \BibitemOpen
  \bibfield  {author} {\bibinfo {author} {\bibfnamefont {M.~J.}\ \bibnamefont
  {Graham}} \emph {et~al.},\ }\bibfield  {title} {\bibinfo {title} {{Candidate
  Electromagnetic Counterpart to the Binary Black Hole Merger Gravitational
  Wave Event S190521g}},\ }\href
  {https://doi.org/10.1103/PhysRevLett.124.251102} {\bibfield  {journal}
  {\bibinfo  {journal} {Phys. Rev. Lett.}\ }\textbf {\bibinfo {volume} {124}},\
  \bibinfo {pages} {251102} (\bibinfo {year} {2020})},\ \Eprint
  {https://arxiv.org/abs/2006.14122} {arXiv:2006.14122 [astro-ph.HE]}
  \BibitemShut {NoStop}%
\bibitem [{\citenamefont {Chen}\ \emph {et~al.}(2022)\citenamefont {Chen},
  \citenamefont {Haster}, \citenamefont {Vitale}, \citenamefont {Farr},\ and\
  \citenamefont {Isi}}]{Chen:2020gek}%
  \BibitemOpen
  \bibfield  {author} {\bibinfo {author} {\bibfnamefont {H.-Y.}\ \bibnamefont
  {Chen}}, \bibinfo {author} {\bibfnamefont {C.-J.}\ \bibnamefont {Haster}},
  \bibinfo {author} {\bibfnamefont {S.}~\bibnamefont {Vitale}}, \bibinfo
  {author} {\bibfnamefont {W.~M.}\ \bibnamefont {Farr}},\ and\ \bibinfo
  {author} {\bibfnamefont {M.}~\bibnamefont {Isi}},\ }\bibfield  {title}
  {\bibinfo {title} {{A standard siren cosmological measurement from the
  potential GW190521 electromagnetic counterpart ZTF19abanrhr}},\ }\href
  {https://doi.org/10.1093/mnras/stac989} {\bibfield  {journal} {\bibinfo
  {journal} {Mon. Not. Roy. Astron. Soc.}\ }\textbf {\bibinfo {volume} {513}},\
  \bibinfo {pages} {2152} (\bibinfo {year} {2022})},\ \Eprint
  {https://arxiv.org/abs/2009.14057} {arXiv:2009.14057 [astro-ph.CO]}
  \BibitemShut {NoStop}%
\bibitem [{\citenamefont {{Mukherjee}}\ \emph {et~al.}(2020)\citenamefont
  {{Mukherjee}}, \citenamefont {{Ghosh}}, \citenamefont {{Graham}},
  \citenamefont {{Karathanasis}}, \citenamefont {{Kasliwal}}, \citenamefont
  {{Maga{\~n}a Hernandez}}, \citenamefont {{Nissanke}}, \citenamefont
  {{Silvestri}},\ and\ \citenamefont {{Wandelt}}}]{Mukherjee_GW190521_2020}%
  \BibitemOpen
  \bibfield  {author} {\bibinfo {author} {\bibfnamefont {S.}~\bibnamefont
  {{Mukherjee}}}, \bibinfo {author} {\bibfnamefont {A.}~\bibnamefont
  {{Ghosh}}}, \bibinfo {author} {\bibfnamefont {M.~J.}\ \bibnamefont
  {{Graham}}}, \bibinfo {author} {\bibfnamefont {C.}~\bibnamefont
  {{Karathanasis}}}, \bibinfo {author} {\bibfnamefont {M.~M.}\ \bibnamefont
  {{Kasliwal}}}, \bibinfo {author} {\bibfnamefont {I.}~\bibnamefont
  {{Maga{\~n}a Hernandez}}}, \bibinfo {author} {\bibfnamefont {S.~M.}\
  \bibnamefont {{Nissanke}}}, \bibinfo {author} {\bibfnamefont
  {A.}~\bibnamefont {{Silvestri}}},\ and\ \bibinfo {author} {\bibfnamefont
  {B.~D.}\ \bibnamefont {{Wandelt}}},\ }\bibfield  {title} {\bibinfo {title}
  {{First measurement of the Hubble parameter from bright binary black hole
  GW190521}},\ }\href {https://doi.org/10.48550/arXiv.2009.14199} {\bibfield
  {journal} {\bibinfo  {journal} {arXiv e-prints}\ ,\ \bibinfo {eid}
  {arXiv:2009.14199}} (\bibinfo {year} {2020})},\ \Eprint
  {https://arxiv.org/abs/2009.14199} {arXiv:2009.14199 [astro-ph.CO]}
  \BibitemShut {NoStop}%
\bibitem [{\citenamefont {Chen}\ \emph
  {et~al.}(2024{\natexlab{b}})\citenamefont {Chen}, \citenamefont {Ezquiaga},\
  and\ \citenamefont {Gupta}}]{Chen:2024gdn}%
  \BibitemOpen
  \bibfield  {author} {\bibinfo {author} {\bibfnamefont {H.-Y.}\ \bibnamefont
  {Chen}}, \bibinfo {author} {\bibfnamefont {J.~M.}\ \bibnamefont {Ezquiaga}},\
  and\ \bibinfo {author} {\bibfnamefont {I.}~\bibnamefont {Gupta}},\ }\bibfield
   {title} {\bibinfo {title} {{Cosmography with next-generation gravitational
  wave detectors}},\ }\href {https://doi.org/10.1088/1361-6382/ad424f}
  {\bibfield  {journal} {\bibinfo  {journal} {Class. Quant. Grav.}\ }\textbf
  {\bibinfo {volume} {41}},\ \bibinfo {pages} {125004} (\bibinfo {year}
  {2024}{\natexlab{b}})},\ \Eprint {https://arxiv.org/abs/2402.03120}
  {arXiv:2402.03120 [gr-qc]} \BibitemShut {NoStop}%
\bibitem [{\citenamefont {D\'alya}\ \emph {et~al.}(2018)\citenamefont
  {D\'alya}, \citenamefont {Galg\'oczi}, \citenamefont {Dobos}, \citenamefont
  {Frei}, \citenamefont {Heng}, \citenamefont {Macas}, \citenamefont
  {Messenger}, \citenamefont {Raffai},\ and\ \citenamefont
  {de~Souza}}]{Dalya:2018cnd}%
  \BibitemOpen
  \bibfield  {author} {\bibinfo {author} {\bibfnamefont {G.}~\bibnamefont
  {D\'alya}}, \bibinfo {author} {\bibfnamefont {G.}~\bibnamefont {Galg\'oczi}},
  \bibinfo {author} {\bibfnamefont {L.}~\bibnamefont {Dobos}}, \bibinfo
  {author} {\bibfnamefont {Z.}~\bibnamefont {Frei}}, \bibinfo {author}
  {\bibfnamefont {I.~S.}\ \bibnamefont {Heng}}, \bibinfo {author}
  {\bibfnamefont {R.}~\bibnamefont {Macas}}, \bibinfo {author} {\bibfnamefont
  {C.}~\bibnamefont {Messenger}}, \bibinfo {author} {\bibfnamefont
  {P.}~\bibnamefont {Raffai}},\ and\ \bibinfo {author} {\bibfnamefont {R.~S.}\
  \bibnamefont {de~Souza}},\ }\bibfield  {title} {\bibinfo {title} {{GLADE: A
  galaxy catalogue for multimessenger searches in the advanced
  gravitational-wave detector era}},\ }\href
  {https://doi.org/10.1093/mnras/sty1703} {\bibfield  {journal} {\bibinfo
  {journal} {Mon. Not. Roy. Astron. Soc.}\ }\textbf {\bibinfo {volume} {479}},\
  \bibinfo {pages} {2374} (\bibinfo {year} {2018})},\ \Eprint
  {https://arxiv.org/abs/1804.05709} {arXiv:1804.05709 [astro-ph.HE]}
  \BibitemShut {NoStop}%
\bibitem [{\citenamefont {Del~Pozzo}\ \emph {et~al.}(2018)\citenamefont
  {Del~Pozzo}, \citenamefont {Berry}, \citenamefont {Ghosh}, \citenamefont
  {Haines}, \citenamefont {Singer},\ and\ \citenamefont
  {Vecchio}}]{DelPozzo:2018dpu}%
  \BibitemOpen
  \bibfield  {author} {\bibinfo {author} {\bibfnamefont {W.}~\bibnamefont
  {Del~Pozzo}}, \bibinfo {author} {\bibfnamefont {C.~P.}\ \bibnamefont
  {Berry}}, \bibinfo {author} {\bibfnamefont {A.}~\bibnamefont {Ghosh}},
  \bibinfo {author} {\bibfnamefont {T.~S.~F.}\ \bibnamefont {Haines}}, \bibinfo
  {author} {\bibfnamefont {L.~P.}\ \bibnamefont {Singer}},\ and\ \bibinfo
  {author} {\bibfnamefont {A.}~\bibnamefont {Vecchio}},\ }\bibfield  {title}
  {\bibinfo {title} {{Dirichlet Process Gaussian-mixture model: An application
  to localizing coalescing binary neutron stars with gravitational-wave
  observations}},\ }\href {https://doi.org/10.1093/mnras/sty1485} {\bibfield
  {journal} {\bibinfo  {journal} {Mon. Not. Roy. Astron. Soc.}\ }\textbf
  {\bibinfo {volume} {479}},\ \bibinfo {pages} {601} (\bibinfo {year}
  {2018})},\ \Eprint {https://arxiv.org/abs/1801.08009} {arXiv:1801.08009
  [astro-ph.IM]} \BibitemShut {NoStop}%
\bibitem [{\citenamefont {Soares-Santos}\ \emph {et~al.}(2019)\citenamefont
  {Soares-Santos} \emph {et~al.}}]{DES:2019_HO_DarkSirens}%
  \BibitemOpen
  \bibfield  {author} {\bibinfo {author} {\bibfnamefont {M.}~\bibnamefont
  {Soares-Santos}} \emph {et~al.} (\bibinfo {collaboration} {DES, LIGO
  Scientific, Virgo}),\ }\bibfield  {title} {\bibinfo {title} {{First
  Measurement of the Hubble Constant from a Dark Standard Siren using the Dark
  Energy Survey Galaxies and the LIGO/Virgo Binary\textendash{}Black-hole
  Merger GW170814}},\ }\href {https://doi.org/10.3847/2041-8213/ab14f1}
  {\bibfield  {journal} {\bibinfo  {journal} {Astrophys. J. Lett.}\ }\textbf
  {\bibinfo {volume} {876}},\ \bibinfo {pages} {L7} (\bibinfo {year} {2019})},\
  \Eprint {https://arxiv.org/abs/1901.01540} {arXiv:1901.01540 [astro-ph.CO]}
  \BibitemShut {NoStop}%
\bibitem [{\citenamefont {Bera}\ \emph {et~al.}(2020)\citenamefont {Bera},
  \citenamefont {Rana}, \citenamefont {More},\ and\ \citenamefont
  {Bose}}]{Bera:2020_DarkSirens_CrossCorr}%
  \BibitemOpen
  \bibfield  {author} {\bibinfo {author} {\bibfnamefont {S.}~\bibnamefont
  {Bera}}, \bibinfo {author} {\bibfnamefont {D.}~\bibnamefont {Rana}}, \bibinfo
  {author} {\bibfnamefont {S.}~\bibnamefont {More}},\ and\ \bibinfo {author}
  {\bibfnamefont {S.}~\bibnamefont {Bose}},\ }\bibfield  {title} {\bibinfo
  {title} {{Incompleteness Matters Not: Inference of $H_0$ from Binary Black
  Hole\textendash{}Galaxy Cross-correlations}},\ }\href
  {https://doi.org/10.3847/1538-4357/abb4e0} {\bibfield  {journal} {\bibinfo
  {journal} {Astrophys. J.}\ }\textbf {\bibinfo {volume} {902}},\ \bibinfo
  {pages} {79} (\bibinfo {year} {2020})},\ \Eprint
  {https://arxiv.org/abs/2007.04271} {arXiv:2007.04271 [astro-ph.CO]}
  \BibitemShut {NoStop}%
\bibitem [{\citenamefont {Mukherjee}\ \emph
  {et~al.}(2021{\natexlab{b}})\citenamefont {Mukherjee}, \citenamefont
  {Wandelt}, \citenamefont {Nissanke},\ and\ \citenamefont
  {Silvestri}}]{Mukherjee:2020_DarkSirens_Clustering}%
  \BibitemOpen
  \bibfield  {author} {\bibinfo {author} {\bibfnamefont {S.}~\bibnamefont
  {Mukherjee}}, \bibinfo {author} {\bibfnamefont {B.~D.}\ \bibnamefont
  {Wandelt}}, \bibinfo {author} {\bibfnamefont {S.~M.}\ \bibnamefont
  {Nissanke}},\ and\ \bibinfo {author} {\bibfnamefont {A.}~\bibnamefont
  {Silvestri}},\ }\bibfield  {title} {\bibinfo {title} {{Accurate precision
  Cosmology with redshift unknown gravitational wave sources}},\ }\href
  {https://doi.org/10.1103/PhysRevD.103.043520} {\bibfield  {journal} {\bibinfo
   {journal} {Phys. Rev. D}\ }\textbf {\bibinfo {volume} {103}},\ \bibinfo
  {pages} {043520} (\bibinfo {year} {2021}{\natexlab{b}})},\ \Eprint
  {https://arxiv.org/abs/2007.02943} {arXiv:2007.02943 [astro-ph.CO]}
  \BibitemShut {NoStop}%
\bibitem [{\citenamefont {Palmese}\ \emph {et~al.}(2023)\citenamefont
  {Palmese}, \citenamefont {Bom}, \citenamefont {Mucesh},\ and\ \citenamefont
  {Hartley}}]{Palmese:2021_H0_DarkSirens_O3}%
  \BibitemOpen
  \bibfield  {author} {\bibinfo {author} {\bibfnamefont {A.}~\bibnamefont
  {Palmese}}, \bibinfo {author} {\bibfnamefont {C.~R.}\ \bibnamefont {Bom}},
  \bibinfo {author} {\bibfnamefont {S.}~\bibnamefont {Mucesh}},\ and\ \bibinfo
  {author} {\bibfnamefont {W.~G.}\ \bibnamefont {Hartley}},\ }\bibfield
  {title} {\bibinfo {title} {{A Standard Siren Measurement of the Hubble
  Constant Using Gravitational-wave Events from the First Three LIGO/Virgo
  Observing Runs and the DESI Legacy Survey}},\ }\href
  {https://doi.org/10.3847/1538-4357/aca6e3} {\bibfield  {journal} {\bibinfo
  {journal} {Astrophys. J.}\ }\textbf {\bibinfo {volume} {943}},\ \bibinfo
  {pages} {56} (\bibinfo {year} {2023})},\ \Eprint
  {https://arxiv.org/abs/2111.06445} {arXiv:2111.06445 [astro-ph.CO]}
  \BibitemShut {NoStop}%
\bibitem [{\citenamefont {Mo}\ and\ \citenamefont {Chen}(2024)}]{Mo:2024frl}%
  \BibitemOpen
  \bibfield  {author} {\bibinfo {author} {\bibfnamefont {G.}~\bibnamefont
  {Mo}}\ and\ \bibinfo {author} {\bibfnamefont {H.-Y.}\ \bibnamefont {Chen}},\
  }\bibfield  {title} {\bibinfo {title} {{Identifying the hosts of binary black
  hole and neutron star-black hole mergers with next-generation
  gravitational-wave detectors}},\ }\href@noop {} {\bibfield  {journal}
  {\bibinfo  {journal} {arXiv e-prints}\ } (\bibinfo {year} {2024})},\ \Eprint
  {https://arxiv.org/abs/2402.09684} {arXiv:2402.09684 [astro-ph.HE]}
  \BibitemShut {NoStop}%
\bibitem [{\citenamefont {Mo}\ \emph {et~al.}(2025)\citenamefont {Mo},
  \citenamefont {Haster},\ and\ \citenamefont {Katsavounidis}}]{Mo:2024uim}%
  \BibitemOpen
  \bibfield  {author} {\bibinfo {author} {\bibfnamefont {G.}~\bibnamefont
  {Mo}}, \bibinfo {author} {\bibfnamefont {C.-J.}\ \bibnamefont {Haster}},\
  and\ \bibinfo {author} {\bibfnamefont {E.}~\bibnamefont {Katsavounidis}},\
  }\bibfield  {title} {\bibinfo {title} {{On the Use of Galaxy Catalogs in
  Gravitational-wave Parameter Estimation}},\ }\href
  {https://doi.org/10.3847/1538-4357/ad9f36} {\bibfield  {journal} {\bibinfo
  {journal} {Astrophys. J.}\ }\textbf {\bibinfo {volume} {979}},\ \bibinfo
  {pages} {102} (\bibinfo {year} {2025})},\ \Eprint
  {https://arxiv.org/abs/2410.14663} {arXiv:2410.14663 [astro-ph.HE]}
  \BibitemShut {NoStop}%
\bibitem [{\citenamefont {{Mukherjee}}\ \emph {et~al.}(2024)\citenamefont
  {{Mukherjee}}, \citenamefont {{Krolewski}}, \citenamefont {{Wandelt}},\ and\
  \citenamefont {{Silk}}}]{Mukherjee_cross-corr_2024}%
  \BibitemOpen
  \bibfield  {author} {\bibinfo {author} {\bibfnamefont {S.}~\bibnamefont
  {{Mukherjee}}}, \bibinfo {author} {\bibfnamefont {A.}~\bibnamefont
  {{Krolewski}}}, \bibinfo {author} {\bibfnamefont {B.~D.}\ \bibnamefont
  {{Wandelt}}},\ and\ \bibinfo {author} {\bibfnamefont {J.}~\bibnamefont
  {{Silk}}},\ }\bibfield  {title} {\bibinfo {title} {{Cross-correlating Dark
  Sirens and Galaxies: Constraints on H $_{0}$ from GWTC-3 of
  LIGO{\textendash}Virgo{\textendash}KAGRA}},\ }\href
  {https://doi.org/10.3847/1538-4357/ad7d90} {\bibfield  {journal} {\bibinfo
  {journal} {\apj}\ }\textbf {\bibinfo {volume} {975}},\ \bibinfo {eid} {189}
  (\bibinfo {year} {2024})},\ \Eprint {https://arxiv.org/abs/2203.03643}
  {arXiv:2203.03643 [astro-ph.CO]} \BibitemShut {NoStop}%
\bibitem [{\citenamefont {Gair}\ \emph {et~al.}(2023)\citenamefont {Gair} \emph
  {et~al.}}]{Gair:2022zsa}%
  \BibitemOpen
  \bibfield  {author} {\bibinfo {author} {\bibfnamefont {J.~R.}\ \bibnamefont
  {Gair}} \emph {et~al.},\ }\bibfield  {title} {\bibinfo {title} {{The
  Hitchhiker\textquoteright{}s Guide to the Galaxy Catalog Approach for Dark
  Siren Gravitational-wave Cosmology}},\ }\href
  {https://doi.org/10.3847/1538-3881/acca78} {\bibfield  {journal} {\bibinfo
  {journal} {Astron. J.}\ }\textbf {\bibinfo {volume} {166}},\ \bibinfo {pages}
  {22} (\bibinfo {year} {2023})},\ \Eprint {https://arxiv.org/abs/2212.08694}
  {arXiv:2212.08694 [gr-qc]} \BibitemShut {NoStop}%
\bibitem [{\citenamefont {Will}(1998)}]{Will:1997bb}%
  \BibitemOpen
  \bibfield  {author} {\bibinfo {author} {\bibfnamefont {C.~M.}\ \bibnamefont
  {Will}},\ }\bibfield  {title} {\bibinfo {title} {{Bounding the mass of the
  graviton using gravitational wave observations of inspiralling compact
  binaries}},\ }\href {https://doi.org/10.1103/PhysRevD.57.2061} {\bibfield
  {journal} {\bibinfo  {journal} {Phys. Rev. D}\ }\textbf {\bibinfo {volume}
  {57}},\ \bibinfo {pages} {2061} (\bibinfo {year} {1998})},\ \Eprint
  {https://arxiv.org/abs/gr-qc/9709011} {arXiv:gr-qc/9709011} \BibitemShut
  {NoStop}%
\bibitem [{\citenamefont {Blanchet}(2014)}]{Blanchet:2013haa}%
  \BibitemOpen
  \bibfield  {author} {\bibinfo {author} {\bibfnamefont {L.}~\bibnamefont
  {Blanchet}},\ }\bibfield  {title} {\bibinfo {title} {{Gravitational Radiation
  from Post-Newtonian Sources and Inspiralling Compact Binaries}},\ }\href
  {https://doi.org/10.12942/lrr-2014-2} {\bibfield  {journal} {\bibinfo
  {journal} {Living Rev. Rel.}\ }\textbf {\bibinfo {volume} {17}},\ \bibinfo
  {pages} {2} (\bibinfo {year} {2014})},\ \Eprint
  {https://arxiv.org/abs/1310.1528} {arXiv:1310.1528 [gr-qc]} \BibitemShut
  {NoStop}%
\bibitem [{\citenamefont {Grohs}\ and\ \citenamefont
  {Fuller}(2023)}]{Grohs:2023voo}%
  \BibitemOpen
  \bibfield  {author} {\bibinfo {author} {\bibfnamefont {E.}~\bibnamefont
  {Grohs}}\ and\ \bibinfo {author} {\bibfnamefont {G.~M.}\ \bibnamefont
  {Fuller}},\ }\bibinfo {title} {{Big Bang Nucleosynthesis}},\ in\ \href
  {https://doi.org/10.1007/978-981-19-6345-2_127} {\emph {\bibinfo {booktitle}
  {{Handbook of Nuclear Physics}}}},\ \bibinfo {editor} {edited by\ \bibinfo
  {editor} {\bibfnamefont {I.}~\bibnamefont {Tanihata}}, \bibinfo {editor}
  {\bibfnamefont {H.}~\bibnamefont {Toki}},\ and\ \bibinfo {editor}
  {\bibfnamefont {T.}~\bibnamefont {Kajino}}}\ (\bibinfo  {publisher} {Springer
  Nature Singapore},\ \bibinfo {address} {Singapore},\ \bibinfo {year} {2023})\
  pp.\ \bibinfo {pages} {3713--3733},\ \Eprint
  {https://arxiv.org/abs/2301.12299} {arXiv:2301.12299 [astro-ph.CO]}
  \BibitemShut {NoStop}%
\bibitem [{\citenamefont {Aghanim}\ \emph {et~al.}(2020)\citenamefont {Aghanim}
  \emph {et~al.}}]{Planck:2018vyg}%
  \BibitemOpen
  \bibfield  {author} {\bibinfo {author} {\bibfnamefont {N.}~\bibnamefont
  {Aghanim}} \emph {et~al.} (\bibinfo {collaboration} {Planck}),\ }\bibfield
  {title} {\bibinfo {title} {{Planck 2018 results. VI. Cosmological
  parameters}},\ }\href {https://doi.org/10.1051/0004-6361/201833910}
  {\bibfield  {journal} {\bibinfo  {journal} {Astron. Astrophys.}\ }\textbf
  {\bibinfo {volume} {641}},\ \bibinfo {pages} {A6} (\bibinfo {year} {2020})},\
  \bibinfo {note} {[Erratum: Astron.Astrophys. 652, C4 (2021)]},\ \Eprint
  {https://arxiv.org/abs/1807.06209} {arXiv:1807.06209 [astro-ph.CO]}
  \BibitemShut {NoStop}%
\bibitem [{\citenamefont {Blanchard}\ \emph {et~al.}(2024)\citenamefont
  {Blanchard}, \citenamefont {H\'eloret}, \citenamefont {Ili\'c}, \citenamefont
  {Lamine},\ and\ \citenamefont {Tutusaus}}]{Blanchard:2022xkk}%
  \BibitemOpen
  \bibfield  {author} {\bibinfo {author} {\bibfnamefont {A.}~\bibnamefont
  {Blanchard}}, \bibinfo {author} {\bibfnamefont {J.-Y.}\ \bibnamefont
  {H\'eloret}}, \bibinfo {author} {\bibfnamefont {S.}~\bibnamefont {Ili\'c}},
  \bibinfo {author} {\bibfnamefont {B.}~\bibnamefont {Lamine}},\ and\ \bibinfo
  {author} {\bibfnamefont {I.}~\bibnamefont {Tutusaus}},\ }\bibfield  {title}
  {\bibinfo {title} {{$\Lambda$CDM is alive and well}},\ }\href
  {https://doi.org/10.33232/001c.117170} {\bibfield  {journal} {\bibinfo
  {journal} {Open J. Astrophys.}\ }\textbf {\bibinfo {volume} {7}},\ \bibinfo
  {pages} {117170} (\bibinfo {year} {2024})},\ \Eprint
  {https://arxiv.org/abs/2205.05017} {arXiv:2205.05017 [astro-ph.CO]}
  \BibitemShut {NoStop}%
\bibitem [{\citenamefont {{Yunes}}\ and\ \citenamefont
  {{Pretorius}}(2009)}]{Yunes_Pretorius_ppE_2009}%
  \BibitemOpen
  \bibfield  {author} {\bibinfo {author} {\bibfnamefont {N.}~\bibnamefont
  {{Yunes}}}\ and\ \bibinfo {author} {\bibfnamefont {F.}~\bibnamefont
  {{Pretorius}}},\ }\bibfield  {title} {\bibinfo {title} {{Fundamental
  theoretical bias in gravitational wave astrophysics and the parametrized
  post-Einsteinian framework}},\ }\href
  {https://doi.org/10.1103/PhysRevD.80.122003} {\bibfield  {journal} {\bibinfo
  {journal} {\prd}\ }\textbf {\bibinfo {volume} {80}},\ \bibinfo {eid} {122003}
  (\bibinfo {year} {2009})},\ \Eprint {https://arxiv.org/abs/0909.3328}
  {arXiv:0909.3328 [gr-qc]} \BibitemShut {NoStop}%
\bibitem [{\citenamefont {{Agathos}}\ \emph {et~al.}(2014)\citenamefont
  {{Agathos}}, \citenamefont {{Del Pozzo}}, \citenamefont {{Li}}, \citenamefont
  {{Van Den Broeck}}, \citenamefont {{Veitch}},\ and\ \citenamefont
  {{Vitale}}}]{Agathos_et_al_TIGER_2014}%
  \BibitemOpen
  \bibfield  {author} {\bibinfo {author} {\bibfnamefont {M.}~\bibnamefont
  {{Agathos}}}, \bibinfo {author} {\bibfnamefont {W.}~\bibnamefont {{Del
  Pozzo}}}, \bibinfo {author} {\bibfnamefont {T.~G.~F.}\ \bibnamefont {{Li}}},
  \bibinfo {author} {\bibfnamefont {C.}~\bibnamefont {{Van Den Broeck}}},
  \bibinfo {author} {\bibfnamefont {J.}~\bibnamefont {{Veitch}}},\ and\
  \bibinfo {author} {\bibfnamefont {S.}~\bibnamefont {{Vitale}}},\ }\bibfield
  {title} {\bibinfo {title} {{TIGER: A data analysis pipeline for testing the
  strong-field dynamics of general relativity with gravitational wave signals
  from coalescing compact binaries}},\ }\href
  {https://doi.org/10.1103/PhysRevD.89.082001} {\bibfield  {journal} {\bibinfo
  {journal} {\prd}\ }\textbf {\bibinfo {volume} {89}},\ \bibinfo {eid} {082001}
  (\bibinfo {year} {2014})},\ \Eprint {https://arxiv.org/abs/1311.0420}
  {arXiv:1311.0420 [gr-qc]} \BibitemShut {NoStop}%
\bibitem [{\citenamefont {Zimmerman}\ \emph {et~al.}(2019)\citenamefont
  {Zimmerman}, \citenamefont {Haster},\ and\ \citenamefont
  {Chatziioannou}}]{Zimmerman:2019wzo}%
  \BibitemOpen
  \bibfield  {author} {\bibinfo {author} {\bibfnamefont {A.}~\bibnamefont
  {Zimmerman}}, \bibinfo {author} {\bibfnamefont {C.-J.}\ \bibnamefont
  {Haster}},\ and\ \bibinfo {author} {\bibfnamefont {K.}~\bibnamefont
  {Chatziioannou}},\ }\bibfield  {title} {\bibinfo {title} {{On combining
  information from multiple gravitational wave sources}},\ }\href
  {https://doi.org/10.1103/PhysRevD.99.124044} {\bibfield  {journal} {\bibinfo
  {journal} {Phys. Rev. D}\ }\textbf {\bibinfo {volume} {99}},\ \bibinfo
  {pages} {124044} (\bibinfo {year} {2019})},\ \Eprint
  {https://arxiv.org/abs/1903.11008} {arXiv:1903.11008 [astro-ph.IM]}
  \BibitemShut {NoStop}%
\bibitem [{\citenamefont {Abbott}\ \emph
  {et~al.}(2021{\natexlab{a}})\citenamefont {Abbott} \emph
  {et~al.}}]{LIGOScientific:2021sio}%
  \BibitemOpen
  \bibfield  {author} {\bibinfo {author} {\bibfnamefont {R.}~\bibnamefont
  {Abbott}} \emph {et~al.} (\bibinfo {collaboration} {LIGO Scientific, VIRGO,
  KAGRA}),\ }\bibfield  {title} {\bibinfo {title} {{Tests of General Relativity
  with GWTC-3}},\ }\href@noop {} {\bibfield  {journal} {\bibinfo  {journal}
  {arXiv e-prints}\ } (\bibinfo {year} {2021}{\natexlab{a}})},\ \Eprint
  {https://arxiv.org/abs/2112.06861} {arXiv:2112.06861 [gr-qc]} \BibitemShut
  {NoStop}%
\bibitem [{\citenamefont {Will}(2018)}]{Will:2018gku}%
  \BibitemOpen
  \bibfield  {author} {\bibinfo {author} {\bibfnamefont {C.~M.}\ \bibnamefont
  {Will}},\ }\bibfield  {title} {\bibinfo {title} {{Solar system versus
  gravitational-wave bounds on the graviton mass}},\ }\href
  {https://doi.org/10.1088/1361-6382/aad13c} {\bibfield  {journal} {\bibinfo
  {journal} {Class. Quant. Grav.}\ }\textbf {\bibinfo {volume} {35}},\ \bibinfo
  {pages} {17LT01} (\bibinfo {year} {2018})},\ \Eprint
  {https://arxiv.org/abs/1805.10523} {arXiv:1805.10523 [gr-qc]} \BibitemShut
  {NoStop}%
\bibitem [{\citenamefont {Cantiello}\ \emph {et~al.}(2018)\citenamefont
  {Cantiello} \emph {et~al.}}]{Cantiello:2018ffy}%
  \BibitemOpen
  \bibfield  {author} {\bibinfo {author} {\bibfnamefont {M.}~\bibnamefont
  {Cantiello}} \emph {et~al.},\ }\bibfield  {title} {\bibinfo {title} {{A
  Precise Distance to the Host Galaxy of the Binary Neutron Star Merger
  GW170817 Using Surface Brightness Fluctuations}},\ }\href
  {https://doi.org/10.3847/2041-8213/aaad64} {\bibfield  {journal} {\bibinfo
  {journal} {Astrophys. J. Lett.}\ }\textbf {\bibinfo {volume} {854}},\
  \bibinfo {pages} {L31} (\bibinfo {year} {2018})},\ \Eprint
  {https://arxiv.org/abs/1801.06080} {arXiv:1801.06080 [astro-ph.GA]}
  \BibitemShut {NoStop}%
\bibitem [{\citenamefont {Abbott}\ \emph
  {et~al.}(2021{\natexlab{b}})\citenamefont {Abbott} \emph
  {et~al.}}]{LIGOScientific:2021qlt}%
  \BibitemOpen
  \bibfield  {author} {\bibinfo {author} {\bibfnamefont {R.}~\bibnamefont
  {Abbott}} \emph {et~al.} (\bibinfo {collaboration} {LIGO Scientific, KAGRA,
  VIRGO}),\ }\bibfield  {title} {\bibinfo {title} {{Observation of
  Gravitational Waves from Two Neutron Star\textendash{}Black Hole
  Coalescences}},\ }\href {https://doi.org/10.3847/2041-8213/ac082e} {\bibfield
   {journal} {\bibinfo  {journal} {Astrophys. J. Lett.}\ }\textbf {\bibinfo
  {volume} {915}},\ \bibinfo {pages} {L5} (\bibinfo {year}
  {2021}{\natexlab{b}})},\ \Eprint {https://arxiv.org/abs/2106.15163}
  {arXiv:2106.15163 [astro-ph.HE]} \BibitemShut {NoStop}%
\bibitem [{\citenamefont {Collaboration}\ \emph {et~al.}(2021)\citenamefont
  {Collaboration}, \citenamefont {Collaboration},\ and\ \citenamefont
  {Collaboration}}]{ligo_scientific_collaboration_and_virgo_2021_5546663}%
  \BibitemOpen
  \bibfield  {author} {\bibinfo {author} {\bibfnamefont {L.~S.}\ \bibnamefont
  {Collaboration}}, \bibinfo {author} {\bibfnamefont {V.}~\bibnamefont
  {Collaboration}},\ and\ \bibinfo {author} {\bibfnamefont {K.}~\bibnamefont
  {Collaboration}},\ }\bibfield  {title} {\bibinfo {title} {{GWTC-3: Compact
  Binary Coalescences Observed by LIGO and Virgo During the Second Part of the
  Third Observing Run — Parameter estimation data release}},\ }\href
  {https://doi.org/10.5281/zenodo.5546663} {10.5281/zenodo.5546663} (\bibinfo
  {year} {2021})\BibitemShut {NoStop}%
\bibitem [{\citenamefont {Ashton}\ \emph {et~al.}(2019)\citenamefont {Ashton}
  \emph {et~al.}}]{Ashton:2018jfp}%
  \BibitemOpen
  \bibfield  {author} {\bibinfo {author} {\bibfnamefont {G.}~\bibnamefont
  {Ashton}} \emph {et~al.},\ }\bibfield  {title} {\bibinfo {title} {{BILBY: A
  user-friendly Bayesian inference library for gravitational-wave astronomy}},\
  }\href {https://doi.org/10.3847/1538-4365/ab06fc} {\bibfield  {journal}
  {\bibinfo  {journal} {Astrophys. J. Suppl.}\ }\textbf {\bibinfo {volume}
  {241}},\ \bibinfo {pages} {27} (\bibinfo {year} {2019})},\ \Eprint
  {https://arxiv.org/abs/1811.02042} {arXiv:1811.02042 [astro-ph.IM]}
  \BibitemShut {NoStop}%
\bibitem [{\citenamefont {Ade}\ \emph {et~al.}(2016)\citenamefont {Ade} \emph
  {et~al.}}]{Planck:2015fie}%
  \BibitemOpen
  \bibfield  {author} {\bibinfo {author} {\bibfnamefont {P.~A.~R.}\
  \bibnamefont {Ade}} \emph {et~al.} (\bibinfo {collaboration} {Planck}),\
  }\bibfield  {title} {\bibinfo {title} {{Planck 2015 results. XIII.
  Cosmological parameters}},\ }\href
  {https://doi.org/10.1051/0004-6361/201525830} {\bibfield  {journal} {\bibinfo
   {journal} {Astron. Astrophys.}\ }\textbf {\bibinfo {volume} {594}},\
  \bibinfo {pages} {A13} (\bibinfo {year} {2016})},\ \Eprint
  {https://arxiv.org/abs/1502.01589} {arXiv:1502.01589 [astro-ph.CO]}
  \BibitemShut {NoStop}%
\bibitem [{\citenamefont {Miller}\ \emph {et~al.}(2015)\citenamefont {Miller},
  \citenamefont {Barsotti}, \citenamefont {Vitale}, \citenamefont {Fritschel},
  \citenamefont {Evans},\ and\ \citenamefont {Sigg}}]{Miller:2014kma}%
  \BibitemOpen
  \bibfield  {author} {\bibinfo {author} {\bibfnamefont {J.}~\bibnamefont
  {Miller}}, \bibinfo {author} {\bibfnamefont {L.}~\bibnamefont {Barsotti}},
  \bibinfo {author} {\bibfnamefont {S.}~\bibnamefont {Vitale}}, \bibinfo
  {author} {\bibfnamefont {P.}~\bibnamefont {Fritschel}}, \bibinfo {author}
  {\bibfnamefont {M.}~\bibnamefont {Evans}},\ and\ \bibinfo {author}
  {\bibfnamefont {D.}~\bibnamefont {Sigg}},\ }\bibfield  {title} {\bibinfo
  {title} {{Prospects for doubling the range of Advanced LIGO}},\ }\href
  {https://doi.org/10.1103/PhysRevD.91.062005} {\bibfield  {journal} {\bibinfo
  {journal} {Phys. Rev. D}\ }\textbf {\bibinfo {volume} {91}},\ \bibinfo
  {pages} {062005} (\bibinfo {year} {2015})},\ \Eprint
  {https://arxiv.org/abs/1410.5882} {arXiv:1410.5882 [gr-qc]} \BibitemShut
  {NoStop}%
\bibitem [{\citenamefont {Pratten}\ \emph {et~al.}(2021)\citenamefont {Pratten}
  \emph {et~al.}}]{Pratten:2020ceb}%
  \BibitemOpen
  \bibfield  {author} {\bibinfo {author} {\bibfnamefont {G.}~\bibnamefont
  {Pratten}} \emph {et~al.},\ }\bibfield  {title} {\bibinfo {title}
  {{Computationally efficient models for the dominant and subdominant harmonic
  modes of precessing binary black holes}},\ }\href
  {https://doi.org/10.1103/PhysRevD.103.104056} {\bibfield  {journal} {\bibinfo
   {journal} {Phys. Rev. D}\ }\textbf {\bibinfo {volume} {103}},\ \bibinfo
  {pages} {104056} (\bibinfo {year} {2021})},\ \Eprint
  {https://arxiv.org/abs/2004.06503} {arXiv:2004.06503 [gr-qc]} \BibitemShut
  {NoStop}%
\bibitem [{\citenamefont {Pratten}\ \emph {et~al.}(2020)\citenamefont
  {Pratten}, \citenamefont {Husa}, \citenamefont {Garcia-Quiros}, \citenamefont
  {Colleoni}, \citenamefont {Ramos-Buades}, \citenamefont {Estelles},\ and\
  \citenamefont {Jaume}}]{Pratten:2020fqn}%
  \BibitemOpen
  \bibfield  {author} {\bibinfo {author} {\bibfnamefont {G.}~\bibnamefont
  {Pratten}}, \bibinfo {author} {\bibfnamefont {S.}~\bibnamefont {Husa}},
  \bibinfo {author} {\bibfnamefont {C.}~\bibnamefont {Garcia-Quiros}}, \bibinfo
  {author} {\bibfnamefont {M.}~\bibnamefont {Colleoni}}, \bibinfo {author}
  {\bibfnamefont {A.}~\bibnamefont {Ramos-Buades}}, \bibinfo {author}
  {\bibfnamefont {H.}~\bibnamefont {Estelles}},\ and\ \bibinfo {author}
  {\bibfnamefont {R.}~\bibnamefont {Jaume}},\ }\bibfield  {title} {\bibinfo
  {title} {{Setting the cornerstone for a family of models for gravitational
  waves from compact binaries: The dominant harmonic for nonprecessing
  quasicircular black holes}},\ }\href
  {https://doi.org/10.1103/PhysRevD.102.064001} {\bibfield  {journal} {\bibinfo
   {journal} {Phys. Rev. D}\ }\textbf {\bibinfo {volume} {102}},\ \bibinfo
  {pages} {064001} (\bibinfo {year} {2020})},\ \Eprint
  {https://arxiv.org/abs/2001.11412} {arXiv:2001.11412 [gr-qc]} \BibitemShut
  {NoStop}%
\bibitem [{\citenamefont {Garc\'\i{}a-Quir\'os}\ \emph
  {et~al.}(2020)\citenamefont {Garc\'\i{}a-Quir\'os}, \citenamefont {Colleoni},
  \citenamefont {Husa}, \citenamefont {Estell\'es}, \citenamefont {Pratten},
  \citenamefont {Ramos-Buades}, \citenamefont {Mateu-Lucena},\ and\
  \citenamefont {Jaume}}]{Garcia-Quiros:2020qpx}%
  \BibitemOpen
  \bibfield  {author} {\bibinfo {author} {\bibfnamefont {C.}~\bibnamefont
  {Garc\'\i{}a-Quir\'os}}, \bibinfo {author} {\bibfnamefont {M.}~\bibnamefont
  {Colleoni}}, \bibinfo {author} {\bibfnamefont {S.}~\bibnamefont {Husa}},
  \bibinfo {author} {\bibfnamefont {H.}~\bibnamefont {Estell\'es}}, \bibinfo
  {author} {\bibfnamefont {G.}~\bibnamefont {Pratten}}, \bibinfo {author}
  {\bibfnamefont {A.}~\bibnamefont {Ramos-Buades}}, \bibinfo {author}
  {\bibfnamefont {M.}~\bibnamefont {Mateu-Lucena}},\ and\ \bibinfo {author}
  {\bibfnamefont {R.}~\bibnamefont {Jaume}},\ }\bibfield  {title} {\bibinfo
  {title} {{Multimode frequency-domain model for the gravitational wave signal
  from nonprecessing black-hole binaries}},\ }\href
  {https://doi.org/10.1103/PhysRevD.102.064002} {\bibfield  {journal} {\bibinfo
   {journal} {Phys. Rev. D}\ }\textbf {\bibinfo {volume} {102}},\ \bibinfo
  {pages} {064002} (\bibinfo {year} {2020})},\ \Eprint
  {https://arxiv.org/abs/2001.10914} {arXiv:2001.10914 [gr-qc]} \BibitemShut
  {NoStop}%
\bibitem [{\citenamefont {Wickramasinghe}\ and\ \citenamefont
  {Ukwatta}(2010)}]{wickramasinghe:2010}%
  \BibitemOpen
  \bibfield  {author} {\bibinfo {author} {\bibfnamefont {T.}~\bibnamefont
  {Wickramasinghe}}\ and\ \bibinfo {author} {\bibfnamefont {T.}~\bibnamefont
  {Ukwatta}},\ }\bibfield  {title} {\bibinfo {title} {An analytical approach
  for the determination of the luminosity distance in a flat universe with dark
  energy},\ }\href {https://doi.org/10.1111/j.1365-2966.2010.16686.x}
  {\bibfield  {journal} {\bibinfo  {journal} {Mon. Not. Roy. Astron. Soc.}\
  }\textbf {\bibinfo {volume} {406}},\ \bibinfo {pages} {548} (\bibinfo {year}
  {2010})},\ \Eprint {https://arxiv.org/abs/1003.0483} {arXiv:1003.0483
  [astro-ph.CO]} \BibitemShut {NoStop}%
\bibitem [{\citenamefont {Adachi}\ and\ \citenamefont
  {Kasai}(2012)}]{adachi:2012}%
  \BibitemOpen
  \bibfield  {author} {\bibinfo {author} {\bibfnamefont {M.}~\bibnamefont
  {Adachi}}\ and\ \bibinfo {author} {\bibfnamefont {M.}~\bibnamefont {Kasai}},\
  }\bibfield  {title} {\bibinfo {title} {An analytical approximation of the
  luminosity distance in flat cosmologies with a cosmological constant},\
  }\href {https://doi.org/10.1143/PTP.127.145} {\bibfield  {journal} {\bibinfo
  {journal} {Progress of Theoretical Physics}\ }\textbf {\bibinfo {volume}
  {127}},\ \bibinfo {pages} {145} (\bibinfo {year} {2012})},\ \Eprint
  {https://arxiv.org/abs/1111.6396} {arXiv:1111.6396 [astro-ph.CO]}
  \BibitemShut {NoStop}%
\bibitem [{\citenamefont {Scott}(2015)}]{ScottKDE}%
  \BibitemOpen
  \bibfield  {author} {\bibinfo {author} {\bibfnamefont {D.~W.}\ \bibnamefont
  {Scott}},\ }\href@noop {} {\emph {\bibinfo {title} {Multivariate Density
  Estimation: Theory, Practice, and Visualization, 2nd edn}}}\ (\bibinfo
  {publisher} {John Wiley and Sons},\ \bibinfo {address} {Hoboken, NJ},\
  \bibinfo {year} {2015})\BibitemShut {NoStop}%
\bibitem [{\citenamefont {Silverman}(1986)}]{SilvermanKDE}%
  \BibitemOpen
  \bibfield  {author} {\bibinfo {author} {\bibfnamefont {B.~W.}\ \bibnamefont
  {Silverman}},\ }\href@noop {} {\emph {\bibinfo {title} {Density estimation
  for statistics and data analysis}}}\ (\bibinfo  {publisher} {Chapman and
  Hall},\ \bibinfo {address} {London},\ \bibinfo {year} {1986})\BibitemShut
  {NoStop}%
\bibitem [{\citenamefont {Virtanen}\ \emph {et~al.}(2020)\citenamefont
  {Virtanen}, \citenamefont {Gommers}, \citenamefont {Oliphant}, \citenamefont
  {Haberland}, \citenamefont {Reddy}, \citenamefont {Cournapeau}, \citenamefont
  {Burovski}, \citenamefont {Peterson}, \citenamefont {Weckesser},
  \citenamefont {Bright}, \citenamefont {{van der Walt}}, \citenamefont
  {Brett}, \citenamefont {Wilson}, \citenamefont {Millman}, \citenamefont
  {Mayorov}, \citenamefont {Nelson}, \citenamefont {Jones}, \citenamefont
  {Kern}, \citenamefont {Larson}, \citenamefont {Carey}, \citenamefont {Polat},
  \citenamefont {Feng}, \citenamefont {Moore}, \citenamefont {{VanderPlas}},
  \citenamefont {Laxalde}, \citenamefont {Perktold}, \citenamefont {Cimrman},
  \citenamefont {Henriksen}, \citenamefont {Quintero}, \citenamefont {Harris},
  \citenamefont {Archibald}, \citenamefont {Ribeiro}, \citenamefont
  {Pedregosa}, \citenamefont {{van Mulbregt}},\ and\ \citenamefont {{SciPy 1.0
  Contributors}}}]{Scipy}%
  \BibitemOpen
  \bibfield  {author} {\bibinfo {author} {\bibfnamefont {P.}~\bibnamefont
  {Virtanen}}, \bibinfo {author} {\bibfnamefont {R.}~\bibnamefont {Gommers}},
  \bibinfo {author} {\bibfnamefont {T.~E.}\ \bibnamefont {Oliphant}}, \bibinfo
  {author} {\bibfnamefont {M.}~\bibnamefont {Haberland}}, \bibinfo {author}
  {\bibfnamefont {T.}~\bibnamefont {Reddy}}, \bibinfo {author} {\bibfnamefont
  {D.}~\bibnamefont {Cournapeau}}, \bibinfo {author} {\bibfnamefont
  {E.}~\bibnamefont {Burovski}}, \bibinfo {author} {\bibfnamefont
  {P.}~\bibnamefont {Peterson}}, \bibinfo {author} {\bibfnamefont
  {W.}~\bibnamefont {Weckesser}}, \bibinfo {author} {\bibfnamefont
  {J.}~\bibnamefont {Bright}}, \bibinfo {author} {\bibfnamefont {S.~J.}\
  \bibnamefont {{van der Walt}}}, \bibinfo {author} {\bibfnamefont
  {M.}~\bibnamefont {Brett}}, \bibinfo {author} {\bibfnamefont
  {J.}~\bibnamefont {Wilson}}, \bibinfo {author} {\bibfnamefont {K.~J.}\
  \bibnamefont {Millman}}, \bibinfo {author} {\bibfnamefont {N.}~\bibnamefont
  {Mayorov}}, \bibinfo {author} {\bibfnamefont {A.~R.~J.}\ \bibnamefont
  {Nelson}}, \bibinfo {author} {\bibfnamefont {E.}~\bibnamefont {Jones}},
  \bibinfo {author} {\bibfnamefont {R.}~\bibnamefont {Kern}}, \bibinfo {author}
  {\bibfnamefont {E.}~\bibnamefont {Larson}}, \bibinfo {author} {\bibfnamefont
  {C.~J.}\ \bibnamefont {Carey}}, \bibinfo {author} {\bibfnamefont
  {{\.I}.}~\bibnamefont {Polat}}, \bibinfo {author} {\bibfnamefont
  {Y.}~\bibnamefont {Feng}}, \bibinfo {author} {\bibfnamefont {E.~W.}\
  \bibnamefont {Moore}}, \bibinfo {author} {\bibfnamefont {J.}~\bibnamefont
  {{VanderPlas}}}, \bibinfo {author} {\bibfnamefont {D.}~\bibnamefont
  {Laxalde}}, \bibinfo {author} {\bibfnamefont {J.}~\bibnamefont {Perktold}},
  \bibinfo {author} {\bibfnamefont {R.}~\bibnamefont {Cimrman}}, \bibinfo
  {author} {\bibfnamefont {I.}~\bibnamefont {Henriksen}}, \bibinfo {author}
  {\bibfnamefont {E.~A.}\ \bibnamefont {Quintero}}, \bibinfo {author}
  {\bibfnamefont {C.~R.}\ \bibnamefont {Harris}}, \bibinfo {author}
  {\bibfnamefont {A.~M.}\ \bibnamefont {Archibald}}, \bibinfo {author}
  {\bibfnamefont {A.~H.}\ \bibnamefont {Ribeiro}}, \bibinfo {author}
  {\bibfnamefont {F.}~\bibnamefont {Pedregosa}}, \bibinfo {author}
  {\bibfnamefont {P.}~\bibnamefont {{van Mulbregt}}},\ and\ \bibinfo {author}
  {\bibnamefont {{SciPy 1.0 Contributors}}},\ }\bibfield  {title} {\bibinfo
  {title} {{{SciPy} 1.0: Fundamental Algorithms for Scientific Computing in
  Python}},\ }\href {https://doi.org/10.1038/s41592-019-0686-2} {\bibfield
  {journal} {\bibinfo  {journal} {Nature Methods}\ }\textbf {\bibinfo {volume}
  {17}},\ \bibinfo {pages} {261} (\bibinfo {year} {2020})}\BibitemShut
  {NoStop}%
\bibitem [{\citenamefont {Turlach}(1999)}]{BWSelection_Turlach}%
  \BibitemOpen
  \bibfield  {author} {\bibinfo {author} {\bibfnamefont {B.}~\bibnamefont
  {Turlach}},\ }\bibfield  {title} {\bibinfo {title} {Bandwidth selection in
  kernel density estimation: A review},\ }\href@noop {} {\bibfield  {journal}
  {\bibinfo  {journal} {Technical Report}\ } (\bibinfo {year}
  {1999})}\BibitemShut {NoStop}%
\bibitem [{\citenamefont {Bashtannyk}\ and\ \citenamefont
  {Hyndman}(2001)}]{BWSelection_Bashtannyk_Hyndman}%
  \BibitemOpen
  \bibfield  {author} {\bibinfo {author} {\bibfnamefont {D.~M.}\ \bibnamefont
  {Bashtannyk}}\ and\ \bibinfo {author} {\bibfnamefont {R.~J.}\ \bibnamefont
  {Hyndman}},\ }\bibfield  {title} {\bibinfo {title} {{Bandwidth selection for
  kernel conditional density estimation}},\ }\href@noop {} {\bibfield
  {journal} {\bibinfo  {journal} {Computational Statistics \& Data Analysis}\
  }\textbf {\bibinfo {volume} {36}},\ \bibinfo {pages} {279} (\bibinfo {year}
  {2001})}\BibitemShut {NoStop}%
\bibitem [{\citenamefont {Mosteller}\ and\ \citenamefont
  {Tukey}(1977)}]{KFoldCV}%
  \BibitemOpen
  \bibfield  {author} {\bibinfo {author} {\bibfnamefont {F.}~\bibnamefont
  {Mosteller}}\ and\ \bibinfo {author} {\bibfnamefont {J.~W.}\ \bibnamefont
  {Tukey}},\ }\href@noop {} {\emph {\bibinfo {title} {Density estimation for
  statistics and data analysis}}}\ (\bibinfo  {publisher} {Addison-Wesley},\
  \bibinfo {address} {Reading, Mass},\ \bibinfo {year} {1977})\BibitemShut
  {NoStop}%
\bibitem [{\citenamefont {James}\ \emph {et~al.}(2023)\citenamefont {James},
  \citenamefont {Witten}, \citenamefont {Hastie}, \citenamefont {Tibshirani},\
  and\ \citenamefont {Taylor}}]{StatisticalLearning}%
  \BibitemOpen
  \bibfield  {author} {\bibinfo {author} {\bibfnamefont {G.}~\bibnamefont
  {James}}, \bibinfo {author} {\bibfnamefont {D.}~\bibnamefont {Witten}},
  \bibinfo {author} {\bibfnamefont {T.}~\bibnamefont {Hastie}}, \bibinfo
  {author} {\bibfnamefont {R.}~\bibnamefont {Tibshirani}},\ and\ \bibinfo
  {author} {\bibfnamefont {J.}~\bibnamefont {Taylor}},\ }\href@noop {} {\emph
  {\bibinfo {title} {An Introduction to Statistical Learning}}}\ (\bibinfo
  {publisher} {Springer},\ \bibinfo {year} {2023})\BibitemShut {NoStop}%
\bibitem [{\citenamefont {Pedregosa}\ \emph {et~al.}(2011)\citenamefont
  {Pedregosa}, \citenamefont {Varoquaux}, \citenamefont {Gramfort},
  \citenamefont {Michel}, \citenamefont {Thirion}, \citenamefont {Grisel},
  \citenamefont {Blondel}, \citenamefont {Prettenhofer}, \citenamefont {Weiss},
  \citenamefont {Dubourg}, \citenamefont {Vanderplas}, \citenamefont {Passos},
  \citenamefont {Cournapeau}, \citenamefont {Brucher}, \citenamefont {Perrot},\
  and\ \citenamefont {Duchesnay}}]{scikit-learn}%
  \BibitemOpen
  \bibfield  {author} {\bibinfo {author} {\bibfnamefont {F.}~\bibnamefont
  {Pedregosa}}, \bibinfo {author} {\bibfnamefont {G.}~\bibnamefont
  {Varoquaux}}, \bibinfo {author} {\bibfnamefont {A.}~\bibnamefont {Gramfort}},
  \bibinfo {author} {\bibfnamefont {V.}~\bibnamefont {Michel}}, \bibinfo
  {author} {\bibfnamefont {B.}~\bibnamefont {Thirion}}, \bibinfo {author}
  {\bibfnamefont {O.}~\bibnamefont {Grisel}}, \bibinfo {author} {\bibfnamefont
  {M.}~\bibnamefont {Blondel}}, \bibinfo {author} {\bibfnamefont
  {P.}~\bibnamefont {Prettenhofer}}, \bibinfo {author} {\bibfnamefont
  {R.}~\bibnamefont {Weiss}}, \bibinfo {author} {\bibfnamefont
  {V.}~\bibnamefont {Dubourg}}, \bibinfo {author} {\bibfnamefont
  {J.}~\bibnamefont {Vanderplas}}, \bibinfo {author} {\bibfnamefont
  {A.}~\bibnamefont {Passos}}, \bibinfo {author} {\bibfnamefont
  {D.}~\bibnamefont {Cournapeau}}, \bibinfo {author} {\bibfnamefont
  {M.}~\bibnamefont {Brucher}}, \bibinfo {author} {\bibfnamefont
  {M.}~\bibnamefont {Perrot}},\ and\ \bibinfo {author} {\bibfnamefont
  {E.}~\bibnamefont {Duchesnay}},\ }\bibfield  {title} {\bibinfo {title}
  {Scikit-learn: Machine learning in {P}ython},\ }\href@noop {} {\bibfield
  {journal} {\bibinfo  {journal} {Journal of Machine Learning Research}\
  }\textbf {\bibinfo {volume} {12}},\ \bibinfo {pages} {2825} (\bibinfo {year}
  {2011})}\BibitemShut {NoStop}%
\bibitem [{\citenamefont {{Hoy}}\ and\ \citenamefont
  {{Raymond}}(2021)}]{PESUMMARY}%
  \BibitemOpen
  \bibfield  {author} {\bibinfo {author} {\bibfnamefont {C.}~\bibnamefont
  {{Hoy}}}\ and\ \bibinfo {author} {\bibfnamefont {V.}~\bibnamefont
  {{Raymond}}},\ }\bibfield  {title} {\bibinfo {title} {{PESUMMARY: The code
  agnostic Parameter Estimation Summary page builder}},\ }\href
  {https://doi.org/10.1016/j.softx.2021.100765} {\bibfield  {journal} {\bibinfo
   {journal} {SoftwareX}\ }\textbf {\bibinfo {volume} {15}},\ \bibinfo {eid}
  {100765} (\bibinfo {year} {2021})},\ \Eprint
  {https://arxiv.org/abs/2006.06639} {arXiv:2006.06639 [astro-ph.IM]}
  \BibitemShut {NoStop}%
\bibitem [{\citenamefont {Messenger}\ and\ \citenamefont
  {Veitch}(2013)}]{Messenger:2012_SelEf}%
  \BibitemOpen
  \bibfield  {author} {\bibinfo {author} {\bibfnamefont {C.}~\bibnamefont
  {Messenger}}\ and\ \bibinfo {author} {\bibfnamefont {J.}~\bibnamefont
  {Veitch}},\ }\bibfield  {title} {\bibinfo {title} {{Avoiding selection bias
  in gravitational wave astronomy}},\ }\href
  {https://doi.org/10.1088/1367-2630/15/5/053027} {\bibfield  {journal}
  {\bibinfo  {journal} {New J. Phys.}\ }\textbf {\bibinfo {volume} {15}},\
  \bibinfo {pages} {053027} (\bibinfo {year} {2013})},\ \Eprint
  {https://arxiv.org/abs/1206.3461} {arXiv:1206.3461 [astro-ph.IM]}
  \BibitemShut {NoStop}%
\bibitem [{\citenamefont {Mandel}\ \emph {et~al.}(2019)\citenamefont {Mandel},
  \citenamefont {Farr},\ and\ \citenamefont {Gair}}]{Mandel:2018_SelEf}%
  \BibitemOpen
  \bibfield  {author} {\bibinfo {author} {\bibfnamefont {I.}~\bibnamefont
  {Mandel}}, \bibinfo {author} {\bibfnamefont {W.~M.}\ \bibnamefont {Farr}},\
  and\ \bibinfo {author} {\bibfnamefont {J.~R.}\ \bibnamefont {Gair}},\
  }\bibfield  {title} {\bibinfo {title} {{Extracting distribution parameters
  from multiple uncertain observations with selection biases}},\ }\href
  {https://doi.org/10.1093/mnras/stz896} {\bibfield  {journal} {\bibinfo
  {journal} {Mon. Not. Roy. Astron. Soc.}\ }\textbf {\bibinfo {volume} {486}},\
  \bibinfo {pages} {1086} (\bibinfo {year} {2019})},\ \Eprint
  {https://arxiv.org/abs/1809.02063} {arXiv:1809.02063 [physics.data-an]}
  \BibitemShut {NoStop}%
\bibitem [{\citenamefont {Vitale}\ \emph {et~al.}(2020)\citenamefont {Vitale},
  \citenamefont {Gerosa}, \citenamefont {Farr},\ and\ \citenamefont
  {Taylor}}]{Vitale:2020_selection_effects_review}%
  \BibitemOpen
  \bibfield  {author} {\bibinfo {author} {\bibfnamefont {S.}~\bibnamefont
  {Vitale}}, \bibinfo {author} {\bibfnamefont {D.}~\bibnamefont {Gerosa}},
  \bibinfo {author} {\bibfnamefont {W.~M.}\ \bibnamefont {Farr}},\ and\
  \bibinfo {author} {\bibfnamefont {S.~R.}\ \bibnamefont {Taylor}},\ }\bibfield
   {title} {\bibinfo {title} {{Inferring the properties of a population of
  compact binaries in presence of selection effects}},\ }\href@noop {}
  {\bibfield  {journal} {\bibinfo  {journal} {arXiv}\ } (\bibinfo {year}
  {2020})},\ \Eprint {https://arxiv.org/abs/2007.05579} {arXiv:2007.05579
  [astro-ph.IM]} \BibitemShut {NoStop}%
\bibitem [{\citenamefont {The~LIGO}\ and\ \citenamefont
  {collaborations}(2021)}]{GWTC-3_LVK_Cosmic_Expansion}%
  \BibitemOpen
  \bibfield  {author} {\bibinfo {author} {\bibfnamefont {V.}~\bibnamefont
  {The~LIGO}}\ and\ \bibinfo {author} {\bibfnamefont {K.}~\bibnamefont
  {collaborations}},\ }\bibfield  {title} {\bibinfo {title} {Data distribution
  of constraints on the cosmic expansion history from the gwtc-3},\ }\href
  {https://doi.org/10.5281/zenodo.5645777} {10.5281/zenodo.5645777} (\bibinfo
  {year} {2021})\BibitemShut {NoStop}%
\bibitem [{LIG(2025)}]{LIGOScientific:2025slb}%
  \BibitemOpen
  \bibfield  {title} {\bibinfo {title} {{GWTC-4.0: Updating the
  Gravitational-Wave Transient Catalog with Observations from the First Part of
  the Fourth LIGO-Virgo-KAGRA Observing Run}},\ }\href@noop {} {\bibfield
  {journal} {\bibinfo  {journal} {arXiv e-prints}\ } (\bibinfo {year}
  {2025})},\ \Eprint {https://arxiv.org/abs/2508.18082} {arXiv:2508.18082
  [gr-qc]} \BibitemShut {NoStop}%
\bibitem [{\citenamefont {Evans}\ \emph {et~al.}(2021)\citenamefont {Evans}
  \emph {et~al.}}]{Evans:2021gyd}%
  \BibitemOpen
  \bibfield  {author} {\bibinfo {author} {\bibfnamefont {M.}~\bibnamefont
  {Evans}} \emph {et~al.},\ }\bibfield  {title} {\bibinfo {title} {{A Horizon
  Study for Cosmic Explorer: Science, Observatories, and Community}},\
  }\href@noop {} {\bibfield  {journal} {\bibinfo  {journal} {arXiv e-prints}\ }
  (\bibinfo {year} {2021})},\ \Eprint {https://arxiv.org/abs/2109.09882}
  {arXiv:2109.09882 [astro-ph.IM]} \BibitemShut {NoStop}%
\bibitem [{\citenamefont {Maggiore}\ \emph {et~al.}(2020)\citenamefont
  {Maggiore} \emph {et~al.}}]{Maggiore:2019uih}%
  \BibitemOpen
  \bibfield  {author} {\bibinfo {author} {\bibfnamefont {M.}~\bibnamefont
  {Maggiore}} \emph {et~al.},\ }\bibfield  {title} {\bibinfo {title} {{Science
  Case for the Einstein Telescope}},\ }\href
  {https://doi.org/10.1088/1475-7516/2020/03/050} {\bibfield  {journal}
  {\bibinfo  {journal} {JCAP}\ }\textbf {\bibinfo {volume} {3}},\ \bibinfo
  {pages} {50}},\ \Eprint {https://arxiv.org/abs/1912.02622} {arXiv:1912.02622
  [astro-ph.CO]} \BibitemShut {NoStop}%
\bibitem [{\citenamefont {P\"urrer}\ and\ \citenamefont
  {Haster}(2020)}]{Purrer:2019jcp}%
  \BibitemOpen
  \bibfield  {author} {\bibinfo {author} {\bibfnamefont {M.}~\bibnamefont
  {P\"urrer}}\ and\ \bibinfo {author} {\bibfnamefont {C.-J.}\ \bibnamefont
  {Haster}},\ }\bibfield  {title} {\bibinfo {title} {{Gravitational waveform
  accuracy requirements for future ground-based detectors}},\ }\href
  {https://doi.org/10.1103/PhysRevResearch.2.023151} {\bibfield  {journal}
  {\bibinfo  {journal} {Phys. Rev. Res.}\ }\textbf {\bibinfo {volume} {2}},\
  \bibinfo {pages} {023151} (\bibinfo {year} {2020})},\ \Eprint
  {https://arxiv.org/abs/1912.10055} {arXiv:1912.10055 [gr-qc]} \BibitemShut
  {NoStop}%
\bibitem [{\citenamefont {Hassan}\ and\ \citenamefont
  {Rosen}(2012)}]{Hassan:2011zd}%
  \BibitemOpen
  \bibfield  {author} {\bibinfo {author} {\bibfnamefont {S.~F.}\ \bibnamefont
  {Hassan}}\ and\ \bibinfo {author} {\bibfnamefont {R.~A.}\ \bibnamefont
  {Rosen}},\ }\bibfield  {title} {\bibinfo {title} {{Bimetric Gravity from
  Ghost-free Massive Gravity}},\ }\href
  {https://doi.org/10.1007/JHEP02(2012)126} {\bibfield  {journal} {\bibinfo
  {journal} {JHEP}\ }\textbf {\bibinfo {volume} {2}},\ \bibinfo {pages}
  {126}},\ \Eprint {https://arxiv.org/abs/1109.3515} {arXiv:1109.3515 [hep-th]}
  \BibitemShut {NoStop}%
\end{thebibliography}%

\end{document}